%% file: arxiv.tex
\documentclass[twocolumn]{aastex701}

\input{commands}

\usepackage{amsmath}
\usepackage[thinc]{esdiff} 
\usepackage{xifthen}
\usepackage{txfonts}
\usepackage{xfrac}
\usepackage{enumitem}
\usepackage{hyperref}
\usepackage{makecell}
\usepackage[nolist]{acronym}

\begin{document}

\title{Constraints on Supermassive Black Hole Binaries from \ecgedit{the Lack of Resolvable Sources} in the NANOGrav 15-yr Dataset}

\input{authors_arxiv.tex}







\begin{abstract}
Pulsar timing arrays (PTAs) have found evidence of a gravitational wave background (GWB) consistent with a population of supermassive black hole binaries (SMBHBs), but have not yet resolved a continuous wave (CW) from an individual binary. 
We present the first SMBHB population constraints incorporating PTA upper limits on CWs, based on the NANOGrav 15-year (NG15) data.
Using semi-analytic models to generate simulated populations, we calculate each sample's GWB likelihood by free spectral comparison and CW nondetection (nCW) likelihoods by (1) mapping each realization's highest CW signal-to-noise ratio (S/N) to a detection probability and (2) counting the fraction of realizations without a binary close enough that NG15 would have detected it.
CW nondetection favors populations in which many lower-mass sources, rather than a few ultramassive ones, dominate the GWB.
This information is \ecgcirc{complementary} to that provided by the GWB amplitude, breaking the degeneracy between the masses and number of binaries required to reproduce the observed spectrum.
Relative to GWB-only constraints, the joint posteriors favor higher galaxy merger rates, lower $\mmbulge$ scatter, shorter binary lifetimes, and steeper small-separation hardening.
The most detectable sources in the S/N-weighted joint posterior have chirp masses $\sim 10^{9.6}\,\msol$, luminosity distances $\sim 1300\,\tr{Mpc}$, and frequencies $\sim 6\,\tr{nHz}$, while the nCW-only constraints favor lower frequencies, coinciding with the NG15 4\,nHz candidate. 
These constraints will tighten if CW nondetection continues.

\end{abstract}


\section{Introduction} \label{sec:intro}

Supermassive black holes (SMBHs) reside in the centers of galaxies \citep{Richstone+1998_SMBHs}, become binary systems when galaxies merge \citep{Lacey+1993_galaxymergers}, co-evolve with their host galaxies \citep{kh2013}, and transfer energy to their environments, possibly reaching mpc separations and emitting gravitational waves (GWs) \citep{BBR_1980, merritt+2005}. 
While this picture is well-motivated and formalized, open questions remain, including how many and how massive the SMBH binaries (SMBHBs) are throughout cosmic time and how they harden through their final $\sim$parsec of separation. 
Pulsar timing array (PTA) measurements and upper limits on gravitational wave signals offer a novel means of probing these questions \ecgedit{\citep{burkespolaor+2019, taylor2021}}.

If SMBHBs are able to reach mpc separations, they would emit GWs that drive their final inspiral until merger, with their stochastic superposition creating a nanohertz gravitational wave background (GWB). PTAs have found evidence in the correlations of pulsar time delays for a GWB \citep{ng+23_gwb, epta+23_gwb, ppta+23_gwb, cpta+23, mpta+2025} with a characteristic strain spectrum consistent with this SMBHB scenario \citep{ng+23_astro, epta+23_individuals}.

The number density of SMBHBs contributing to the GWB depends on the distribution of galaxies and their pair fraction \citep{chen+2019}. 
The galaxy stellar mass function (GSMF) is well-constrained at low redshifts ($z \lesssim 1$) and intermediate stellar masses ($10^9  \lesssim M_\star /\msol \lesssim 10^{11}$), where galaxy surveys provide mass-complete samples \citep{leja+2020, driver+2022}.  However, discrepancies remain in the low-number, high-mass tail, where previous Schechter mass functions may have \ecgedit{under-counted ultramassive ($M_\star \gtrsim 10^{11.5}\,\msol$) galaxies due to both cosmic variance and underestimated bright-end mass-to-light ratios} \citep{liepold+ma2024}. These discrepancies significantly impact the GWB, which is dominated by the most massive sources \citep{ng+23_astro}.

The galaxies' central SMBH masses are linked to their stellar masses and stellar velocity dispersions, indicating co-evolution between SMBHs and their host galaxies \citep{kh2013, mm2013}. 
These relations must be extrapolated from the local universe (within $\sim 100\,\tr{Mpc}$), where it is possible to calculate the black holes' masses from observations of their sphere of influence. 
There are discrepancies between the masses predicted via velocity dispersion and bulge mass, 
\ecgedit{
which grow at high redshifts \citep{matt+2023} and for disk-dominated, low-mass galaxies \citep{huber+2025}.
The relations may also break down during mergers themselves, with broad-line active galactic nuclei (AGN) close pairs hosting SMBHs more massive than standard $\mmbulge$ expectations \citep{huber+2026}.
}
The redshift evolution introduces further uncertainty in both the $\mmbulge$ relation \citep{matt+2026evol} and bulge fraction, which also relies on measurements of local galaxies \citep{Lang+2014, Bluck+2014}. 
Additionally, intrinsic scatter in the $\mmbulge$ relation can inflate the GWB through Eddington-type bias \citep{lauer+2007}, preferentially shifting SMBHs up in mass because low-mass galaxies are far more numerous \citep{matt+2026scatter}.

Not all the binaries created by galaxy mergers and co-evolved with their host galaxies will produce GWs. 
To do so, they must also decrease in separation, or harden, to about a mpc, at which point the binaries' orbits induce strong enough perturbations in spacetime for GWs to extract significant energy \citep{peters1964, BBR_1980}. 
The hardening mechanisms between galaxy-merger and GW emission have been studied theoretically, but their efficiencies remain uncertain and cannot be resolved by direct electromagnetic observations. 
First, the SMBHs sink to the center of their post-merger galaxies by dynamical friction, which can be calculated analytically for simplified stellar systems \citep{chandrasekhar1943} or by N-body simulations, which show discrepancies depending on the stellar and gaseous environments  \citep{boylankolchin+2008, ostriker1999}. Dynamical friction becomes inefficient at $\sim$pc separations, at which point individual stellar scattering events can be more efficient. However, these interactions are limited by the loss cone refill rate \citep{binney+tremaine1987}. To bridge the gap from dynamical friction to GWs within a Hubble time, some additional energy extraction is required, with leading candidates including viscous drag from a circumbinary disk \citep{siwek+2023} or triple interactions \citep{kelley+2017, bonetti+2018}. 
The GWB and single-source detection rate can help answer this ``final parsec problem'' \citep{mm2003_finalparsec} by constraining the population of binaries that must have successfully crossed the final parsec \citep{harris+2026}.


\citet[hereafter NG15astro]{ng+23_astro} used the GWB spectrum, as measured by \citet{ng+23_gwb}, in agreement with other international PTAs  \citep{epta+23_gwb, ppta+23_gwb, cpta+23, mpta+2025} to constrain the GSMF, the $\mmbulge$ relation, and a phenomenological model for binary hardening. They found that the GWB sits at the high end of astrophysical expectations, favoring some combination of a high galaxy number density or merger rates, SMBH masses above typical astrophysically motivated priors, and a rapid binary hardening, moving especially quickly through the viscous drag and stellar scattering regime. However, these parameters are strongly degenerate.

\ecgedit{
The amplitude can be raised through the \emph{mass channel} or the \emph{number channel}. 
The mass channel includes anything leading to more massive black holes, including increasing the $\mmbulge$ relation so that the same galaxies host larger black holes or increasing the GSMF characteristic mass so that there are more galaxies with the highest-mass black holes. 
The number channel includes anything increasing the overall number of binaries contributing to the GWB, either through an increased number of galaxies in the GSMF normalization, a larger pair fraction or shorter merger timescale increasing the galaxy merger rate, faster pre-GW hardening, or longer residence times within the PTA band. 
The GWB spectrum alone cannot separate these channels \citep{spzq2024_where}. 
\ecgcirc{%
Even with realistic simulated data and astrophysically informed priors, GWB-only inference on this model constrains the hardening parameters and GSMF characteristic mass while leaving the $\mmbulge$ parameters prior dominated \citep{laal+2026}.
}
\citet{gardiner+2024}, however, showed that information from individual sources can break these degeneracies, potentially tightening the constraints.
}

\ecgedit{An individual binary loud enough to be distinguished from the background would appear as a continuous gravitational wave (CW), a sinusoidal signal at roughly constant frequency over the observing baseline.
Thus far, no CW has been detected by PTAs. }
\citet[hereafter NG15cw]{ng+23_individuals} conducted a Bayesian search for an individual source by fitting a common-uncorrelated-red-noise (CURN) + CW model to the timing data. They found excess likelihood for the CURN+CW model over a CURN-only model at two frequencies, with both candidates being discounted: the one at 4 nHz became disfavored after incorporating spatial correlations into the likelihood, and the other at 170 nHz was found to be highly dependent on the noise models employed \ecgedit{\citep{ng+26_cnm}}.
\citet{ng+26_targeted} complemented this all-sky analysis with targeted searches toward 114 AGN flagged as SMBHB candidates by electromagnetic periodicity, fixing sky location, luminosity distance, and GW frequency to their EM values to tighten the strain upper limits by a median factor of $\sim2$ relative to the sky-averaged limits. 
Although two candidates retain marginal Bayes factors ($\sim2$--$3$), none yielded a detection.

This nondetection is consistent with the expectation that a CW would be observed $\gtrsim5$ years after the first GWB detection \citep{rosado+2015, kelley+2018}. 
However, \citet{gardiner+2025} showed that many SMBHB population scenarios consistent with the current GWB measurement should have also \ecgedit{produced a CW detection. CW nondetection therefore rules out scenarios that the GWB amplitude alone cannot distinguish} \citep{gardiner+2024, goncharov+2026}.
We leverage this by using \citetalias{ng+23_individuals} Bayesian upper limits on individual sources to fold CW nondetection (nCW) into SMBHB population inference for the first time.
Beyond population constraints, the properties of the most likely CW source—its mass, distance, and sky location—determine which galaxies are viable hosts and thus the prospects for multi-messenger follow-up once a CW is detected. Several works have predicted these properties for the first detectable source \citep{rosado+2015, gardiner+2024, gardiner+2025, houlden+2026}; we update those predictions using the joint GWB and CW upper-limit constraints.

\ecgedit{
The remainder of this paper is organized as follows.
In \secref{sec:methods} we describe our semi-analytic SMBHB populations,
the GWB and nCW likelihood methods, and our two sample libraries and their
priors. In \secref{sec:results} we present the nCW-only posteriors
(\secref{sec:results_cw_post}), the joint CW+GWB posteriors
(\secref{sec:results_joint_post}), the mass and number effects that drive
these constraints (\secref{sec:results_MN}), and the updated predictions
for the binary properties dominating the GWB and CW signals
(\secref{sec:results_props}). We discuss the implications for the
mass--number tension and prospects for future constraints in
\secref{sec:discussion}, and summarize our conclusions in
\secref{sec:conclusion}.}


\section{Methods} \label{sec:methods}
We construct simulated populations of supermassive black hole binaries (SMBHBs) from semi-analytic models, compute their gravitational-wave background (GWB) and continuous-wave (CW) signals, and constrain the model parameters using the NANOGrav 15-year (NG15) GWB free spectrum \citep{ng+23_gwb} and CW upper limits \citepalias{ng+23_individuals}. 

\subsection{SMBHB Populations}
\ecgedit{%
Our populations begin with a semi-analytic model that maps astrophysical parameters to a smooth number density of binaries across mass, mass ratio, redshift, and frequency. 
From each smooth distribution we draw discrete Poisson realizations, representing possible universes consistent with that parameter combination, and sum their strains to get
the GWB spectrum. 
Finally, we assign the loudest binaries sky locations and orientations to compute the S/Ns of each realization's CW candidates.
}

\subsubsection{Semi-Analytic Model}
\ecgedit{%
The semi-analytic model that we use to construct our SMBHB populations involves a (i) GSMF to set the initial distribution of host galaxies, (ii) galaxy pair fraction to determine which contain binaries, (iii) galaxy merger time to set the post-merger redshifts of the galaxies, (iv) bulge fraction that determines the bulge mass from the galaxy stellar mass, (v) relation between black hole mass and bulge mass ($\mmbulge$) that determines the masses of the black hole binaries residing in each galaxy, and (vi) hardening model that determines the final redshifts at which each binary reaches the PTA frequencies. 
There are six free parameters subject to observational constraints: two in the GSMF, two in the $\mmbulge$ relation, and two in the hardening model.
Uncertainties in the other model components are absorbed by these six free parameters.
}
The following paragraphs detail the free parameters we aim to constrain, while a detailed description of the full model can be found in \citetalias{ng+23_astro}.

\paragraph{\bf GSMF} The free parameters of the GSMF, which takes a Schechter form following \citet{chen+2019}, are its present-day normalization, $\psi_0$, which gives the overall normalization as a phenomenological function of redshift,
\begin{equation}
    \label{eq:gsmf_norm}
    \log_{10}\lr{\gsmfNormTot / \tr{Mpc}^{-3}} =  \psi_0 + \psi_z \cdot z, 
\end{equation}
and present-day characteristic mass, which similarly determines the overall characteristic mass where the number density drops off,
\begin{equation}
    \label{eq:gsmf_mass}
    \log_{10}\lr{\gsmfMassTot / \msol} =  m_{\psi,0} + m_{\psi,z} \cdot z.
\end{equation}
The normalization and characteristic mass, in combination with the low-mass-end slope, 
 \begin{equation}
 \label{eq:gsmf_slope}
    \alpha_\psi = 1 + \alpha_{\psi,0} + \alpha_{\psi,z} \cdot z
 \end{equation}
determine the initial distribution of galaxies as the total number density per decade of stellar mass $\mstar$, following the single Schechter function,
\begin{equation}
  \label{eq:gsmf_schechter}
    \gsmffunc(\mstar, z) = \ln(10)\gsmfNormTot\cdot \lrs{\frac{\mstar}{\gsmfMassTot} }^{\alpha_\psi} \exp \lr{-\frac{\mstar}{\gsmfMassTot} }.
\end{equation}
While $\psi_0$ and $m_{\psi,0}$ vary from $-3.5$ to $-1.5$ and 10.5 to 12.5, respectively, the remaining variables are fixed to the values listed in \citetalias{ng+23_astro} Table B1.

\paragraph{\bf $\mathbf{\mmbulge}$ Relation} The free parameters describing the $\mmbulge$ relation are the dimensionless mass amplitude $\mmbamp$ and the scatter $\mmbscatter$. 
\ecgedit{%
The bulge mass is calculated from the galaxy stellar mass assuming a bulge fraction of 0.615, based on empirical observations from \citet{Lang+2014} and \citet{Bluck+2014}. The black hole mass then follows from the bulge mass via a power law with index $\mmbplaw$
}
and random scatter drawn from a normal distribution of standard deviation $\mmbscatter$,
\begin{equation}
\label{eq:mmbulge_relation}
\log_{10}\! \lr{\mbh/\msol} = \mmbamp + \mmbplaw \log_{10}\!\scale{\mbulge}{10^{11} \, \msol} + \mathcal{N}\lr{0, \mmbscatter}.   
\end{equation}
$\mmbplaw$ is fixed to 1.10, $\mmbamp$ varies from 7.6 to 9.0, and $\mmbscatter$ varies from 0.0 to 0.9 dex.

\paragraph{\bf Hardening} The free parameters in the phenom hardening model are total binary lifetime $\thardf$ and small-separation hardening index $\hardnuinner$. These describe a phenomenological hardening component that encapsulates all non-GW sources of hardening, such as dynamical friction, stellar scattering, viscous drag, and triple interactions. These environmental effects are well described by a double power-law \citep{kelley+2017},
\begin{equation}
\label{eq:hard_phenom}
    \frac{d a}{d t}\bigg|_\tr{phenom} = \harddadtnorm \cdot \lr[1-\hardnuinner]{\frac{a}{\hardrchar}} \cdot \lr[\hardnuinner-\hardnuouter]{1 + \frac{a}{\hardrchar}}
\end{equation}
where $\hardnuinner$ sets the hardening rate in the ``inner'' regime, and $\hardnuouter$ sets the hardening rate in the ``outer'' regime, separated by a characteristic separation $\hardrchar=100\,\tr{pc}$. 
The phenom component is normalized by the constant $\harddadtnorm<0$ such that the total binary lifetime 
\ecgedit{
$\thardf$, from initial separation $\hardainit=10^{4}\,\tr{pc}$ to the innermost stable circular orbit $\hardaisco$, is
}
\begin{equation}
    \thardf = \int_{\hardainit}^{\hardaisco} \left(\frac{da}{dt}\bigg|_\tr{phenom} + \frac{da}{dt}\bigg|_\tr{gw} \right)^{-1}da,
\end{equation}
where the overall hardening rate also includes a GW component determined by the binaries' mass and separation,
\begin{equation}
    \label{eq:gw_hard}
    \frac{d a}{d t}\bigg|_\tr{gw} = -\frac{64 \, G^3}{5 \, c^5} \frac{m_1 \, m_2 \, M}{a^3}. 
\end{equation}
We vary $\thardf$ from 0.1 to 11.0 Gyr and $\hardnuinner$ from $-1.5$ to 0.0. While $\hardnuouter=+2.5$ is fixed, the large-separation hardening is sped up by slowing the inner regime, and vice versa, for fixed binary lifetime 
\ecgedit{
\citep[see][for further discussion about the hardening parameterization and alternative functional forms]{blecha2026}.
}

\begin{deluxetable}{ccc}
    \tablecaption{Uniform distributions of each free parameter.}
    \label{tab:free_params}
    \tablehead{
    \colhead{Parameter} 
    & \colhead{Symbol} 
    & \colhead{Prior Range}    }
    \startdata
         GSMF Normalization & $\gsmfnorm$ & $\uniform{-3.5}{-1.5} $ \\
         GSMF Characteristic Mass & $\gsmfmass$ & $\uniform{10.5}{12.5} $ \\
         $\mmbulge$ Amplitude & $\mmbamp$ & $\uniform{7.6}{ 9.0}$ \\
         $\mmbulge$ Scatter & $\mmbscatter$ & $\uniform{0.0}{ 0.9}\,\tr{dex}$ \\
         Phenom Binary Lifetime & $\thardf$ & $\uniform{0.1}{11.0}\,\tr{Gyr}$ \\
         Phenom Small-Separation Index & $\hardnuinner$ & $\uniform{-1.5}{0.0}$ \\
        \enddata
\end{deluxetable}

\subsubsection{Discrete Realizations}
We refer to a ``sample'' as a single selection of parameter values, which determines the comoving volumetric number density of SMBHBs $\frac{\partial^3 \eta}{\partial M \partial q \partial z}$ and thus the differential number of SMBHBs per total mass $M$, mass ratio $q$, redshift $z$ (at the time of GW emission), and log rest-frame orbital frequency $\ln f_p$ \citep{sesana+2008}: 
\begin{equation}
    \label{eq:number_density_to_number_frequency}
    \diffp{N}{{M} {q} {z} {\ln f_p}} = \diffp{\ndens}{{M} {q} {z}} \diffp{t}{{\ln f_p}} \diffp{z}{{t}} \diffp{V_c}{{z}}. 
\end{equation}
\ecgedit{Integrating the differential number over each bin gives that bin's expectation value,
\begin{equation}
    \label{eq:expectation}
    \langle N(M,q,z,f)\rangle 
    = \int_\tr{bin} \diffp{N}{{M'}{q'}{z'}{\ln f_p'}} 
    \, dM' \, dq' \, dz' \, d\ln f_p', 
\end{equation}
where the bins in rest-frame orbital frequency correspond to observed GW frequencies $f=2 f_p/(1+z)$ (assuming circular orbits, so the GW frequency is twice the orbital frequency).
This integral is evaluated with the trapezoid rule in $M$, $q$, and $z$, and a left-edge Riemann sum in $\ln f_p$, so that the 
frequency bins match the PTA observing band rather than collapsing to bin centers.
}
Then, to generate a population of binaries that could exist in a single universe, we draw the discrete number of binaries in each $(M,q,z,f)$ bin from a Poisson distribution with mean equal to the expectation value,
\begin{equation}
    \label{eq:number_sampling}
    N(M,\!q,\!z,\!f) = 
     \mathcal{P}\,\big( \langle N(M,\!q,\!z,\!f) \rangle \big).
\end{equation}
Each Poisson draw (over all $M,q,z,f$ bins) represents one ``realization'' of the sample. 

\ecgedit{Each binary emits a sky- and polarization-averaged GW strain amplitude, assuming circular orbits \citep{Finn+2000}
\begin{equation}
    \label{eq:strain_amp}
    \hscirc^2(f_p) = \frac{32}{5 c^8} \,  \frac{\left(G\mchirp\right)^{10/3}}{\distcom^2} \lr[4/3]{2\pi f_p},
\end{equation} 
where $\mchirp \equiv M q^{3/5}/(1+q)^{6/5}$ is the chirp mass and $\distcom$ the comoving distance.
Summing the strain from every binary in a realization gives the characteristic strain of the GWB, 
\begin{equation}
\label{eq:char_strain_bg}
    \hc^2(f) = \sum_{M,q,z} N(M,q,z,f) \hscirc^2(f_p) \frac{ f}{\Delta f}.
\end{equation}
}

\subsubsection{Single Sources}
To analyze the single source detectability, we generate $N_\tr{skies}$ random ``skies'' for each realization. In a given ``sky'' we assign the loudest source at each frequency a colatitude $\theta$, right ascension $\phi$, inclination $\iota$, polarization $\psi$, and initial phase $\Phi_0$, all of which are drawn from 
\ecgedit{
isotropic distributions, uniform in $\cos \theta$, $\phi$, $\cos \iota$, $\psi$, and $\Phi_0$.
} 
These parameters, in combination with the chirp mass, strain amplitude, and frequency, determine the signal, $\signal$, defined as in \citetalias{ng+23_individuals}. Then their signal-to-noise ratios (S/Ns) are \citep{ellis+2012snr}
\begin{equation}
    \label{eq:snr}
    \snr  = \sqrt{\innerprod{\signal}{\signal}}
\end{equation}
where the parenthetical notation refers to the noise-weighted inner product
    \begin{equation}
    \label{eq:innerprod}
        ( a| b) = a^T \mathbf{C}^{-1} b
    \end{equation}
for the noise covariance matrix $\mathbf{C}$, as calculated in \citet{becsy+2022_quickcw}, \ecgcirc{where the GWB is treated as CURN such that $\mathbf{C}$ is block-diagonal across pulsars.} 
All sources except the one in question are included as noise sources in $\mathbf{C}$, in addition to pulsar red noise and measurement noise matching the NG15 timing models \citep{ng+23_timing}. 
This S/N calculation corresponds to the ``optimal S/N'', $\sigsig$, in \citet{gardiner+2025}. Although the other detection statistics presented in that work better correlate with a Bayesian analysis of an individual simulated dataset by incorporating information about the noise in the data, we do not require the true detectability for an individual realization when making model constraints. Instead, we select the quickest calculation that represents the typical detectability of a CW under a given model.

\subsection{Constraints}

In order to calculate new constraints on the free parameters, we need some priors, likelihoods, and means of combining them to get the posteriors. The GWB likelihood matches that of \citetalias{ng+23_astro}, described in \secref{sec:methods_gwb_like}. We examine two methods of nCW likelihood calculation, our fiducial S/N approach in \secref{sec:methods_cw_sn} and an additional luminosity distance approach in \secref{sec:methods_cw_dl}. These likelihoods are evaluated directly at gridpoints sampled from two sets of priors: the uniform library and the chain library, described in \secref{sec:methods_priors}. Each library has an appropriate means of weighting its distribution by the likelihoods to calculate posteriors, following the approach in \citetalias{ng+23_astro} Appendix C, \ecgedit{obviating the need for} any Monte Carlo sampling.


\subsubsection{GWB Likelihood} \label{sec:methods_gwb_like}
The GWB likelihood is the probability of the observed times of arrival data, $\data$, given a sample with holodeck parameters, $\Theta$. As in \citetalias{ng+23_astro}, we use \texttt{ceffyl} \citep{lamb2023rapid} posteriors on the free spectrum, $\log_{10} \rho_i$ as an intermediate data product, where $\rho_i^2 = \Phi(f_i)/T_\mathrm{obs}$ 
is the power in time residuals at frequency $f_i$, instead of directly fitting a model to the TOAs. Marginalizing over \texttt{ceffyl} posteriors, and multiplying over the first $N_f=5$ frequency bin components, gives the total marginalized GWB likelihood 
\begin{equation}
    \bglike(\data | \Theta) \propto \prod_{i=1}^{N_f} \int d(\log_{10} \rho_i) p(\log_{10} \rho_i | \data)p(\log_{10} \rho_i | \Theta).
\end{equation}
\ecgcirc{%
To calculate the model-predicted probability distribution, we assume the power is log-normally distributed, i.e., Gaussian in $\log_{10}\rho_i$ with mean and standard deviation set by the median and scatter of $\log_{10}\rho_i$ over many realizations of a given sample.
This form neglects the non-Gaussian tail of the underlying characteristic strain distribution \citep{lamb+2026}, but analytic treatments find the approximation sufficient for population-level inference at current PTA sensitivity \citep{spz2025_distribution, xue+2025, iemoto+2026}.
}

\ecgcirc{This likelihood is used implicitly in our chain library by drawing samples from the \citetalias{ng+23_astro} Markov Chain Monte Carlo (MCMC) chain, as described in \secref{sec:methods_priors}. We select this approach for direct comparison to the GWB constraints published in \citetalias{ng+23_astro}.}

\subsubsection{nCW--S/N Likelihood} \label{sec:methods_cw_sn}

\begin{figure}
    \centering
    \includegraphics[width=\linewidth]{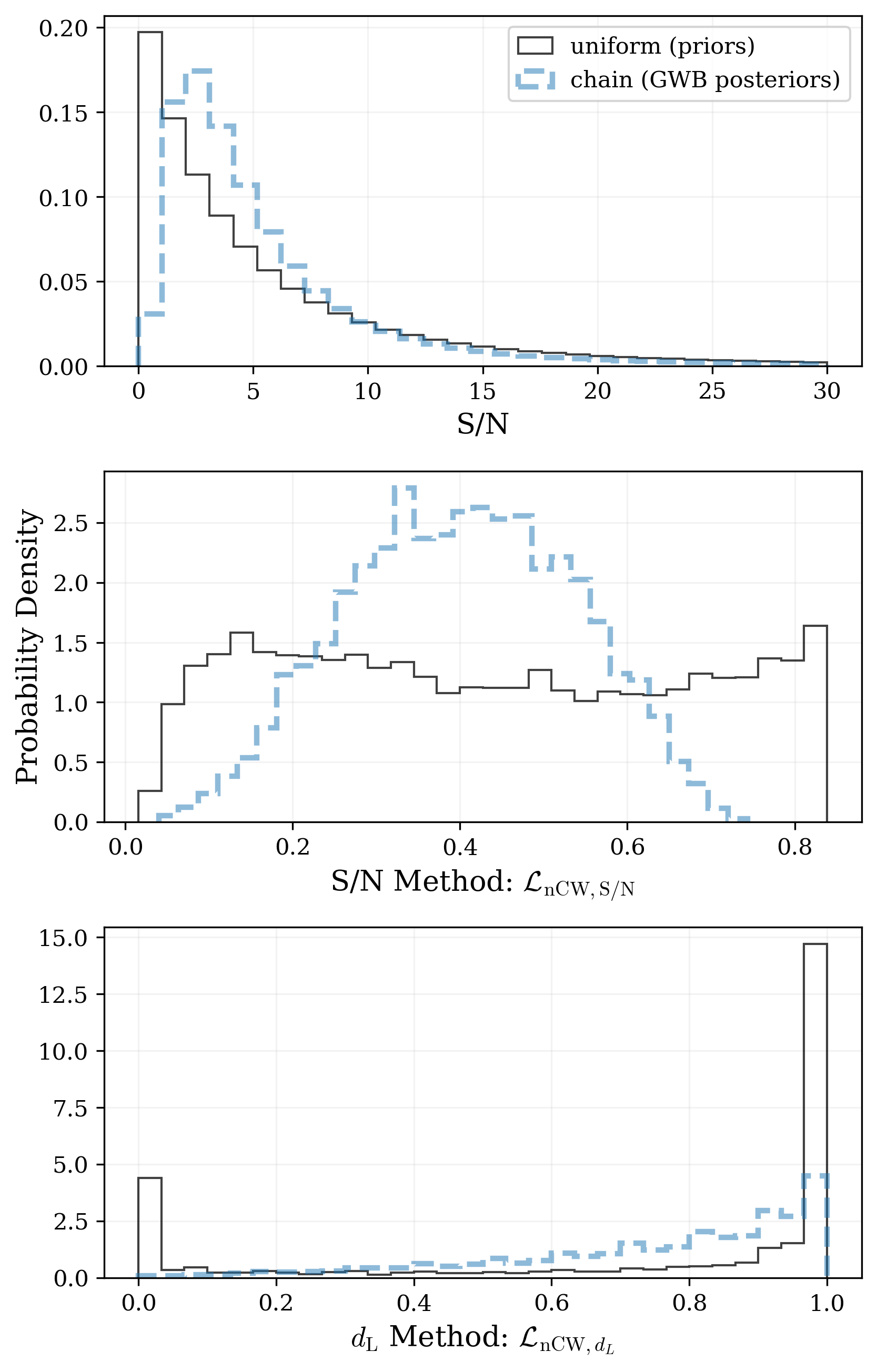}
    \caption{
    \ecgedit{
    Maximum CW S/N (top panel), corresponding $\snlike$ distributions (middle panel), and $\dllike$ distributions (bottom panel) for the uniform prior (solid black) and GWB posterior chain (dashed blue) libraries. 
    The maximum S/Ns represent the most detectable source of each realization and sky assignment, and then the corresponding probabilities are averaged across realizations to get sample likelihoods. 
    The uniform library contains more samples with low S/Ns because these low S/Ns often correspond to low GWB amplitudes, disfavored by the GWB posteriors. This corresponds to a bimodal likelihood distribution in $\snlike$ due to the sigmoid nature of the S/N--detection probability mapping. 
    While the chain library has fewer low S/Ns ($\lesssim1$), it also has fewer very high ($\gtrsim12$) S/N samples since these high S/Ns tend to result from the \ecgedit{extremely low GWBs that provide little noise competition,} or extremely large black hole numbers/masses that overestimate both the GWB and CW amplitudes. Thus the chain library likelihoods are concentrated at more intermediate values.
    The luminosity distance likelihoods also have a bimodal uniform library distribution, and fewer low-likelihood scenarios in the chain library, but both are skewed towards higher likelihoods than $\snlike$.
    }
    }
    \label{fig:snr_likes}
\end{figure}

The first nCW likelihood method uses the \citetalias{ng+23_individuals} global Bayes factor as our data.
\ecgedit{%
For a single realization $i$ and sky placement $j$, we need the probability of \textit{not} detecting a CW above this Bayes factor threshold, $\bfthresh$. 
Following \citet{gardiner+2025}, we map the highest CW S/N in each realization and sky to a detection probability via the detection fraction $\tr{DF}(\snr,\bfthresh)$: the fraction of cases, among many simulated datasets containing a source of the given S/N, that yield a Bayes factor above $\bfthresh$. 
The probability of nondetection is then
\begin{equation} \label{eq:cw_prob_real}
    p(\bfthresh|\Theta_{i,j}) = 1 - \tr{DF}(\snr_{i,j}, \bfthresh).
\end{equation}
}

\ecgedit{%
With a nondetection probability for every $\Theta_{i,j}$, we take the mean over realizations and skies to get the sample likelihood,
\begin{equation} \label{eq:cw_prob_mean}
    \snlike(\bfthresh | \Theta) = \frac{1}{N_\tr{skies} N_\tr{reals}} \sum_{i=1}^{N_\tr{reals}} \sum_{j=1}^{N_\tr{skies}} \left( 1 - \tr{DF}(\snr_{i,j}, \bfthresh) \right).
\end{equation}
This average is the fraction of possible universes consistent with model $\Theta$ in which no CW would have been detected above the \citetalias{ng+23_individuals} Bayes factor threshold, averaged over Poisson realizations and source orientations---that is, the probability of the observed nondetection given the model.
}

We calibrate our detection fractions using the same CW detection pipeline as in \citetalias{ng+23_individuals}, \texttt{QuickCW} \citep{becsy+2022_quickcw}, which performs a Markov Chain Monte Carlo (MCMC) fit of a CW+CURN model to the timing data. \texttt{QuickCW} is faster than traditional pipelines like \texttt{enterprise} \citep{enterprise} due to a custom likelihood method that pre-computes inner products for each set of ``shape'' parameters (those involved in the noise covariance matrices) to rapidly explore the remaining ``projection'' parameters, like sky position, inclination, and initial phase. However, it is still too slow to run on the 50 million simulated realization-sky combinations necessary for model constraints, necessitating the S/N--DF mapping first described in \citet{gardiner+2025}. 

The mapping between S/N and DF follows a Gaussian cumulative distribution function \ecgedit{of the form 
\begin{equation} \label{eq:gcdf}
    \tr{DF}(\snr, \bfthresh) = \frac{1}{2} \left[ 1+ \tr{erf} \left( \frac{m \cdot \snr - \log_{10} \bfthresh}{\sigma \sqrt{2}} \right) \right],
\end{equation}
}%
where the slope $m$ and width $\sigma$ are calibrated on $\sim2000$ simulated datasets and their \texttt{QuickCW} Bayes factors. 
We use a Bayes factor threshold of $\bfthresh = 1.3282$, which is the marginal likelihood ratio between the CW+CURN and CURN models in \citetalias{ng+23_individuals}, marginalized over all frequencies up to 30 nHz. 
We exclude frequencies above 30 nHz because even when there is some statistical preference for the CW model, the astrophysical prior odds for such a high-frequency CW detection are negligible.
The fit yields $m=0.03468$ and $\sigma=0.1242$, with a reduced $\chi^2=0.912$ and a 50\% detection fraction at $\snr_{50}=3.554$. 
This maps the S/N distributions in the top panel of \figref{fig:snr_likes} to the $\snlike$ distributions in the middle panel, for the uniform (solid black) and chain (dashed blue) libraries, described in \secref{sec:methods_priors}.

\subsubsection{nCW--$d_\tr{L}$ Likelihood} \label{sec:methods_cw_dl}
The second nCW likelihood uses information from the effective luminosity distance radius $R_{\text{eff}}$ which describes the frequency and chirp mass dependent single source exclusion volume at 95\% credible level (detailed in \citetalias{ng+23_individuals}) as a lower bound on the distance of detectability of CW strains. We then infer parameter likelihoods proportional to the sum:
\begin{equation}
    \dllike (\data| \Theta) \propto \frac{1}{N_{\rm reals}}\sum^{N_{\rm reals}}_{i=1}\left(1 - \rm{DP}_i \right),
    \label{eq:cw_prob_dl}
\end{equation}
where the detection probability (DP) in a given realization is a binary value:
\begin{equation}
    \rm{DP}_i =
    \begin{cases}
        1 & \text{if there exists a binary with } d_L < R_{\rm eff} (\mathcal{M}_c, f) \\
        0 & \text{otherwise}.
    \end{cases}
\end{equation}
If a given realization contains an SMBHB located at a luminosity distance closer than the upper threshold set by \citetalias{ng+23_individuals}, it counts as a detection and contributes 0 to the sum. Otherwise, it contributes 1. 
This contribution is averaged over the number of realizations, providing a term proportional to the likelihood. 

We opted to use $R_\text{eff}$, instead of the more commonly used upper limits on the strain amplitude, as those are dominated by the least sensitive part of the sky and not a realistic measurement of detectability. 
Using $R_\text{eff}$ circumvents this issue by effectively volume-averaging the sensitivity across the sky, naturally giving more weight to more sensitive areas \citepalias{ng+23_individuals}.
\ecgedit{%
The resulting $\dllike$ distributions for both sample libraries are shown
in the bottom panel of \figref{fig:snr_likes}.
}

\subsubsection{Priors and Posteriors} \label{sec:methods_priors}

\begin{figure}
    \centering
    \includegraphics[width=\linewidth]{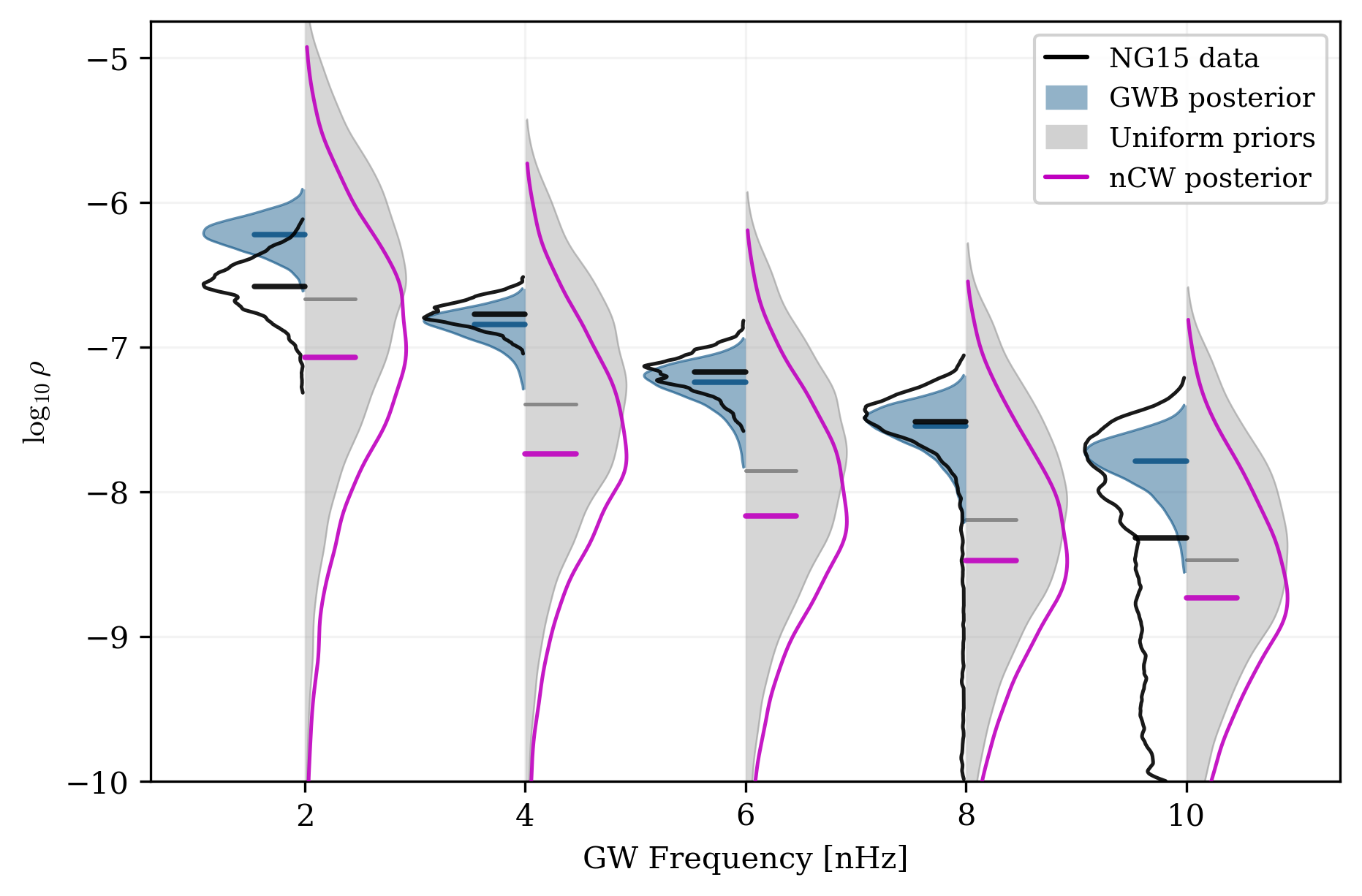}
    \caption{
    \ecgedit{
    Gravitational wave background, \ecgcirc{represented by the square root of the timing-residual power $\rho$,} of the NG15 data (left-hand violins outlined in black), the uniform priors (right-hand violins shaded gray), the GWB posteriors (left-hand violins shaded blue), and the nCW posteriors (right-hand violins outlined in pink). Solid horizontal lines mark the medians of each distribution. The NG15 data refers to the ceffyl KDE representation of the DMGP free spectral fit (see \citet{ng+23_gwb}). The GWB posteriors make up the chain library of samples. They are close to the data, but overshoot the lowest frequency bin amplitude. The uniform priors serve as the uniform library of samples, and weighting them by their S/N-method nCW likelihoods gives the nCW posteriors. The weak nCW constraints leave the nCW-only posteriors only slightly lower in amplitude than the uniform priors.
    }
    }
    \label{fig:violins}
\end{figure}

We generate two sets of samples, referred to as the \textit{uniform} library and \textit{chain} library. Each contains $N_\tr{samps}=10,000$ samples of varying free parameters, $N_\tr{reals}=100$ realizations per sample, and $N_\tr{skies}=50$ per realization. 
\ecgedit{%
\figref{fig:violins} shows the free spectra of each library marginalized over all samples, with the chain library shaded in blue in the left-hand violins and the uniform library shaded in gray on the right-hand violins. 
The free spectral fit to the NG15 data, using the DMGP model described in \citet{ng+23_gwb}, is outlined in black on the left-hand sides, and the CW-weighted uniform library is outlined in pink on the right-hand sides. 
}

The uniform library is necessary to isolate the GWB and nCW constraints before combining them into a joint posterior. The samples are drawn from the uniform distributions listed in \tabref{tab:free_params} with Latin hypercube sampling, so that they optimally explore the six-dimensional parameter space for the given number of samples. Because the samples are uniformly distributed in each parameter, the posteriors are just proportional to the likelihoods,
\begin{eqnarray} \label{eq:uniform_posteriors}
    \pi(\Theta | \data)_\gwb &\propto& \gwblike \\
    \pi(\Theta | \data)_\tr{CW} &\propto& \cwlike \label{eq:uniform_cw} \\
    \pi(\Theta | \data)_\tr{joint} &\propto& \gwblike \cdot \cwlike \label{eq:uniform_joint}.   
\end{eqnarray}
Either $\snlike$ or $\dllike$ can be used for the nCW likelihood in \refeq{eq:uniform_cw}. 

\ecgedit{
Although both nCW and GWB likelihoods depend on each realization's background, realization-by-realization fluctuations are already captured by the population standard deviation in $\bglike$, so that the variation in $\bglike$ between realizations is smooth.
Therefore, we can treat the GWB likelihood as constant across realizations of a given sample and factor it out, treating the nCW and GWB likelihoods as approximately independent.
Having verified that the fully coupled per-realization calculation gives negligibly different results from our factorized likelihood approach, we multiply the nCW and GWB likelihoods for each sample to obtain the joint posteriors in \refeq{eq:uniform_joint}. 
}

\ecgedit{%
While the uniform library is the only dataset that allows for nCW-only constraints, and shows a demonstrable shift from the priors in the free spectra in \figref{fig:violins} and posteriors in \secref{sec:results_cw_post}, the chain library offers better sampling for the GWB and joint posteriors. 
We draw 10,000 samples from the \citetalias{ng+23_astro} MCMC posterior chain, which yields smoother GWB contours than a one-time likelihood calculation at the Latin hypercube gridpoints.
The histogram of these samples, already concentrated in regions of highest GWB likelihoods, gives the GWB-only posteriors. 
These posteriors are generally close to the NG15 data in \figref{fig:violins}, overshooting the first frequency bin because it is difficult to recover both high amplitudes and low-frequency flattening with the same astrophysical model. }

\ecgedit{
We can weight the GWB chain samples by their nCW likelihoods to get the joint posteriors. This reweighting minimally decreases the effective sample size because the nCW constraints are weaker than the GWB ones, demonstrated by the GWB posteriors on the free spectrum being much more concentrated than the nCW posteriors. This creates a slight shift from GWB to joint posteriors (see \secref{sec:results_joint_post}), which still benefits from the focus on high-likelihood regions.} An nCW-only constraint is not possible with the GWB chain due to low effective sample sizes in regions with high nCW likelihood but low GWB likelihood. 


\ecgedit{%
The two libraries also differ in their CW detectability.
The uniform library distribution of S/N in \figref{fig:snr_likes} peaks low, at $\snr \sim 1.0$. It maps to a bimodal distribution of likelihoods due to the sigmoid nature of Bayesian detection probability: most samples exist clearly in the detection or nondetection regimes for our broad distribution of models. 
}While the two nCW likelihood methods produce different distributions, they both show this bimodality. When concentrating the samples towards areas of high GWB likelihood in the chain library, the S/N distribution shifts higher, peaking around 2--3 instead of $<1$, and the low-likelihood spike disappears because samples that produce a loud GWB tend to also have better CW candidates.



\section{Results} \label{sec:results}
This work presents the first nCW-only and joint CW+GWB constraints on GSMF, $\mmbulge$ relation, and binary evolution models based on the NANOGrav 15-year GWB free spectrum and Bayesian upper limit on individual sources. 
The nCW-only posteriors are based on source S/Ns (\refeq{eq:cw_prob_mean}) or luminosity distances (\refeq{eq:cw_prob_dl}) for a uniformly distributed library of samples, presented and compared to the GWB posteriors in \secref{sec:results_cw_post}.
Our joint posteriors are calculated using the GWB chain as the priors, reweighted by the two nCW likelihood methods, presented in \secref{sec:results_joint_post}. 
We investigate the physical explanations for these constraints in terms of the typical binary numbers and masses in \secref{sec:results_MN}. 
Finally, we show the distributions of binary properties dominating the background and single sources in \secref{sec:results_props}.

\subsection{CW-Only Posteriors} \label{sec:results_cw_post}

\begin{figure*}
    \centering
    \includegraphics[width=\linewidth]{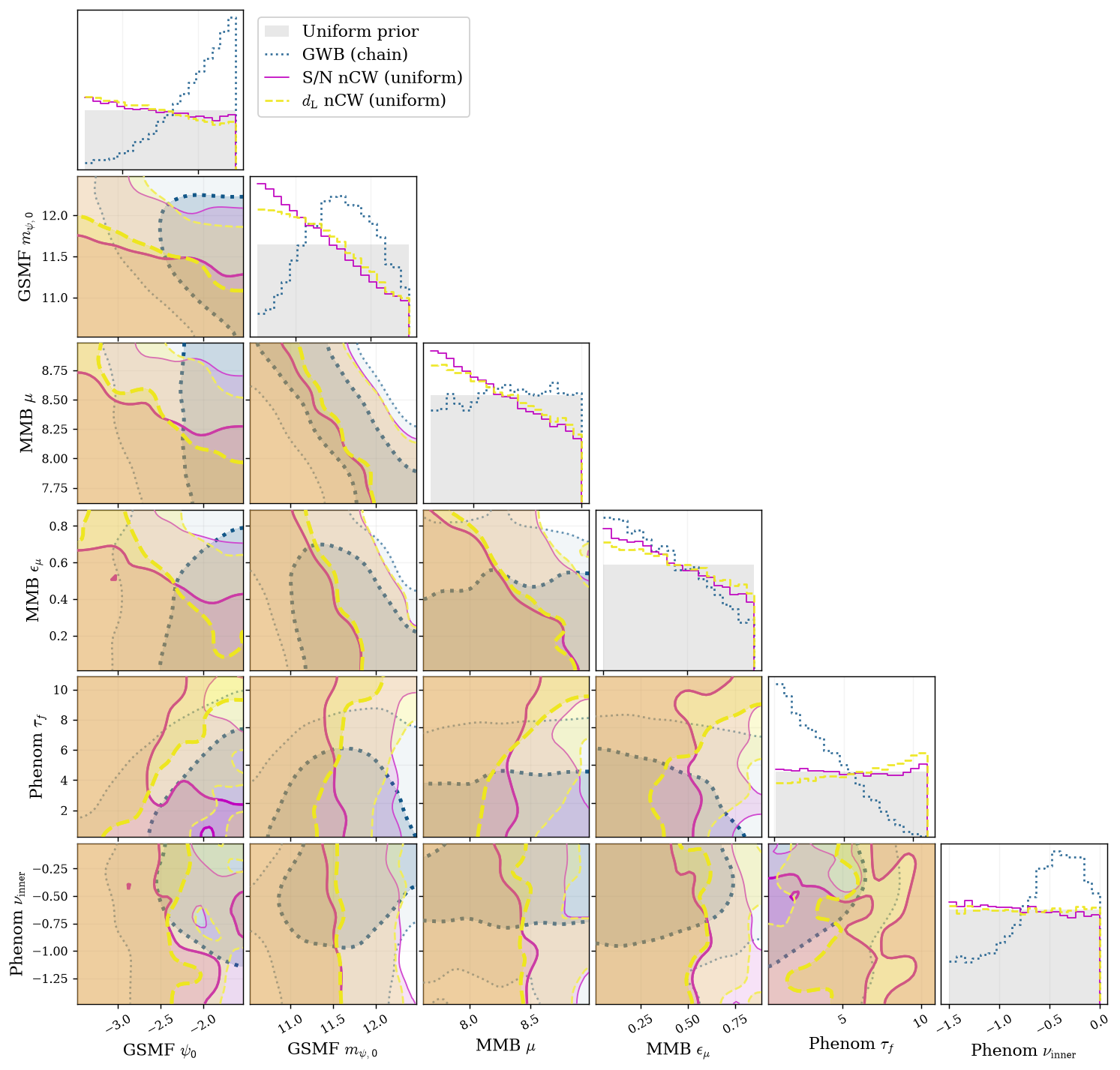}
    \caption{Corner plot comparing nCW-only posteriors to GWB-only posteriors \ecgcirc{on the model parameters summarized in \tabref{tab:free_params}, assuming uniform priors.}  The nCW-only posteriors by the S/N method are represented by solid magenta lines and the $\dlum$ method by dashed yellow lines. The GWB-only posteriors are marked by blue dotted lines \ecgcirc{and the uniform priors are shaded gray}. 
    \ecgedit{The 2D contours enclose the 68\% and 95\% credible regions of the Gaussian-smoothed weighted histograms (kernel width $7.5\%$ of each parameter range, which inflates the credible-region areas by $\leq2.5\%$).
    } 
    The nondetection of CWs favors cases that limit opportunities for high-mass outliers: low GSMF normalization and characteristic mass, low $\mmbulge$ amplitude and scatter. 
    The hardening parameter impacts are more subtle, and distinct between methods, because they impact CW detectability by shifting when and how many sources reach the PTA band. 
    \ecgedit{
    When the hardening time is shorter ($\thardf \lesssim 5\,\tr{Gyr}$), CWs have a loud GWB to contend with, while at very long hardening times ($\thardf \gtrsim 9\,\tr{Gyr}$), even though GWB noise is low, it is unlikely for any CW to rise above pulsar and measurement noise. 
    Thus the $\snlike$ marginal dips to a shallow minimum near $\sim6$--$8\,\tr{Gyr}$ and rises toward both prior edges, more steeply at \ecgcirc{long} lifetimes. 
    The 1D marginal understates this structure because it averages over $\hardnuinner$. In combination, the two hardening parameters create a boomerang-shaped zone of avoidance visible in the magenta 2D contours, explored in \secref{sec:results_MN} and \figref{fig:hard_contours}.
    When the low GWB is not incorporated via an S/N, the long hardening times become even more favored by the $\dlum$ method. 
    In the mass-sensitive pairs, e.g., $\mmbamp$ versus $\gsmfmass$, the nCW exclusion boundary runs parallel to the diagonal GWB degeneracy ridge, placing an upper limit on the shared mass combination.
    In the pairs combining a mass and number parameter, e.g., $\mmbamp$ versus $\gsmfnorm$, the one-sided nCW bounds run perpendicular to the one-sided GWB bounds, providing \ecgcirc{complementary} information with a localized region of overlap.
    }
    }
    \label{fig:cw_corner}
\end{figure*}

The Bayesian upper limits on individual CW sources in \citetalias{ng+23_individuals} offer new constraints on SMBHB populations, distinct from GWB-based constraints. We find these constraints using two different likelihood methods: $\snlike$, which maps CW S/Ns to a probability of not having a Bayes factor exceeding that of \citetalias{ng+23_individuals}, and $\dllike$, which uses the fraction of realizations without binaries inside the exclusion volume $R_\tr{eff}$ around Earth. The posteriors on our six free parameters (GSMF normalization $\gsmfnorm$, GSMF characteristic mass $\gsmfmass$, $\mmbulge$ amplitude $\mmbamp$, $\mmbulge$ scatter $\mmbscatter$, binary lifetime $\thardf$, and small-separation hardening index $\hardnuinner$) by the S/N method in magenta and the $\dlum$ method in yellow are presented in \figref{fig:cw_corner}.

\begin{figure}
    \centering
    \includegraphics[width=0.99\linewidth]{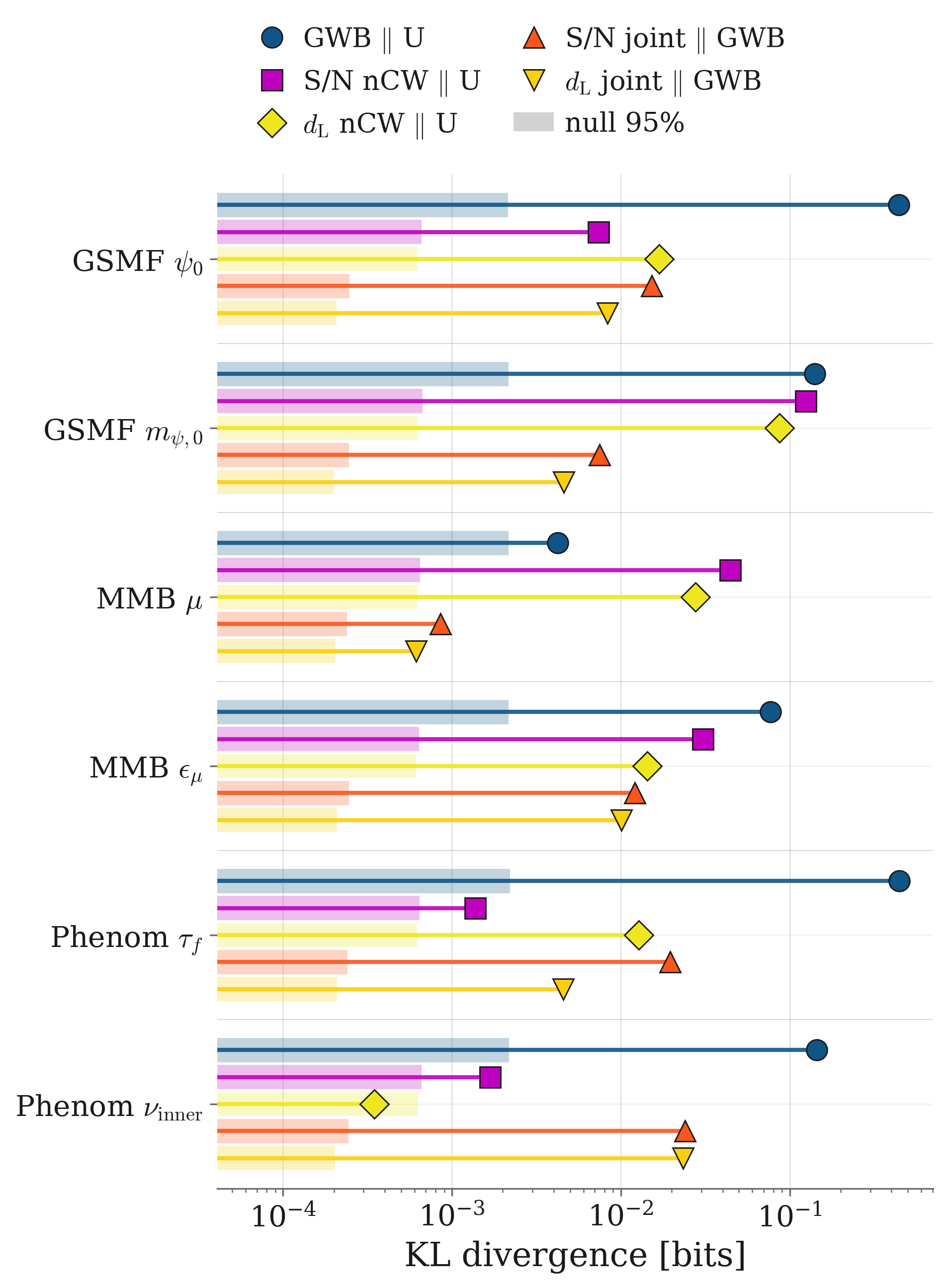}
    \caption{ \ecgcirc{
    KL divergence $\kl$ of each 1D marginal posterior from its reference distribution: GWB-only posteriors from uniform priors (blue circles), S/N nCW-only posteriors from uniform priors (magenta squares), $\dlum$ nCW-only posteriors from uniform priors (light yellow diamonds), reweighted S/N joint posteriors from GWB-only posteriors (orange triangles), and reweighted $\dlum$ joint posteriors from GWB-only posteriors (dark yellow inverted triangles). The shaded band along each line extends to the 95th percentile of the corresponding null distribution. 
    We generate the $\kl (\ncw \| \tr{U})$ and $\kl (\tr{joint} \| \gwb)$ null distributions by shuffling the sample likelihood weights because their reference ($Q$) and updated ($P$) distributions use the same samples. 
    We generate the $\kl (\gwb \| \tr{U})$ null distributions by resampling within the uniform prior bounds, because the GWB posteriors use a different sample library than the uniform priors.
    }}
    \label{fig:kl}
\end{figure}

\ecgcirc{%
The change from priors to nCW and GWB posteriors is quantified by the Kullback--Leibler (KL) divergence \citep{kl_divergence},
\begin{equation}
    \kl(P\| Q) = \sum_x P(x) \log_2 \frac{P(x)}{Q(x)}\quad [\tr{bits}],
\end{equation}
which measures the information gained in moving from a reference distribution $Q$ to an updated distribution $P$, here the 1D marginals over each free parameter $x$, as shown in \figref{fig:kl}. 
The color conventions match \figref{fig:cw_corner} and \figref{fig:joint_corner}: $\kl(\tr{S/N}\ \ncw\| \tr{U})$ in magenta with square markers and 
$\kl(\dlum\ \ncw\| \tr{U})$ in yellow with diamond markers, where U denotes the uniform priors. 
The GSMF and $\mmbulge$ parameters fall well above the 95th percentile region of their null distributions (shaded band), generated by randomly shuffling the $\cwlike$ sample weights. The hardening parameters, whose nCW constraints largely average out in the 1D marginalization, sit closer to their null distributions.
}

CW nondetection favors low characteristic mass, low $\mmbulge$ amplitude, and low $\mmbulge$ scatter. 
These corners of parameter space minimize the opportunity for massive outliers. 
Low GSMF normalization also improves nCW likelihoods, but to a lesser degree than the $\mmbulge$ and characteristic mass parameters. 
\ecgedit{%
All four parameters can raise or lower the GWB amplitude, but they distribute the added strain differently. 
Increasing $\gsmfmass$, $\mmbamp$, or $\mmbscatter$ preferentially strengthens the high-mass outliers, while increasing $\gsmfnorm$ scales up the number of all masses proportionally, so the numerous lower-mass sources reliably boost the background noise that a CW must overcome, while the Poisson odds of a loud outlier improve only through higher expectation values in the sparse high-mass tail.
}
%

Long hardening times help avoid CW detection slightly by preferentially stalling the most massive sources, whose dwell times within the PTA band are fixed entirely by general relativity.
\ecgedit{%
This 1D trend understates the structure in the hardening parameters, however, because it marginalizes over $\hardnuinner$. In 2D, CW nondetection disfavors an intermediate band at $\thardf \sim 6$--$8\,\tr{Gyr}$ and moderate $\hardnuinner$, and allows both the short-lifetime, steep $\hardnuinner$ region and a narrow corner at the longest lifetimes and flattest $\hardnuinner$, leaving a boomerang-shaped zone of avoidance between the allowable regions. 
This effect is a result of our model parameterization, but likely matches the sign of physical mechanisms discussed in \secref{sec:discussion_MN}.
}%
\secref{sec:results_MN} further explores how these parameters impact the CW candidates and GWB competition through their effect on the populations' mass and number distributions.  

The constraints agree between nCW likelihood methods for the majority of parameters, with a minor difference in the strength of their binary lifetime trends. 
$\dllike$ shows a stronger preference for longer hardening times because the only sources that reach the PTA band with large $\thardf$ tend to be low in mass (as explained in \secref{sec:results_MN}). 
\ecgedit{%
These lower masses correspond to a smaller exclusion volume; that is, they can exist at shorter distances without surpassing CW upper limits. 
Meanwhile, the decreased number of massive sources reaching the PTA band also lowers the GWB noise, and the noise reduction partially offsets the weaker CW signals so that $\snlike$ changes little with $\thardf$. 
The overall agreement between the two methods, which use different data products ($\bfthresh$ and $R_\tr{eff}$) and independent calculations, indicates that the constraints reflect the underlying CW nondetection, not the details of the likelihood construction.
}

\ecgcirc{%
We compare the nCW KL divergences to $\kl(\gwb \| \tr{U})$ in blue with circle markers. Its null distributions require resampling from the uniform prior ranges, rather than reshuffling weights, because the GWB posteriors come from a separate library of samples. For most parameters, the GWB posteriors diverge further from the uniform priors than the nCW posteriors, but the nCW divergences are comparable for $\gsmfmass$ and an order of magnitude larger for $\mmbamp$, as these two parameters are especially influential in determining CW detectability. %
}

\ecgedit{
The nCW-only and GWB-only (blue dotted lines) posteriors often differ because populations producing louder GWBs---through more massive and/or more numerous binaries---also tend to produce more detectable CWs, disfavored by nCW likelihoods. 
The exception is populations whose GWB amplitude comes from a large number of lower-mass sources, which can match the observed GWB and the CW nondetection. 
The differences in nCW and GWB constraints shape the contour geometries. 
In pairs of mass-dominant parameters, e.g., $\mmbamp$ versus $\gsmfmass$, the diagonal GWB degeneracy ridge is parallel to the one-sided CW exclusion boundary, which limits the combinations of parameters to the lower-mass side of the GWB ridge. 
}

\ecgedit{
In pairs combining mass- and number-dominant parameters, e.g., $\mmbamp$ versus $\gsmfnorm$, the one-sided nCW boundary runs perpendicular to the one-sided GWB boundary, demonstrating that the nCW information is orthogonal to the GWB information. That is, in these combinations, the GWB more strongly constrains the number parameter while the CW nondetection more strongly constrains the mass parameter, confining the jointly allowed region to a corner. 
}

\ecgedit{
Similar orthogonality appears in $\thardf$ versus $\gsmfnorm$, where the two bounds run perpendicular across different diagonals: the GWB disallows the low $\gsmfnorm$, long $\thardf$ combination while the CW nondetection disallows the high $\gsmfnorm$, long $\thardf$ combination, leaving low $\thardf$ as the overlapping allowed region. 
All three examples demonstrate that CW nondetection encodes new information, distinct from that of the GWB.
}

\subsection{Joint Posteriors} \label{sec:results_joint_post}

\begin{figure*}
    \centering
    \includegraphics[width=\textwidth]{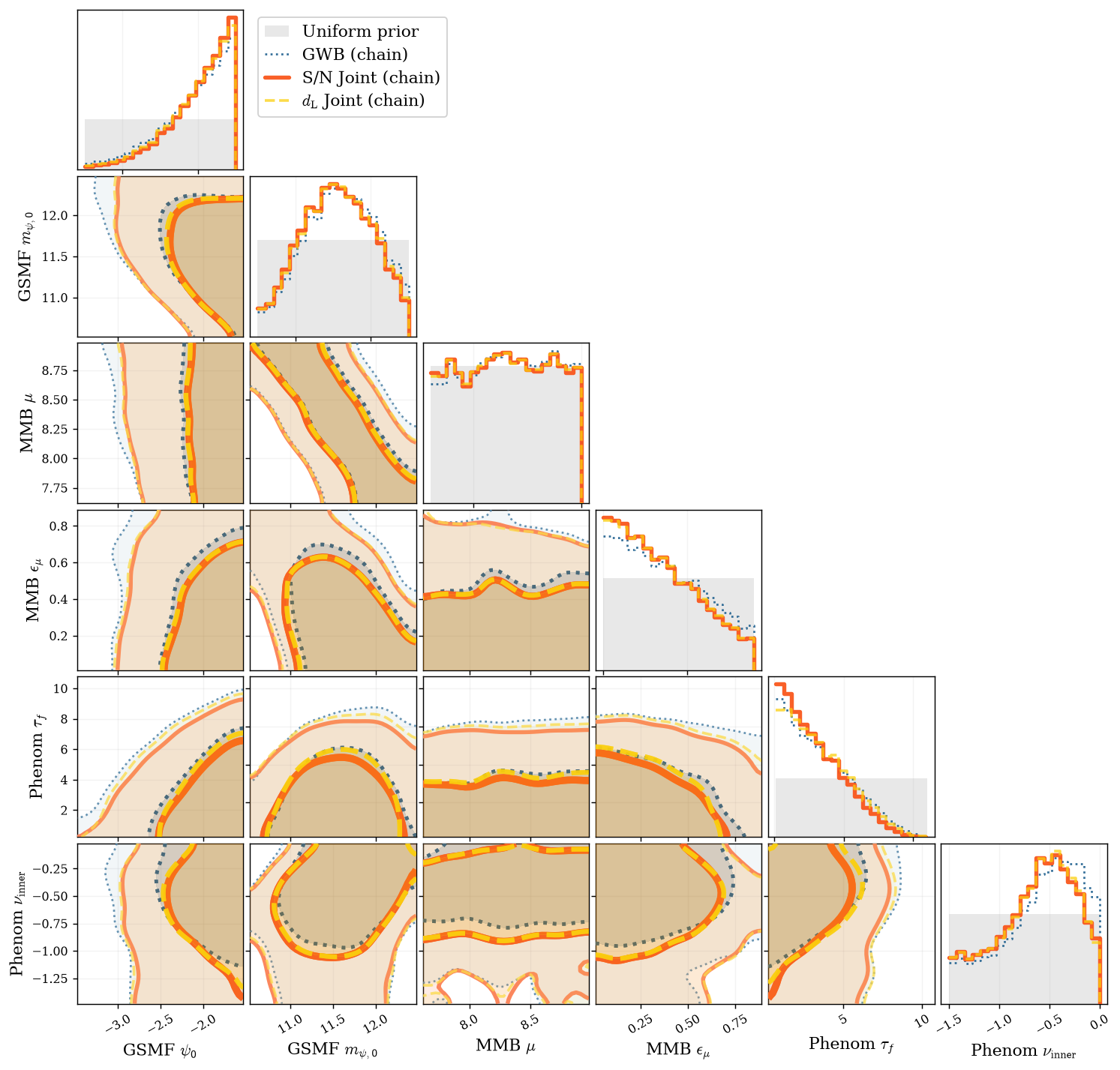}
    \caption{Corner plot of the GWB and joint posteriors. The GWB posteriors (blue dotted) are a histogram of the chain library samples drawn from the \citetalias{ng+23_astro} posteriors, \ecgcirc{based on uniform priors (shaded gray).} We calculate the joint posteriors by reweighting the GWB chain samples by the S/N method nCW likelihoods (orange solid) and by the $\dlum$ method nCW likelihoods (dark yellow dashed). The 2D contours have the same smoothing (kernel width 7.5\% of each parameter range) as in \figref{fig:cw_corner}. In many cases, the nCW reweighting pushes the posteriors towards their prior bounds, narrowing the constraints. The joint posteriors are shifted towards higher GSMF normalization, lower $\mmbulge$ scatter, shorter binary lifetime, and steeper (more negative) small-separation hardening index. 
    }
    \label{fig:joint_corner}
\end{figure*}

The nCW constraints differ from the GWB constraints but are generally weaker, so combining the two gives a joint posterior slightly shifted from the GWB posteriors. Because this shift is marginal, the chain library, made up of samples from the \citetalias{ng+23_astro} GWB posteriors, is ideal for calculating joint posteriors with sampling focused on regions of highest probability. The chain samples make up the GWB posterior distributions marked by blue-dotted lines in \figref{fig:joint_corner}. The joint posteriors, calculated by weighting the distributions by their nCW likelihoods, are shown in orange solid lines for the S/N method and dark yellow dashed lines for the $\dlum$ method. 
\ecgcirc{Their KL divergences, relative to the GWB-only reference distributions, are shown in \figref{fig:kl} with the same color conventions: $\kl (\tr{S/N\ joint}\| \gwb)$ in orange with triangle markers and $\kl (\dlum\ \tr{joint} \| \gwb)$ in dark yellow with inverted triangle markers.}

For many parameters, the joint constraints are tighter, pushing the 68\% credible region of the 2D contours \ecgedit{up against the prior boundaries.} 
For example, all 2D contours containing $\gsmfnorm$ are pushed towards the $\gsmfnorm$ upper limit, binary lifetime compresses against its lower limit, and $\mmbulge$ scatter shifts closer to its lower limit. 

Curiously, the joint posterior favors higher $\gsmfnorm$ than the GWB-only posterior, despite the nCW-only posteriors trending in the opposite direction. 
This occurs in the multidimensional parameter space because most low $\gsmfnorm$ samples \textit{undershoot} the GWB, so the only samples with strong GWB agreement are in the high-mass tail, where CW nondetection probability is low. 
Thus, the high $\bglike$, low $\gsmfnorm$ samples are negated by low $\cwlike$. 
On the other hand, about half of the high $\gsmfnorm$ samples \textit{overshoot} the GWB, while those that agree with it tend to also have stronger $\cwlike$, with fewer massive outliers, such that CW nondetection amplifies the likelihood of the samples in best agreement with the GWB at high $\gsmfnorm$. 
In short, when going from GWB posteriors to CW-weighted joint posteriors, the anti-correlation in GWB and nCW agreement decreases the likelihood at low $\gsmfnorm$, and the positive correlation in GWB and nCW agreement increases the likelihood at high $\gsmfnorm$. 
\ecgedit{
A similar scenario appears for $\thardf$: nCW-only shows a slight preference for long lifetimes, while the joint posteriors prefer shorter lifetimes than GWB-only. }

Although the joint contours are tighter in most cases, a few open up small new allowable regions of parameter space.  This includes lower $\gsmfmass$ when plotted against $\mmbscatter$ and steeper $\hardnuinner$ versus any parameter. The shift in $\hardnuinner$ corresponds to a more negative peak in the 1D $\hardnuinner$ posterior distribution, implying slower small-separation hardening and faster large-separation hardening. See \secref{sec:results_MN} for an explanation of how steeper $\hardnuinner$ causes a larger number of binaries and lower CW detectability. 
\ecgcirc{
In \figref{fig:kl}, the narrowing and shifting from GWB-only to joint constraints are reflected by divergences $\kl (\tr{joint}\| \gwb)$ at least an order of magnitude larger than the corresponding null 95th percentile for every parameter except $\mmbamp$. 
}

\subsection{Mass and Number Effects} \label{sec:results_MN}

The impact of each astrophysical parameter on CW detectability can be understood in terms of the SMBHBs' numbers and masses, with the CW S/N maximized for high masses and low numbers. To describe these properties of the populations, we define two metrics: the strain-weighted average chirp mass $\msw$ and the participation ratio $\npar$. 
The strain-weighted average chirp mass, $\msw$, is calculated at each frequency by averaging over $M,q,z$ bins,
\begin{equation} \label{eq:Msw}
\mathcal\msw(f) = \frac{\sum_{M,q,z} N(M,q,z,f)\,\hscirc^2(M,q,z,f)\,\mathcal{M}_c(M,q)}{\sum_{M,q,z} N(M,q,z,f)\,\hscirc^2(M,q,z,f)}.
\end{equation}

The participation ratio, $\npar$, represents the inverse sum of source weights squared,
\begin{equation}
    \npar (f) \equiv \frac{1}{\sum_i w_i^2} 
\end{equation}
where the weight $w_i$ of each source $i$ in a given frequency bin represents its normalized contribution to the background at that frequency, 
\begin{equation} \label{eq:weights}
w_i = \frac{h_\tr{s,circ,i}^2}{\sum_j h_\tr{s,circ,j}^2}.
\end{equation}
This metric describes how democratic the GWB is. One can imagine the two extreme cases: if every binary contributes evenly to the background, then $\npar$ would equal the total number of sources, and if the signal was dominated entirely by a single source, then $\npar$ would approach 1. Thus, it is not just the number of binaries, but rather the number that contribute significantly to the background. We use the median $\npar$ and median $\msw$ for 100 Poisson realizations to represent each sample.

\begin{figure}
    \centering
    \includegraphics[width=\linewidth]{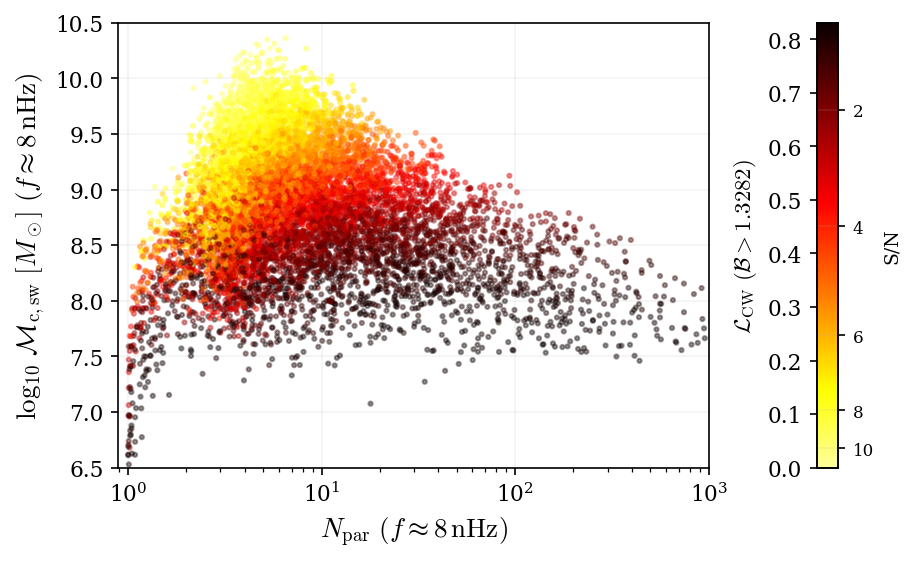}
    \caption{CW nondetection likelihood dependence on strain-weighted average chirp mass and participation ratio. 
    \ecgedit{%
    The $\sim$8\,nHz $\msw$ median across all realizations of a given sample is plotted against the corresponding median $\sim$8\,nHz $\npar$, colored by the sample's likelihood $\snlike$ and the S/N that maps to it. 
    This frequency is high enough that the $\msw$ trends in \figref{fig:hard_contours} are easily decipherable, and low enough that the PTA data is not dominated by noise, detecting a CW is feasible, and $\npar$ trends are also clear. 
    S/N is maximized, and thus $\snlike$ minimized, for high $\msw$ and low $\npar$ because increasing $\msw$ boosts both the GWB and CW amplitudes, while increasing $\npar$ distributes the same amplitude among a larger number of sources.
    }
    }
    \label{fig:snr_Lcw_MN}
\end{figure}

The optimal scenario for continuous wave detection involves the GWB being dominated by a small number of ultramassive sources. 
This allows for more high-strain outliers, improving the odds of one being near enough to detect, while minimizing background noise. 
\ecgcirc{
\citet{houlden+2026} independently confirm the mass dependence, forecasting CW detection fractions for a 9.0-yr MeerKAT PTA to rise from 10\% to 27\% when they increase their characteristic black hole mass (similar to our $\gsmfmass$) from $10^{8.7}\,\msol$ to $10^{9.3}\,\msol$.
} %
The mass--number interplay is apparent in \figref{fig:snr_Lcw_MN}, which contains a scatter plot of median $\msw$ versus $\npar$ at 8 nHz, for each sample, colored by the sample's $\snlike$. The colorbar is also labeled with the S/Ns that correspond to those nondetection likelihoods via \refeq{eq:gcdf} and \refeq{eq:cw_prob_mean}. 

The samples fill a triangular region of this parameter space, where the highest masses, $\log_{10} \msw /\msol \gtrsim 9.5$, correspond to $\npar \lesssim 25$ because masses this extreme will always dominate the background, if they exist, and will always be few in number relative to low-to-intermediate mass SMBHBs. 
Intermediate likelihoods and S/Ns trace a diagonal path from the bottom left, where the masses and number of sources dominating the GWB are both low, to the top right, where both metrics are high. 

Detectable CWs ($\snlike \lesssim 0.25$, $\snr \gtrsim 6$), plotted in yellow, can be found for $\msw$ as low as $\sim 10^{8.5}\, \msol$ if the participation is extremely low ($\npar \lesssim 3$). If the strain-weighted chirp mass is very high ($\msw \gtrsim 10^{9.5} \msol$), detectable CWs can correspond to participation ratios as high as $\npar \sim 20$.  
At the highest participation ratios ($\npar \gtrsim 300$), the CWs are generally not detectable ($\snlike \gtrsim 0.7$, $\snr \lesssim 2.0$); this region lacks any samples with $\msw \gtrsim 10^9 \msol$.

\begin{figure*}
    \centering
    \includegraphics[width=0.49\linewidth]{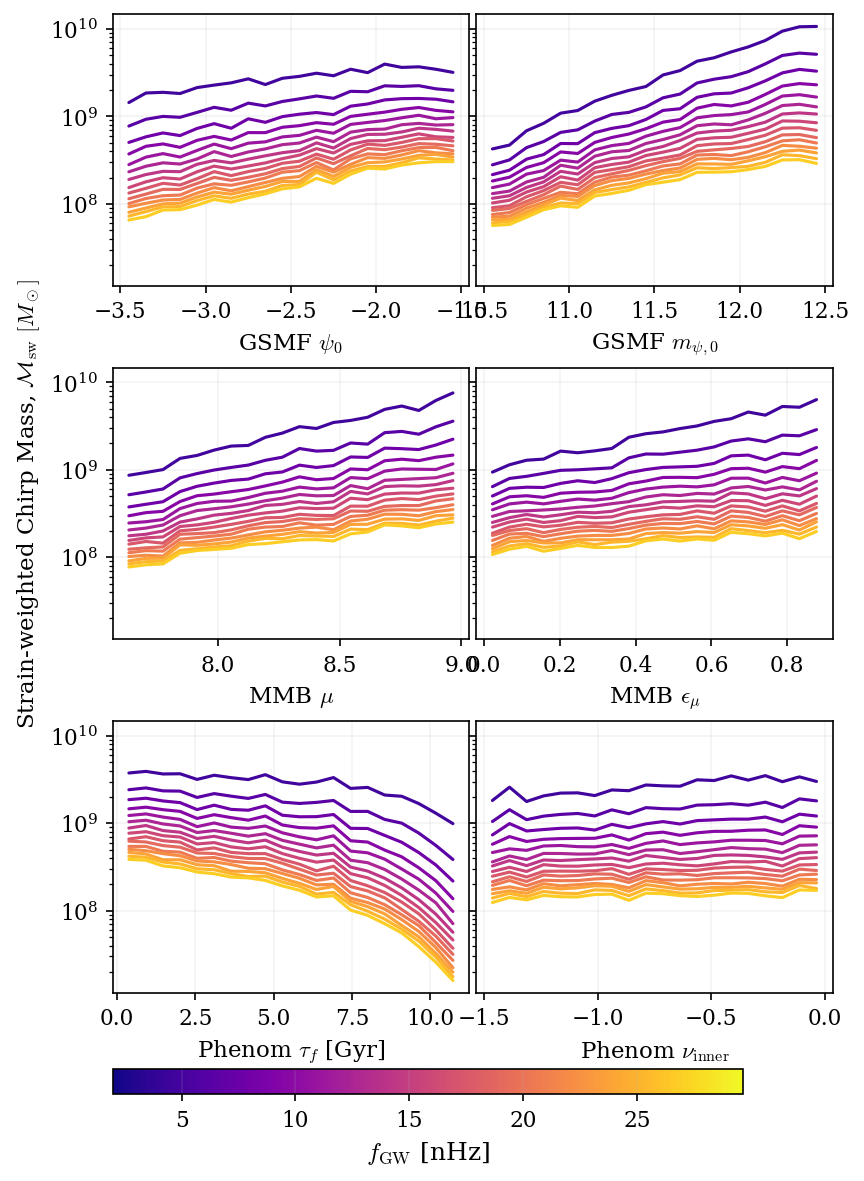}
    \includegraphics[width=0.49 \linewidth]{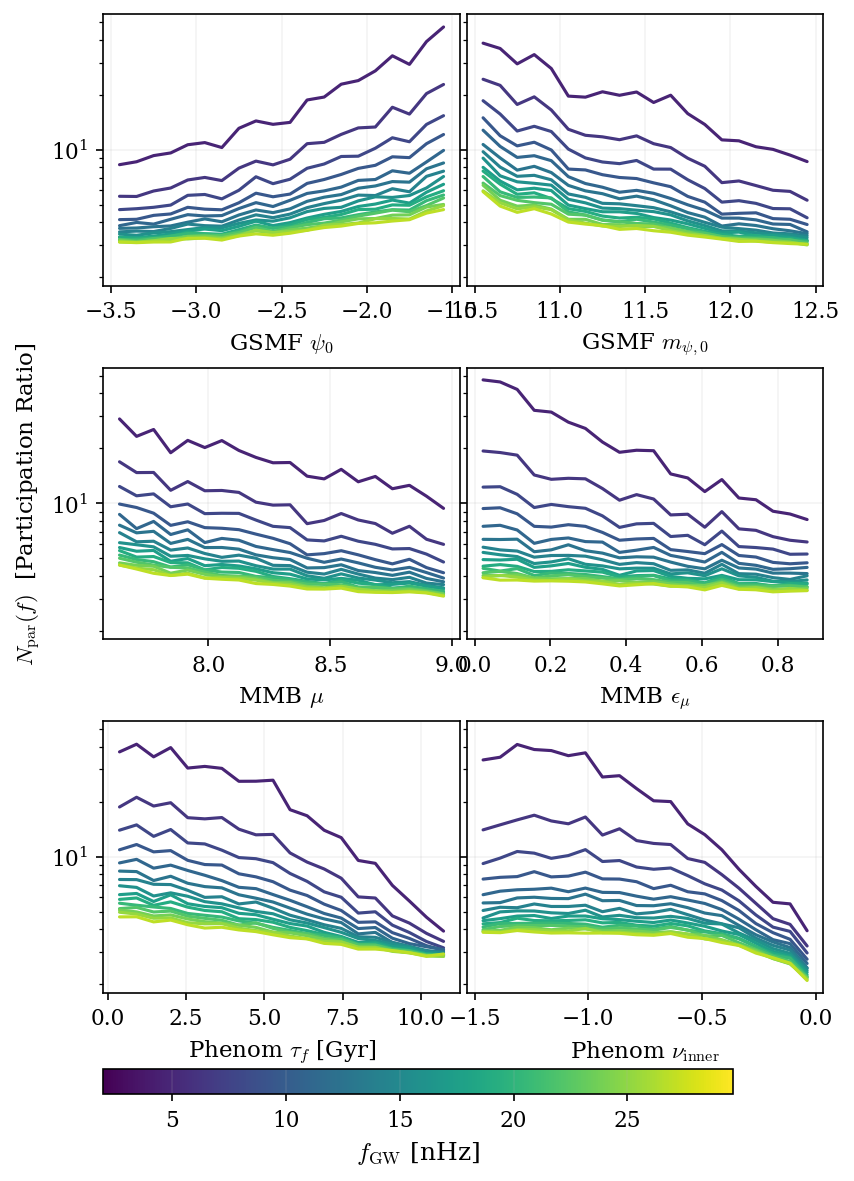}
    \caption{
    Strain-weighted mass $\msw (f)$ and participation ratio $\npar (f)$ as a function of each free parameter. 
    The six panels on the left show median $\msw$ across samples and realizations plotted against the GSMF parameters $\gsmfnorm$ and $\gsmfmass$ in the top row, $\mmbulge$ relation parameters $\mmbamp$ and $\mmbscatter$ in the middle row, and hardening parameters $\thardf$ and $\hardnuinner$ in the bottom row, colored by frequency for the first 15 frequency bins (2--30 nHz).
    The six panels on the right show median $\npar$ across samples plotted against the same free parameters, for the same first 15 frequency bins.
    }
    \label{fig:MN_params}
\end{figure*}

The mass and number metrics serve as a bridge between the SMBHB model parameters and CW detectability. 
\ecgedit{The strain-weighted mass slopes positively with all four GSMF and $\mmbulge$ parameters:} (1) $\gsmfmass$ directly increases the number of high-mass sources, (2) $\mmbamp$ directly increases all binaries' masses, (3) $\mmbscatter$ preferentially scatters binaries to \ecgedit{higher masses \citep{lauer+2007}}, and (4) $\gsmfnorm$ improves the odds of Poisson sampling sources in any bin, including the high-mass tail, which only subtly raises $\msw$ because the number of lower-mass sources is also increased. 

These trends are demonstrated in the first two columns of \figref{fig:MN_params} with the plasma color map. Each line represents the median $\msw$ across samples and realizations at the frequency marked by the colorbar. The trends for the GSMF and $\mmbulge$ parameters are generally consistent across all frequencies, with some being less apparent at high frequencies because so few sources dwell there. Increasing the binary lifetime tends to decrease $\msw$ through a mass filtration effect described below, especially at high frequencies where massive sources are already rare due to their more energetic GW emission \ecgedit{driving them rapidly through high frequencies to coalescence. }

The impacts of each parameter on $\npar$ are shown in the last two columns of \figref{fig:MN_params} with the viridis color map.
\ecgedit{The slope of $\npar$ is positive only for $\gsmfnorm$ because generating more binaries overall through a higher Schechter mass function normalization creates a more democratic signal, spread among more sources. }
\ecgedit{The slopes are negative for 
$\gsmfmass$, $\mmbamp$, and $\mmbscatter$, which all allow for} more ultramassive sources that can dominate the GWB over the large number of weaker sources. 

\ecgedit{$\npar$ versus $\thardf$ also has a negative slope because fewer sources reach the PTA band when the binary lifetime is longer.} 
Flatter $\hardnuinner$ (for fixed total lifetime) also makes large-separation hardening less efficient, so more sources stall in the kpc-to-pc regime, \ecgedit{before speeding up in} the pc-to-mpc regime. Thus the dwell times in the PTA band are shorter, thereby decreasing $\npar$ through two means. \ecgedit{The per-frequency lines show that $\npar$ depends more strongly on the model at low frequencies because high-frequency} sources are rare regardless of the SMBHB model, due to the rapid GW hardening rate near coalescence. 
%

\begin{figure}
    \centering
    \includegraphics[width=\linewidth]{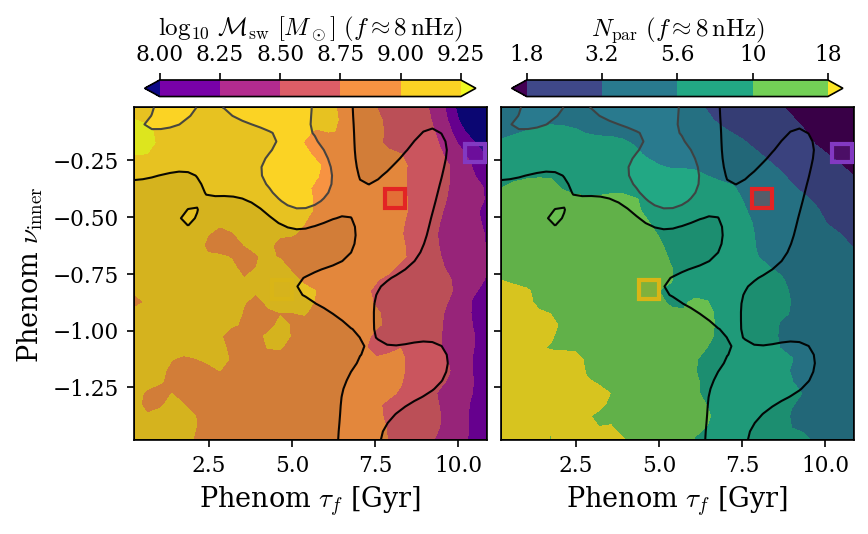}
    \caption{
    2D contours of the strain-weighted chirp mass $\msw$ (left-hand panel) and the participation ratio $\npar$ (right-hand panel) for the small-separation hardening index $\hardnuinner$ versus the total binary lifetime $\thardf$, marginalized over GSMF and $\mmbulge$ parameters. 
    \ecgedit{%
    Both are shown at $\sim$8\,nHz as in \figref{fig:snr_Lcw_MN}.
    As $\thardf$ increases, the dominant masses decrease because massive binaries' more efficient GW hardening requires stretching their phenomenological phase more to match the same fixed total lifetime, preferentially stalling them at large separations.} 
    As both parameters increase towards the top right, the number decreases through a combination of pre-PTA-band stalling and in-PTA-band efficiency. 
    These combined effects create the boomerang-shaped zone of avoidance apparent in the $\hardnuinner$ versus $\thardf$ panel of \figref{fig:cw_corner}, where masses are high enough and numbers low enough that a CW likely would have already been detected. 
    The black lines and shaded regions overlay the S/N method nCW-only posterior contours from \figref{fig:cw_corner}, and the colored boxes mark three example cases whose evolution and spectra are shown in \figref{fig:hard_spec}.
    }
    \label{fig:hard_contours}
\end{figure}

\begin{figure}
    \centering
    \includegraphics[width=\linewidth]{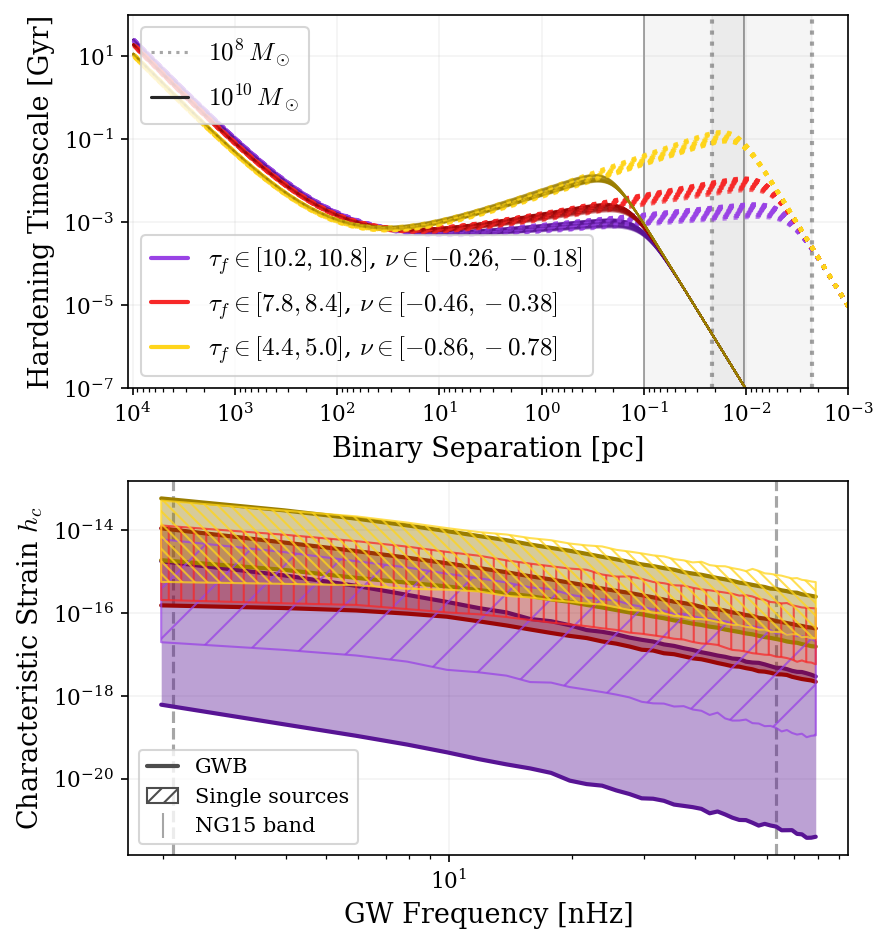}
    \caption{
    Hardening timescale evolution (top) and characteristic strain spectra (bottom) for the three example hardening cases marked by boxes in \figref{fig:hard_contours}. The cases are (1) within the CW-allowed short-lifetime contours (yellow; $\thardf \sim 4.7$\,Gyr, $\hardnuinner \sim -0.8$), (2) inside the zone of avoidance (red; $\thardf \sim 8$\,Gyr, $\hardnuinner \sim -0.4$), and (3) in the CW-allowed long-lifetime corner (purple; $\thardf \sim 10.5$\,Gyr, $\hardnuinner \sim -0.2$). 
    \textit{Top:} Hardening timescale $\tau \equiv a/\dot{a}$ versus binary separation for $10^{10}\,\msol$ (solid) and $10^8\,\msol$ (dotted) binaries; the gray bands mark the separations that produce NG15 PTA-band frequencies for each mass with the same linestyle conventions. 
    \textit{Bottom:} 
    \ecgedit{%
    68\% intervals on the GWB characteristic strain $h_c(f)$ (darker shading, solid outlines) 
    and the loudest single sources (lighter hatching). 
    The 68\% intervals span all realizations and samples inside each hardening case box; the GWB $h_c(f)$ sums all sources except the single loudest per frequency per realization, and the hatched bands show those loudest sources.
    Dashed vertical lines bound the NG15 band.  
    The red case offers the best CW detection prospects because its CW candidates stand above a background weakened by mass filtration and reduced source counts, while the yellow case has comparably loud CWs washed out by a louder GWB, and the purple case has CWs exceeding the GWB but below the PTA noise floor.
    }
    }
    \label{fig:hard_spec}
\end{figure}

The impacts of the binary hardening model on which sources reach and emit in the PTA band prove to be somewhat more complex. 
To make sense of them, we show the 2D dependence of $\msw$ (left-hand panel) and $\npar$ (right-hand panel) for the small-separation hardening index $\hardnuinner$ versus the binary lifetime $\thardf$ in \figref{fig:hard_contours}. 
In both panels, there are three boxed regions of parameter space that we use as examples to help explain the boomerang-shaped zone of avoidance in the nCW posteriors of \figref{fig:cw_corner}. 


We consider these example cases as they increase in both $\hardnuinner$ and $\thardf$ towards the top right of the contour plots. The first, in yellow, has central values of both parameters ($\thardf \in [4.4, 5.0]\,\tr{Gyr}, \hardnuinner \in [-0.86, -0.78]$), falling in the 68\% nCW credible region; the second, in red, is in the zone of avoidance disallowed by CW upper limits ($\thardf \in [7.8, 8.4]\,\tr{Gyr}, \hardnuinner \in [-0.46, -0.38]$); and the third, in purple, is back in the CW-allowed posteriors in the top right corner of $\hardnuinner$ versus $\thardf$ ($\thardf \in [10.2, 10.8]\,\tr{Gyr}, \hardnuinner \in [-0.26, -0.18]$). 

The hardening timescale evolution and characteristic strain spectra of the samples in each of these case boxes are shown in \figref{fig:hard_spec}, in the same colors. 
The top panel plots their hardening timescales, $\tau \equiv a / \dot{a}$ as a function of separation for $10^8\,\msol$ (dotted) and $10^{10} \,\msol$ (solid) binaries, and the same linestyles outlining gray vertical bands show where these $10^8\,\msol$ and $10^{10}\,\msol$ binaries would be emitting within the NG15 PTA band. 
For the characteristic strain spectrum in the bottom panel, the 68\% interval of CW candidates' (the loudest source at each frequency in each realization) characteristic strain are hatched; the 68\% interval of the remaining background sources is shaded in a darker version of each color and bounded by solid lines. 

In the first scenario, the background is loud with a hardening time short enough that many binaries can reach the PTA band, including high-mass binaries, with $\npar \sim 7.5$ and $\log_{10} \msw/\msol \,\sim9.2$. The top panel shows that this intermediate $\hardnuinner$ gives relatively long timescales at the beginning of the PTA band (shaded in gray for a $10^{10} \msol$ case with solid lines and $10^8\msol$ case with dotted lines). This long dwell time keeps the number of sources contributing to the low-frequency GWB noise high, such that the GWB 68\% interval on characteristic strain in the bottom panel is typically louder than those of the CW candidates. Thus, CW detectability is low, and $\snlike$ and $\dllike$ are high. 

Moving to the red square at $\hardnuinner\sim-0.4$, $\thardf \sim 8 \,\tr{Gyr}$ decreases both $\msw$ and $\npar$. This combination of changes primarily increases the kpc hardening timescales that take up most of the binaries' lifetimes, while decreasing the $\sim$pc-scale timescales. 
Flattening $\hardnuinner$ decreases the number of sources reaching the PTA band because the faster small-separation hardening requires slower large-separation hardening, so binaries of all masses are delayed, with fewer reaching PTA-band separations and frequencies. 
Longer $\thardf$ also means that fewer sources evolve quickly enough to reach the PTA frequency band, but unlike $\hardnuinner$, this also causes a shift in typical mass. 
Increasing binary lifetime preferentially filters out the most massive sources because the GW part of their lifetime is fixed to a shorter timescale than that of lower-mass binaries, due to more efficient GW emission. Thus, the variable phenom part must increase more to reach the same total lifetime, so the strain-weighted mass decreases to $\log_{10} \msw /\msol \sim 8.8$. 
Although this is specific to our phenom hardening model, it may represent a physical effect: 
if hardening is generally less efficient, e.g., by slower loss-cone refilling, weaker circumbinary disk interactions, or delayed dynamical friction, this could disproportionately affect the most massive sources because they have more orbital energy requiring extraction through these environmental interactions. 
The characteristic strain spectrum shows that, while both GWB and CW candidate strains are lower, the CWs are often louder than the background at the crucial low-frequency end of this range, causing higher S/Ns.
Our lack of CW detection implies low likelihood of this scenario.

\ecgedit{
Increasing both parameters even further to the purple square in the top right of \figref{fig:hard_contours} decreases $\npar$ to $\sim$1--2, meaning only a couple of sources dominate the signal due to large-separation stalling and short small-separation dwell times. 
Although the ultramassive black holes are in the GW regime for the whole PTA band, many binaries with masses below $10^{10}\,\msol$ remain in the $\hardnuinner$ regime for much of the PTA band, allowing the short dwell times to limit their number. 
With such low numbers, the background offers little noise competition, demonstrated by its very low amplitude in the bottom panel of \figref{fig:hard_spec}. 
The longest $\thardf$ decreases $\log_{10} \msw / \msol$ to $\sim 8.2$ through the preferential filtering of the most massive sources.  
Even without competition from the background, the CW amplitudes of such low-mass sources are too weak to compete with PTA noise, including pulsar red noise and measurement white noise. 
}

\ecgedit{%
The third case is an extreme evolution scenario that is unlikely to reproduce the observed GWB because even if the other model components raise the masses and number of binaries, they cannot produce a large background if they do not harden to the GW regime. 
Still, this scenario is consistent with the lack of CW detection.}
Combining the GWB constraints with this CW analysis disallows the low GWB amplitude scenario. Thus, the region to the bottom left of the avoidance zone is the CW-favored region that shifts the joint constraints to lower $\thardf$ and steeper (more negative) $\hardnuinner$, relative to the GWB-only posteriors.

\subsection{Binary Properties} \label{sec:results_props}


\begin{figure}
    \centering
    \includegraphics[width=\linewidth]{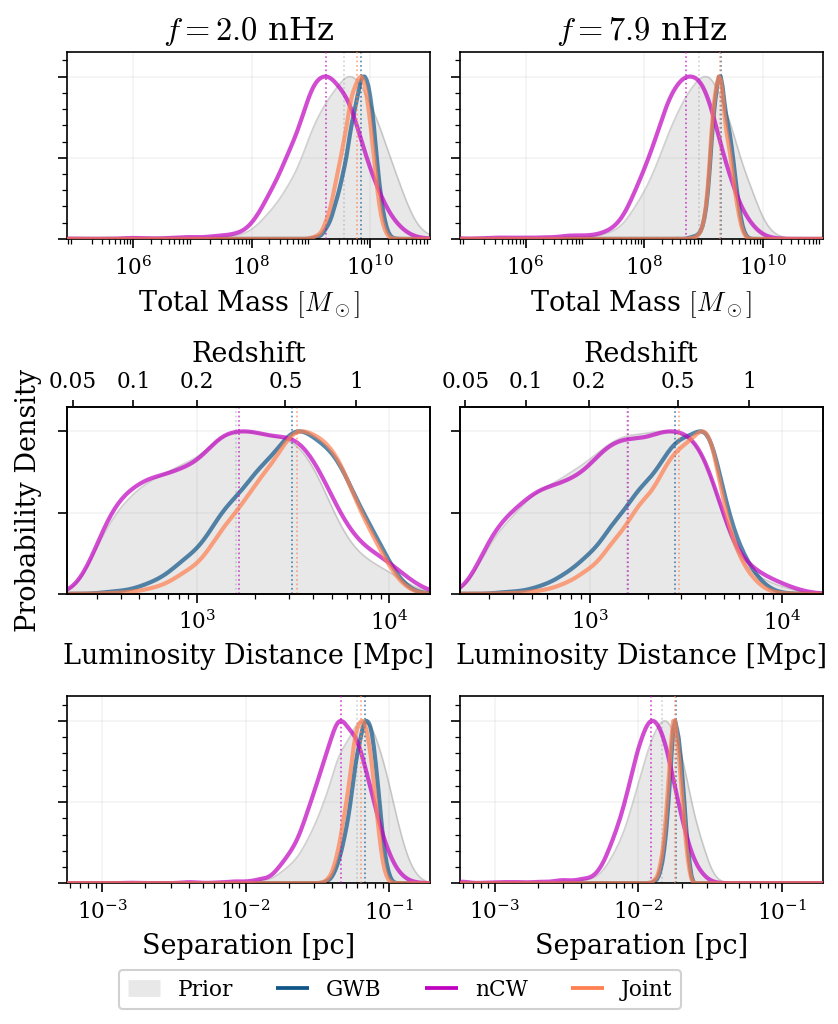}
    \caption{\ecgedit{
    Total mass (top row), luminosity distance (middle row), and separation (bottom row) of the binaries dominating the GWB at $\sim2\,\tr{nHz}$ (left-hand column) and $\sim8\,\tr{nHz}$ (right-hand column). The uniform priors are shaded gray, the GWB posteriors (from the chain library) are blue, the nCW posteriors (from the uniform library, weighted by $\snlike$) are pink, and the joint posteriors (from the chain library, weighted by $\snlike$) are orange. The median of each distribution is marked by a dotted vertical line.
    CW nondetection shifts the masses lower and separations slightly smaller, while the distances remain similar to the priors. The GWB posteriors are narrower, favoring higher masses, larger separations, and further distances, while the joint posteriors remain near them, drawn slightly towards the nCW mass and separation. Massive sources, with large separations, dominate the lowest frequencies. As they evolve rapidly through higher frequencies, their numbers dwindle, and lower masses dominate the signal.}
    }
    \label{fig:bgprops}
\end{figure}

\ecgedit{
The CW upper limits shift expectations for the binary properties dominating the GWB. In \figref{fig:bgprops}, we compare the uniform prior (gray shaded), GWB-only (blue), nCW-only (pink), and joint (orange) posteriors on the strain-weighted average total mass (top row), luminosity distance (middle row), and binary separation (bottom row) of the binaries in the first frequency bin ($\sim 2\,\tr{nHz}$, left-hand column) and fourth frequency bin ($\sim8\,\tr{nHz}$, right-hand column). 
The GWB-only properties are those of the chain library, the nCW-only properties are from the uniform library weighted by $\snlike$, and the joint properties are those of the chain library weighted by $\snlike$. 
}

\ecgedit{
As in \citetalias{ng+23_astro}, the GWB posteriors peak narrowly at high masses, with corresponding large separations. Going from 2\,nHz to 8\,nHz, the median strain-weighted total mass decreases from $\sim10^{9.8}\,\msol$ to $\sim10^{9.3}\,\msol$ because the most massive sources dominating the low-frequency end of the GWB evolve rapidly through high frequencies via strong GW emission. The nCW-only posteriors peak at lower total masses than the priors, with medians near $\sim10^{9.2}\,\msol$ at 2\,nHz and $10^{8.7}\,\msol$ at 8\,nHz, because lower masses improve consistency with the CW nondetection, especially at low frequencies where CWs are otherwise more likely to be observed \citep{rosado+2015,kelley+2018,gardiner+2024}. Thus weighting by $\snlike$ shifts the joint posteriors to visibly lower masses than the GWB posteriors at 2\,nHz and to marginally lower masses at 8\,nHz. The GWB posteriors exclude much of the short-luminosity-distance area spanned by the priors because they rule out the longest hardening times to which these short distances tend to correspond. Meanwhile, the nCW constraints on the hardening model are much weaker, leaving the posteriors on distance very similar to the priors. 
}


\begin{figure}
    \centering
    \includegraphics[width=\linewidth]{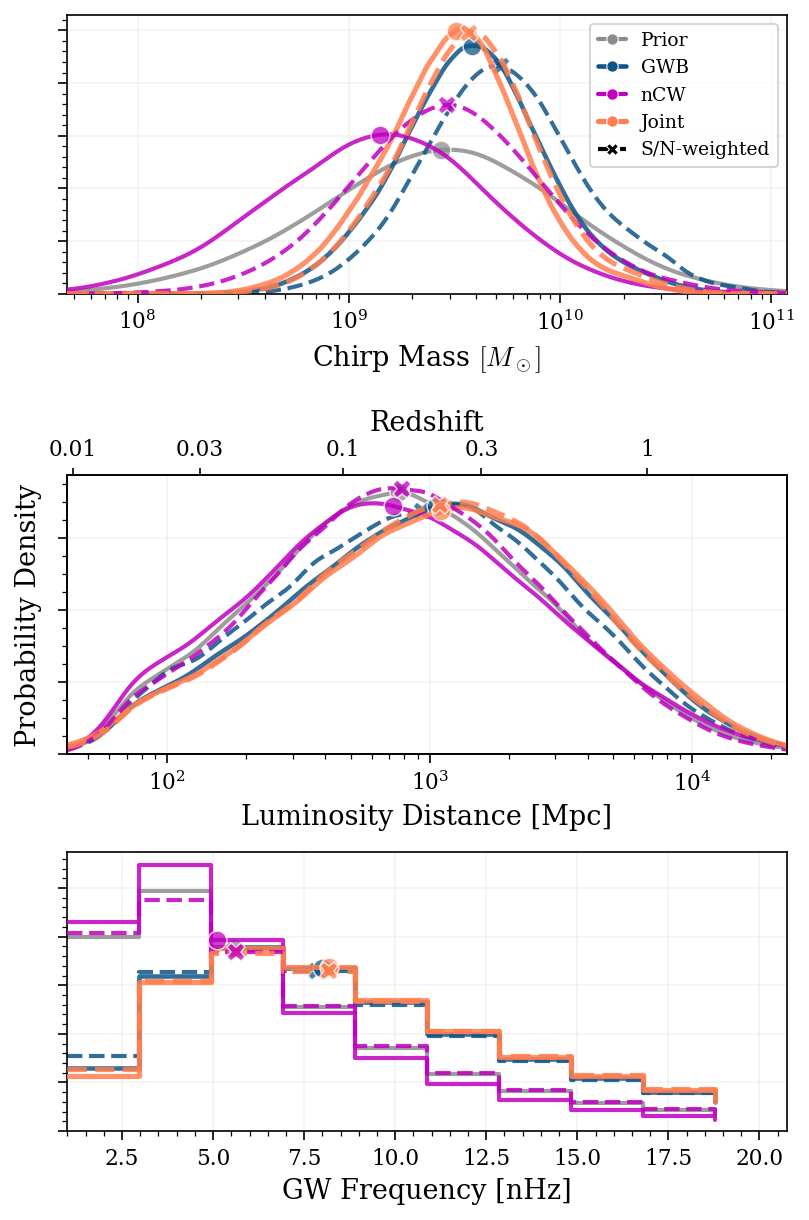}
    \caption{
    \ecgedit{\ecgcirc{Rest-frame} chirp mass (top), luminosity distance (middle), and observed gravitational wave frequency (bottom) of the most likely CW detections. 
    The uniform priors (shaded gray) represent the properties of the highest S/N CW from each sky in each realization of the uniform library; and the nCW-only (pink) represent the same sources, weighted by $\snlike$. 
    The GWB-only lines (blue) contain the most detectable CW in each sky realization of the GWB chain library; and the joint posteriors (orange) weight that distribution by $\snlike$. 
    The dashed lines additionally weight each CW candidate by its S/N. 
    The solid-line posteriors, with circles marking their peaks, convey the CW properties of the most likely scenarios, as informed by the GWB free spectra, CW nondetection, or both. 
    Meanwhile, the S/N weighted (dashed lines with X's at the peaks) describe only the most detectable sources that could exist under any of these constraints. 
    }
    }
    \label{fig:ssprops}
\end{figure}

The CW properties that are most likely to be detected can be similarly informed by the data. In \figref{fig:ssprops}, the posteriors on rest-frame chirp mass (top), luminosity distance (middle), and GW frequency (bottom) of the max-S/N source in each sky of each realization are marked by solid lines, following the same color conventions of GWB in blue, nCW in pink, and joint in orange. 
This produces a data-driven estimate of the CW candidates that are most likely to exist. 
To also consider which of those we are most likely to detect, we weight them by their S/Ns, resulting in the dashed-line distributions. 

The nCW posteriors favor substantially lower masses than the GWB; their expected chirp mass peaks at $10^{9.20}\,\msol$, with S/N-weighting shifting the peak to $10^{9.47}\,\msol$. 
In comparison, the GWB favors higher masses, peaking at $10^{9.59}\,\msol$ based only on the data constraints and $10^{9.70}\,\msol$ with S/N-weighting. Combining the data in the joint posteriors yields the most-favored chirp mass of $\sim 10^{9.54}\,\msol$. 
\ecgcirc{ Our joint-posterior chirp mass predictions are comparable to earlier expectations, including pre-detection forecasts favoring $\mchirp \sim 10^9$--$10^{10}\,\msol$ \citep{rosado+2015, mingarelli+2017}. More recent works incorporate GWB information: \citet{becsy+2022_realistic} predict a median rest-frame chirp mass of $10^{9.5}\,\msol$ and \citet{gardiner+2025} find a 68\% credible interval of $10^{9.2}$--$10^{9.8}\,\msol$, both in close agreement with our predictions despite differing methods. 
As with the strain-weighted background properties, the CWs favor closer distances \ecgcirc{than the 
GWB-only posteriors}, primarily as a result of their hardening model constraints. 
}

\ecgedit{The CW upper limits favor candidates in the second frequency bin ($\sim4\,\tr{nHz}$), regardless of S/N-weighting. 
While this is the same bin that the uniform priors favor, that preference is strengthened by the nCW likelihoods. 
In contrast, the GWB-only posteriors peak at $\sim6$\,nHz, with their median at $\sim8$\,nHz, suppressing the lowest two bins through two related effects.
First, detectable 2--4\,nHz CWs arise primarily in samples with weak backgrounds, which are down-weighted by $\bglike$. 
Second, samples with a GWB matching NG15 have significant GWB noise contributing to the noise covariance matrix in our S/N calculation, \refeq{eq:snr}, primarily impacting low frequencies where the GWB dominates the noise budget. 
As a result, the max S/N candidates are more likely to be at higher frequencies, up to 6--8\,nHz, beyond which pulsar and measurement noise overpower them.}

\ecgcirc{ %
While those two effects partially explain why the GWB-informed CW predictions of \citet{gardiner+2025} have higher frequencies than prior expectations, our GWB and joint posteriors still sit at the low end of their $4$--$12$\,nHz range. That work conditioned their populations on a power-law fit to the NG15 GWB, where high-amplitude, high-frequency sources could improve the apparent agreement between the data and simulated power-law fits. We instead weight randomly drawn populations by their free-spectrum GWB and nCW likelihoods, eliminating the bias of a power-law assumption. Since the NG15 free spectrum is not well fit by a power law \citep{ng+23_astro}, these $6$--$8$\,nHz estimates are a more reliable prediction.  
}

The nCW peak in the second frequency bin also aligns with the 4\,nHz candidate in \citetalias{ng+23_individuals}, which had a Bayes factor of $\sim 4$ in favor of the CW+CURN model over the CURN only model. This may lend some credence to the possibility of a real source there, but to let that candidate properly inform our constraints, the full \citetalias{ng+23_individuals} posteriors would need to be incorporated in the likelihood calculation \citep{turgeon+2026} as opposed to our methods using only a global Bayes factor or exclusion volume. 

The S/N-weighted joint posteriors (dashed orange lines) are the most informed estimates of CW properties. 
Based on those distributions, the most detectable CW sources in binary populations consistent with the GWB and CW upper limits have a chirp mass of $10^{9.58}\,\msol$, luminosity distance of $\sim1300$\,Mpc, and GW frequency in the third bin, $\sim6$\,nHz. \ecgedit{The shift in chirp mass relative to the GWB-only expectations demonstrates that CW upper limits contain constraining power unexploited by previous GWB-only analyses.}


\section{Discussion} \label{sec:discussion}

\ecgedit{
Our results demonstrate that CW upper limits add information \ecgcirc{complementary} to the GWB, favoring populations in which the observed amplitude arises from many lower-mass binaries rather than a few ultramassive ones.
We place these constraints in the context of the broader tension between the GWB amplitude and expectations from electromagnetic (EM) observations in \secref{sec:discussion_MN}. Then, we consider how future data will sharpen this picture in \secref{sec:discussion_future}.
}

\subsection{The Mass--Number Tension} \label{sec:discussion_MN}
Best estimates for the GSMF, $\mmbulge$ relation, and binary hardening are informed by EM observations, the GWB, and now also CW upper limits. 
A tension exists between these probes: the GWB amplitude predicted from the present-day abundance of SMBHs, derived from local scaling relations and galaxy catalogs, falls short of that observed by PTAs, even under optimistic assumptions of equal-mass mergers, short delay times, and negligible accretion after binary formation \citep{spzq2024_where, izquierdo+2022}. The observed GWB must therefore be inflated relative to these EM-based expectations through higher masses or larger numbers of SMBHBs. 

\paragraph{\bf{The Mass Channel}}
EM arguments tend to favor the mass channel. 
\citet{liepold+ma2024} propose that the Schechter GSMFs miss ultramassive galaxies due to cosmic variance in the low number density, high-mass tail. They show that a GSMF recalibrated to the volume-limited MASSIVE survey predicts a higher GWB, consistent with the PTA measurement. 
Recent dynamical detections of additional $\gtrsim 10^{10}\,\msol$ black holes in brightest cluster galaxies lend support to this enhanced high-mass tail \citep{deNicola+2025}.
From a different starting point, \citet{spzq2024_where} argue that the only way to recover the observed GWB with a total SMBH mass budget consistent with the integrated quasar luminosity function \citep{soltan1982} is for the GWB to be dominated by ultramassive black holes. 
These arguments are limited by systematic uncertainty in the massive-end black hole mass function, which remains sensitive to the choice of scaling relation. The primary tracers, stellar velocity dispersion and stellar mass, often disagree, with even the sign of the offset changing with redshift \citep{matt+2023, spz2025_uncertainties}.
%


\paragraph{\bf{The Number Channel}}
GW arguments, including this work, point instead towards the number channel. 
We demonstrate that CW upper limits favor large numbers over high masses, because higher masses correlate with better CW detection prospects. 
This is independently corroborated by the smooth spectral shape of the GWB; \citet{spz2025_distribution} demonstrate that a background dominated by more massive sources and a lower source count would produce greater frequency-bin-to-bin excursions. 
The number channel initially appears more EM-constrained, with the GSMF normalizations directly measured by mass-complete surveys \citep{leja+2020, driver+2022}, and the galaxy merger rates that set the binary abundance being limited by observed galaxy pairs \citep{conselice+2022}.
Despite this, there is room for error in the galaxy merger rate.

In our model, the pair fraction and galaxy merger times are fixed, so the GSMF normalization, $\gsmfnorm$, absorbs the uncertainty in all three factors; a higher $\gsmfnorm$ is therefore potentially better interpreted as a high SMBHB merger rate than as a literal excess of galaxies.
\ecgedit{%
This merger rate retains a factor-of-several freedom through uncertainties in the merger time. 
Our model infers this rate by dividing the observed pair fraction by an assumed merger time \citep{chen+2019}, and simulations show that merger times are shorter at higher redshift \citep{snyder+2017}, implying higher merger rates than one would naively expect from pair fraction observations alone \citep{conselice+2022}.
The merger rate also depends on binary mass ratio. In our model, the merger time scales inversely with mass ratio, suppressing unequal-mass pairs.
Taking the alternate approach of extracting merger rates directly from simulations produces a larger number of unequal-mass binaries \citep{rodriguezgomez+2015}. 
While this difference is unimportant for the GWB, which major mergers dominate, it strongly affects the total binary abundance.
}%
Even with some freedom in the merger rate, a semi-empirical model of SMBH demographics finds that merger rates high enough to fully source the GWB would overproduce the local black hole mass function \citep{lapi+2026}. 
Thus, whether this freedom suffices to explain the high GWB amplitude remains contested.






\paragraph{\bf{Binary Hardening}}
Binary hardening offers a third knob, the least EM-constrained, for setting the SMBHB distribution.
To isolate the nCW constraints, this work applies uniform or GWB-informed priors to all six parameters equally, but if the tighter EM priors on the GSMF and $\mmbulge$ relation were imposed, the burden of amplifying the PTA-band number shifts onto the hardening model.
\ecgedit{%
In our model, this corresponds to joint constraints with a shorter binary lifetime and steeper $\hardnuinner$ than in the GWB-only constraints. Efficient hardening from galactic to $\sim$pc scales delivers more binaries to the PTA band within a Hubble time, while slower hardening at small separations sustains long dwell times, and thus high source counts, within the band.  
Notably, the joint preference for steep $\hardnuinner$ moves towards the value predicted by N-body simulations of stellar scattering, which find an approximately constant hardening rate $\frac{d}{dt}(\frac{1}{a})$ \citep{quinlan1996, sesana+khan2015, khan+2018}, corresponding to $\hardnuinner=-1$ in our parameterization \citep{blecha2026}. 
This appears in conflict with the \citetalias{ng+23_astro} interpretation of the low-frequency GWB flattening as evidence of rapid small-separation hardening, but this turnover could still be explained by eccentricity moving power up to the higher harmonics \citep[e.g.,][]{enoki+2007, taylor+2016}.
}

\ecgedit{%
Mapping the binary lifetime preference onto physical mechanisms requires more consideration. 
In our fixed-lifetime parameterization, increasing $\thardf$ preferentially stalls the most massive binaries because their GW-dominated phase is the shortest, so their phenom component stretches furthest to reach the same total lifetime.
Nature provides a corresponding mechanism: \citet{yu2002} finds the longest hardening bottlenecks in luminous core galaxies because (1) their low central densities slow stellar diffusion into the loss cone, and (2) they host the most massive black holes, which experience a smaller relative energy extraction per stellar interaction.   
N-body stellar-scattering simulations reach the same conclusion, finding total binary mass among the strongest predictors of coalescence time, which increases by an order of magnitude between $10^7\msol$ and $10^{10}\msol$ \citep{holleybockelmann+2025}.
}

\ecgedit{%
Dynamical friction generally runs in the other direction. Chandrasekhar drag strengthens with the mass of the inspiraling body \citep{chandrasekhar1943}, so at large separations the most massive systems should sink fastest. 
Gaseous drag can shorten this stage further, though predominantly in the lower-mass, gas-rich hosts.
\citet{Mayer+2007} find that the massive nuclear gas disk formed in a merger of two Milky-Way-mass spirals drives a pair of $2.6\times 10^6\,\msol$ black holes into a gravitationally bound pair in under a million years, through drag against the gas rather than the stars. 
\citet{khan+2016} recover a similarly rapid inspiral for higher-mass ($10^8\,\msol$) black holes merging at redshift $z\sim3$, but in their scenario the rapid dynamical friction hardening is driven by high stellar densities found in such high-redshift galaxies, as opposed to gas. 
Whether these speed-ups extend to the higher-mass, lower-redshift ($z\lesssim1$, see \figref{fig:bgprops}) binaries that dominate the GWB is unclear. 
Given that the most massive hosts are expected to be gas-poor early types, it is plausible that the most massive binaries stall, a scenario that would support the large number of lower-mass binaries favored by our CW upper limits. 
}

\ecgedit{Our phenom model cannot resolve these questions on its own; by construction, its normalization ties PTA-band hardening to binary evolution at much larger separations, and the GWB spectrum changes little across wide ranges of $\hardnuinner$ \citep{blecha2026}. Physically motivated parameterizations, such as the \citet{blecha2026} inside-out hardening model (BIO-hardening), used in conjunction with CW nondetections, offer a path toward constraining hardening efficiency and how it depends on binary mass.}

\subsection{Future Prospects} \label{sec:discussion_future}
CW upper limits add a new piece of information to resolve the tension between the high GWB amplitude and expected SMBHB mass functions, but this puzzle is far from solved. While the field of CW astronomy is new and the constraints are still weak relative to the large accumulation of EM information, we demonstrate that upper limits currently favor the number channel over the mass channel for amplifying the GWB, and this argument will be strengthened if nondetection continues. 

On the other hand, if the GWB is produced by SMBHBs, then eventually CW detections are inevitable, and may even involve multiple sources, with new techniques for multi-CW modeling being developed for both PTAs and LISA \citep{BayesHopper, gundersen+cornish2025, criswell+2026}. In that scenario, instead of having upper limits rule out high-mass outliers, we will be able to infer the regions of parameter space that allow for our particular catalog of detections \citep{tomson+2026}. At that point, it will be necessary to replace our factorized likelihood approach with a full joint likelihood that can capture the interdependency of the GWB and multi-CW fits. \ecgcirc{This approach becomes feasible with flow-based emulators \citep{laal+2025}, which run faster and better represent the full shape of the GWB probability distribution.
\citet{laal+2026} use this machinery for GWB-only population inference with simulated data, where they find degeneracies in their parameter recoveries that CW information could help break.
} %


\section{Conclusions} \label{sec:conclusion}
In this work, we present the first constraints on SMBHB populations that incorporate current PTA upper limits on CWs. Our conclusions are summarized as follows:

\begin{enumerate}
    \item The nondetection of CWs favors SMBHB populations with many lower-mass sources over few ultramassive ones. This can come about through some combination of low GSMF characteristic mass, low $\mmbulge$ relation amplitude, and/or low $\mmbulge$ relation scatter to keep typical masses low, and high galaxy merger rate (absorbed by our GSMF normalization) to raise the number of contributing sources. Efficient hardening through the kpc-to-pc regime combined with longer dwell times in the late stellar scattering/circumbinary disk regime offer another means of increasing the number of binaries, without as strong EM limitations as the GSMF.
    
    \item CW upper limits disfavor an intermediate zone in our binary evolution model where total binary lifetimes around $\sim 6$--$8\,\tr{Gyr}$ combined with fairly efficient small-separation hardening ($-0.5 \lesssim \hardnuinner \lesssim -0.2$) \ecgedit{weaken the GWB---through the mass filtration imposed by long lifetimes and the reduced source counts imposed by both parameters---enough that a CW would likely have been detectable above it. CW nondetection instead favors the broad region below and left of this zone, spanning the short-to-intermediate lifetimes and steeper $\hardnuinner$, where efficient large-separation hardening allows many sources across all masses to contribute to the background. A second, smaller allowed corner appears at the longest lifetimes and flattest $\hardnuinner$, where CWs are too weak to detect even without much GWB competition. The GWB posteriors exclude this corner, so the joint posteriors favor shorter lifetime and steeper $\hardnuinner$ than the GWB-only posteriors of \citetalias{ng+23_astro}.} 

    \item Weighting the GWB posterior chain of samples by the nCW likelihoods slightly tightens the constraints on most parameters in our joint posterior. 
    The joint posteriors push several parameters tighter against their prior edges: toward higher GSMF normalization, shorter binary lifetime, and lower $\mmbulge$ scatter. 
    The nCW constraints are weaker than those of the GWB, but they add \ecgcirc{complementary} information, breaking the mass--number degeneracy. \ecgcirc{These constraints could} strengthen if nondetection continues.

    \item The binary properties of detectable CWs most consistent with our current CW upper limits, weighted by their S/Ns, are chirp masses of $\mchirp \sim 10^{9.5}\,\msol$, luminosity distances of $d_L \sim 700\,\tr{Mpc}$, and frequencies of $\sim$4\,nHz, coinciding with the \citetalias{ng+23_individuals} low-frequency candidate. 
    In contrast, the S/N-weighted GWB posteriors favor more massive ($\mchirp \sim 10^{9.7}\,\msol$), more distant ($d_L \sim 1200\,\tr{Mpc}$), higher-frequency ($\sim$6\,nHz) sources, and the joint posteriors fall \ecgcirc{near} the GWB predictions, reflecting the relative weakness of the current nCW constraints ($\mchirp \sim 10^{9.6}\,\msol$, $d_L \sim 1300\,\tr{Mpc}$, $f \sim 6\,\tr{nHz}$).
    
\end{enumerate}

\begin{acknowledgments}
\input{acks}
The authors disclose the use of Claude (Anthropic; Claude Opus and Claude Fable model family, 2025--2026) as a supporting tool for code development, debugging, and language editing during manuscript preparation. All analyses, results, and scientific conclusions were verified by the authors.
\end{acknowledgments}
\software{ \texttt{astropy} \citep{astropy},
\texttt{ceffyl} \citep{lamb2023rapid},
\texttt{cython} \citep{cython2011},
\texttt{holodeck} \citep{holodeck},
\texttt{kalepy} \citep{kalepy},
\texttt{matplotlib} \citep{matplotlib2007},
\texttt{numpy} \citep{numpy2011},
\texttt{pint} \citep{pint2018, pint2024},
\texttt{pta\_replicator} \citep{pta_replicator},
\texttt{QuickCW} \citep{becsy+2022_quickcw},
\texttt{scipy} \citep{2020SciPy-NMeth}
}

\begin{contribution}
E.C.G. was the primary lead for this project. E.C.G., K.G., and L.Z.K. conceptualized this project, with advising and computational support from L.B. E.C.G. conducted the analyses and manuscript writing, with ongoing input from K.G., for all components of this work except the $\dlum$ nCW method, which was developed and written up by P.X.T. and B.B. J.W.L. and S.R.T. contributed suggestions during the editing phase. Additional NANOGrav members are listed in alphabetical order; each contributed to the collaboration-wide endeavors underpinning this work, including pulsar timing and data processing, the NG15 GWB and CW analyses, and software development.
\end{contribution}


\bibliography{refs}{}
\bibliographystyle{aasjournalv7}



\end{document}

%% file: commands.tex
\usepackage{amsmath}
\usepackage{enumitem, amssymb}
\usepackage[thinc]{esdiff} 
\usepackage{pifont}
\usepackage[dvipsnames]{xcolor}

\newlist{todolist}{itemize}{2}
\setlist[todolist]{label=$\square$}
\newcommand{\tr}[1]{\textrm{#1}}

\newcommand{\ecgedit}[1]{\textcolor{Black}{{#1}}}
\newcommand{\ecgcirc}[1]{\textcolor{Black}{{#1}}}

\newcommand{\snr}{\tr{S}/\tr{N}}

\newcommand{\sigsig}{\snr_\tr{s}}

\newcommand{\bglike}{\mathcal{L}_\tr{GWB}}
\newcommand{\gwblike}{\mathcal{L}_\tr{GWB}}
\newcommand{\cwlike}{\mathcal{L}_\tr{nCW}}
\newcommand{\snlike}{\mathcal{L}_\tr{nCW, S/N}}
\newcommand{\dllike}{\mathcal{L}_{\tr{nCW}, d_L}}
\newcommand{\msw}{\mathcal{M}_{\tr{c}, \tr{sw}}}

\newcommand{\npar}{N_\tr{par}}

\newcommand{\kl}{D_\tr{KL}}
\newcommand{\gwb}{\tr{GWB}}
\newcommand{\ncw}{\tr{nCW}}

\newcommand{\bfthresh}{\overline{\bfact}}

\newcommand{\uniform}[2]{\mathcal{U}(#1, #2)}

\newcommand{\gsmfNormTot}{\Psi_0}
\newcommand{\gsmfnorm}{\psi_0}

\newcommand{\gsmfMassTot}{M_\psi}
\newcommand{\gsmfmass}{m_{\psi,0}}

\newcommand{\mmbamp}{\mu}
\newcommand{\mmbplaw}{\alpha_\mu}
\newcommand{\mmbscatter}{\epsilon_\mu}

\newcommand{\hardrchar}{a_c}
\newcommand{\hardainit}{a_\tr{init}}
\newcommand{\hardaisco}{a_\tr{isco}}
\newcommand{\hardnuinner}{\nu_\tr{inner}}
\newcommand{\hardnuouter}{\nu_\tr{outer}}
\newcommand{\harddadtnorm}{H_a}
\newcommand{\thardf}{\tau_f}

\newcommand{\mbh}{M_\tr{BH}}
\newcommand{\mbulge}{M_\tr{bulge}}
\newcommand{\mmbulge}{{\mbh\tr{--}\mbulge}}

\newcommand{\gsmffunc}{\Psi}

\newcommand{\msol}{\tr{M}_{\odot}}

\newcommand{\ndens}{\eta}

\newcommand{\mstar}{m_{\star1}}

\newcommand{\mchirp}{\mathcal{M}_\tr{c}}     
\newcommand{\distcom}{d_c}   
\newcommand{\dlum}{d_L}   

\newcommand{\hc}{h_\tr{c}}

\newcommand{\hscirc}{h_\tr{s,circ}}

\newcommand{\fo}[1][]{
    \ifthenelse{\equal{#1}{}}{
        f_\tr{o}
    }{
        {f_{\tr{o}, #1}}
    }
}

\newcommand{\logfo}[1][]{
    \ifthenelse{\equal{#1}{}}{
        \log_{10}f_\tr{o}
    }{
        \log_{10}{f_{\tr{o}, #1}}
    }
}

\newcommand{\lr}[2][]{
    \ifthenelse{\equal{#1}{}}{
        {\left(#2\right)}
    }{
        {\left(#2\right)}^{#1}
    }
}

\newcommand{\lrs}[2][]{
    \ifthenelse{\equal{#1}{}}{
        {\left[#2\right]}
    }{
        {\left[#2\right]}^{#1}
    }
}

\newcommand{\scale}[3][]{
    \ifthenelse{\equal{#1}{}}{
        \lr{ \frac{#2}{#3} }
    }{
        {\lr[#1]{ \frac{#2}{#3} }}
    }
}

\newcommand{\innerprod}[2]{( #1 | #2 )}

\newcommand{\signal}{\boldsymbol{s}}
\newcommand{\data}{\boldsymbol{d}}

\newcommand{\bfact}{\mathcal{B}}

\newcommand{\figref}[1]{Fig.~\ref{#1}}
\newcommand{\secref}[1]{Section~\ref{#1}}
\newcommand{\refeq}[1]{{Eq.~(\ref{#1})}}
\newcommand{\tabref}[1]{{Table~\ref{#1}}}

\defcitealias{ng+23_astro}{NG15astro}
\defcitealias{ng+23_individuals}{NG15cw}

%% file: authors_arxiv.tex
\author[0000-0002-8857-613X]{Emiko C. Gardiner}
\affiliation{Department of Astronomy, University of California, Berkeley, 501 Campbell Hall \#3411, Berkeley, CA 94720, USA}
\email{}
\author[0000-0002-1146-0198]{Kayhan G\"{u}ltekin}
\affiliation{Department of Astronomy and Astrophysics, University of Michigan, Ann Arbor, MI 48109, USA}
\email{}
\author{Philippe Turgeon}
\affiliation{Institute for Gravitational Wave Astronomy and School of Physics and Astronomy, University of Birmingham, Edgbaston, Birmingham B15 2TT, UK}
\email{}
\author[0000-0003-0909-5563]{Bence B\'{e}csy}
\affiliation{Institute for Gravitational Wave Astronomy and School of Physics and Astronomy, University of Birmingham, Edgbaston, Birmingham B15 2TT, UK}
\email{}
\author[0000-0002-2183-1087]{Laura Blecha}
\affiliation{Physics Department, University of Florida, Gainesville, FL 32611, USA}
\email{}
\author[0000-0001-7544-7876]{Nikita Agarwal}
\affiliation{Department of Physics and Astronomy, West Virginia University, P.O. Box 6315, Morgantown, WV 26506, USA}
\affiliation{Center for Gravitational Waves and Cosmology, West Virginia University, Chestnut Ridge Research Building, Morgantown, WV 26505, USA}
\email{}
\author[0000-0001-5134-3925]{Gabriella Agazie}
\affiliation{Center for Gravitation, Cosmology and Astrophysics, Department of Physics and Astronomy, University of Wisconsin-Milwaukee,\\ P.O. Box 413, Milwaukee, WI 53201, USA}
\email{}
\author[0000-0002-8935-9882]{Akash Anumarlapudi}
\affiliation{Department of Physics and Astronomy, University of North Carolina, Chapel Hill, NC 27599, USA}
\email{}
\author[0000-0003-0638-3340]{Anne M. Archibald}
\affiliation{Newcastle University, NE1 7RU, UK}
\email{}
\author[0009-0008-6187-8753]{Zaven Arzoumanian}
\affiliation{X-Ray Astrophysics Laboratory, NASA Goddard Space Flight Center, Code 662, Greenbelt, MD 20771, USA}
\email{}
% ...

%% file: acks.tex
The work of A.As., R.B., R.C., X.S., J.T., and D.W.\ is partly supported by the George and Hannah Bolinger Memorial Fund in the College of Science at Oregon State University.
A.As.\ gratefully acknowledges the support of the Moore Foundation.
L.B.\ acknowledges support from the National Science Foundation under award AST-2509457.
P.R.B.\ is supported by the Science and Technology Facilities Council, grant number ST/W000946/1.
S.B.\ gratefully acknowledges the support of a Sloan Fellowship, and the support of NSF under award \#1815664.
M.C.\ acknowledges support by the European Union (ERC, MMMonsters, 101117624).
S.C.\ received support from an NSF Astronomy and Astrophysics Grant (AAG) award number 2511105 during this work.
Support for this work was provided by the NSF through the Grote Reber Fellowship Program administered by Associated Universities, Inc./National Radio Astronomy Observatory.
Pulsar research at UBC is supported by an NSERC Discovery Grant and by CIFAR.
K.C.\ and M.S.T.\ are supported by a UBC Four Year Fellowship.
M.E.D.\ acknowledges support from the Naval Research Laboratory by NASA under contract S-15633Y.
T.D.\ and M.T.L.\ received support from an NSF Astronomy and Astrophysics Grant (AAG) award number 2511107 during this work.
E.C.F.\ is supported by NASA under award number 80GSFC24M0006.
E.C.G.\ is supported by a UC Berkeley Graduate Division Mentored Research Award.
K.A.G.\ and S.R.T.\ acknowledge support from an NSF CAREER award \#2146016.
D.C.G.\ is supported by NSF Astronomy and Astrophysics Grant (AAG) award \#2406919.
A.D.J.\ acknowledges support from the Caltech and Jet Propulsion Laboratory President's and Director's Research and Development Fund.
A.D.J.\ acknowledges support from the Sloan Foundation.
N.La.\ was supported by the Vanderbilt Initiative in Data Intensive Astrophysics (VIDA) Fellowship.
Part of this research was carried out at the Jet Propulsion Laboratory, California Institute of Technology, under a contract with the National Aeronautics and Space Administration (80NM0018D0004).
D.R.L.\ and M.A.M.\ are supported by NSF \#1458952.
M.A.M.\ is supported by NSF \#2009425.
C.M.F.M.\ was supported in part by the National Science Foundation under Grants No.\ NSF PHY-1748958 and NASA LPS 80NSSC24K0440. C.M.F.M.\ also thanks the Center for Computational Astrophysics (CCA) of the Flatiron Institute for support. The Flatiron Institute is supported by the Simons Foundation.
A.Mi.\ acknowledges support from a Royal Society University Research Fellowship (URF-R1-251896).
The Dunlap Institute is funded by an endowment established by the David Dunlap family and the University of Toronto.
K.D.O.\ was supported in part by NSF Grant No.\ 2207267.
T.T.P.\ acknowledges support from the Extragalactic Astrophysics Research Group at E\"{o}tv\"{o}s Lor\'{a}nd University, funded by the E\"{o}tv\"{o}s Lor\'{a}nd Research Network (ELKH), which was used during the development of this research.
P.P.\ and S.R.T.\ acknowledge support from NSF AST-2007993.
H.A.R.\ is supported by NSF Partnerships for Research and Education in Physics (PREP) award No.\ 2216793.
S.M.R.\ and I.H.S.\ are CIFAR Fellows.
Portions of this work performed at NRL were supported by ONR 6.1 basic research funding.
J.D.R.\ also acknowledges support from start-up funds from Texas Tech University.
S.C.S.\ and S.J.V.\ are supported by NSF award PHY-2011772.
J.S.\ is supported by an NSF Astronomy and Astrophysics Postdoctoral Fellowship under award AST-2202388, and acknowledges previous support by the NSF under award 1847938.
J.P.W.V.\ acknowledges support from NSF AccelNet award No.~2114721.
O.Y.\ is supported by the National Science Foundation Graduate Research Fellowship under Grant No.\ DGE-2139292.

%% file: refs.bib
@ARTICLE{ng+23_astro,
	Author = {{Agazie}, G. and {Anumarlapudi}, A. and {Archibald}, A. M. and others},
        title = "{The NANOGrav 15 yr Data Set: Constraints on Supermassive Black Hole Binaries from the Gravitational-wave Background}",
      journal = {\apjl},
         year = 2023,
        month = aug,
       volume = {952},
       number = {2},
          eid = {L37},
        pages = {L37},
          doi = {10.3847/2041-8213/ace18b},
archivePrefix = {arXiv},
       eprint = {2306.16220},
 primaryClass = {astro-ph.HE},
       adsurl = {https://ui.adsabs.harvard.edu/abs/2023ApJ...952L..37A},
    shorthand = {NG15astro}
}

@ARTICLE{ng+23_gwb,
	Author = {{Agazie}, G. and {Anumarlapudi}, A. and {Archibald}, A. M. and others},
        title = "{The NANOGrav 15 yr Data Set: Evidence for a Gravitational-wave Background}",
      journal = {\apjl},
         year = 2023,
        month = jul,
       volume = {951},
       number = {1},
          eid = {L8},
        pages = {L8},
          doi = {10.3847/2041-8213/acdac6},
archivePrefix = {arXiv},
       eprint = {2306.16213},
 primaryClass = {astro-ph.HE},
       adsurl = {https://ui.adsabs.harvard.edu/abs/2023ApJ...951L...8A}
}

@article{holodeck,
	Author = {{Kelley}, L. Z. and others},
	Journal = {\apjl},
	Title = {{Enter the \texttt{holodeck}: massive black hole binary population synthesis for gravitational wave calculations}},
         year = {in preparation}
 }

@ARTICLE{ng+23_individuals,
	Author = {{Agazie}, G. and {Anumarlapudi}, A. and {Archibald}, A. M. and others},
        title = "{The NANOGrav 15 yr Data Set: Bayesian Limits on Gravitational Waves from Individual Supermassive Black Hole Binaries}",
      journal = {\apjl},
         year = 2023,
        month = jul,
       volume = {951},
       number = {2},
          eid = {L50},
        pages = {L50},
          doi = {10.3847/2041-8213/ace18a},
archivePrefix = {arXiv},
       eprint = {2306.16222},
 primaryClass = {astro-ph.HE},
       adsurl = {https://ui.adsabs.harvard.edu/abs/2023ApJ...951L..50A},
  shorthand = {NG15cw}
}

@ARTICLE{ng+23_timing,
       author = {{Agazie}, Gabriella and {Alam}, Md Faisal and {Anumarlapudi}, Akash and others},
        title = "{The NANOGrav 15 yr Data Set: Observations and Timing of 68 Millisecond Pulsars}",
      journal = {\apjl},
         year = 2023,
        month = jul,
       volume = {951},
       number = {1},
          eid = {L9},
        pages = {L9},
          doi = {10.3847/2041-8213/acda9a},
archivePrefix = {arXiv},
       eprint = {2306.16217},
 primaryClass = {astro-ph.HE},
       adsurl = {https://ui.adsabs.harvard.edu/abs/2023ApJ...951L...9A}
}

@ARTICLE{epta+23_gwb,
       author = {{Antoniadis}, J. and {Babak}, S. and {Bak Nielsen}, A. -S. and others },
        title = "{The second data release from the European Pulsar Timing Array. I. The dataset and timing analysis}",
      journal = {\aap},
         year = 2023,
        month = oct,
       volume = {678},
          eid = {A48},
        pages = {A48},
          doi = {10.1051/0004-6361/202346841},
archivePrefix = {arXiv},
       eprint = {2306.16224},
 primaryClass = {astro-ph.HE},
       adsurl = {https://ui.adsabs.harvard.edu/abs/2023A&A...678A..48E}
}

@ARTICLE{epta+23_individuals,
       author = {{Antoniadis}, J. and {Arumugam}, P. and {Arumugam}, S. and others},
        title = "{The second data release from the European Pulsar Timing Array IV. Search for continuous gravitational wave signals}",
      journal = {arXiv e-prints},
         year = 2023,
        month = jun,
          eid = {arXiv:2306.16226},
        pages = {arXiv:2306.16226},
          doi = {10.48550/arXiv.2306.16226},
archivePrefix = {arXiv},
       eprint = {2306.16226},
 primaryClass = {astro-ph.HE},
       adsurl = {https://ui.adsabs.harvard.edu/abs/2023arXiv230616226A}
}

@ARTICLE{ppta+23_gwb,
       author = {{Reardon}, Daniel J. and {Zic}, Andrew and {Shannon}, Ryan M. and others},
        title = "{Search for an Isotropic Gravitational-wave Background with the Parkes Pulsar Timing Array}",
      journal = {\apjl},
         year = 2023,
        month = jul,
       volume = {951},
       number = {1},
          eid = {L6},
        pages = {L6},
          doi = {10.3847/2041-8213/acdd02},
archivePrefix = {arXiv},
       eprint = {2306.16215},
 primaryClass = {astro-ph.HE},
       adsurl = {https://ui.adsabs.harvard.edu/abs/2023ApJ...951L...6R}
}

@ARTICLE{cpta+23,
       author = {{Xu}, Heng and {Chen}, Siyuan and {Guo}, Yanjun and others},
        title = "{Searching for the Nano-Hertz Stochastic Gravitational Wave Background with the Chinese Pulsar Timing Array Data Release I}",
      journal = {Research in Astronomy and Astrophysics},
         year = 2023,
        month = jul,
       volume = {23},
       number = {7},
          eid = {075024},
        pages = {075024},
          doi = {10.1088/1674-4527/acdfa5},
archivePrefix = {arXiv},
       eprint = {2306.16216},
 primaryClass = {astro-ph.HE},
       adsurl = {https://ui.adsabs.harvard.edu/abs/2023RAA....23g5024X}
}

@ARTICLE{taylor2021,
       author = {{Taylor}, Stephen R.},
        title = "{The Nanohertz Gravitational Wave Astronomer}",
      journal = {arXiv e-prints},
         year = 2021,
        month = may,
          eid = {arXiv:2105.13270},
        pages = {arXiv:2105.13270},
          doi = {10.48550/arXiv.2105.13270},
archivePrefix = {arXiv},
       eprint = {2105.13270},
 primaryClass = {astro-ph.HE},
       adsurl = {https://ui.adsabs.harvard.edu/abs/2021arXiv210513270T}
}

@ARTICLE{becsy+2022_quickcw,
       author = {{B{\'e}csy}, Bence and {Cornish}, Neil J. and {Digman}, Matthew C.},
        title = "{Fast Bayesian analysis of individual binaries in pulsar timing array data}",
      journal = {\prd},
         year = 2022,
        month = jun,
       volume = {105},
       number = {12},
          eid = {122003},
        pages = {122003},
          doi = {10.1103/PhysRevD.105.122003},
archivePrefix = {arXiv},
       eprint = {2204.07160},
 primaryClass = {gr-qc},
       adsurl = {https://ui.adsabs.harvard.edu/abs/2022PhRvD.105l2003B}
}

@ARTICLE{becsy+2022_realistic,
       author = {{B{\'e}csy}, Bence and {Cornish}, Neil J. and {Kelley}, Luke Zoltan},
        title = "{Exploring Realistic Nanohertz Gravitational-wave Backgrounds}",
      journal = {\apj},
         year = 2022,
        month = dec,
       volume = {941},
       number = {2},
          eid = {119},
        pages = {119},
          doi = {10.3847/1538-4357/aca1b2},
archivePrefix = {arXiv},
       eprint = {2207.01607},
 primaryClass = {astro-ph.HE},
       adsurl = {https://ui.adsabs.harvard.edu/abs/2022ApJ...941..119B}
}

@ARTICLE{kelley+2018,
       author = {{Kelley}, Luke Zoltan and {Blecha}, Laura and {Hernquist}, Lars and {Sesana}, Alberto and {Taylor}, Stephen R.},
        title = "{Single sources in the low-frequency gravitational wave sky: properties and time to detection by pulsar timing arrays}",
      journal = {\mnras},
         year = 2018,
        month = jun,
       volume = {477},
       number = {1},
        pages = {964-976},
          doi = {10.1093/mnras/sty689},
archivePrefix = {arXiv},
       eprint = {1711.00075},
 primaryClass = {astro-ph.HE},
       adsurl = {https://ui.adsabs.harvard.edu/abs/2018MNRAS.477..964K}
}

@ARTICLE{rosado+2015,
       author = {{Rosado}, Pablo A. and {Sesana}, Alberto and {Gair}, Jonathan},
        title = "{Expected properties of the first gravitational wave signal detected with pulsar timing arrays}",
      journal = {\mnras},
         year = 2015,
        month = aug,
       volume = {451},
       number = {3},
        pages = {2417-2433},
          doi = {10.1093/mnras/stv1098},
archivePrefix = {arXiv},
       eprint = {1503.04803},
 primaryClass = {astro-ph.HE},
       adsurl = {https://ui.adsabs.harvard.edu/abs/2015MNRAS.451.2417R}
}

@ARTICLE{gardiner+2024,
       author = {{Gardiner}, Emiko C. and {Kelley}, Luke Zoltan and {Lemke}, Anna-Malin and {Mitridate}, Andrea},
        title = "{Beyond the Background: Gravitational-wave Anisotropy and Continuous Waves from Supermassive Black Hole Binaries}",
      journal = {\apj},
         year = 2024,
        month = apr,
       volume = {965},
       number = {2},
          eid = {164},
        pages = {164},
          doi = {10.3847/1538-4357/ad2be8},
archivePrefix = {arXiv},
       eprint = {2309.07227},
 primaryClass = {astro-ph.HE},
       adsurl = {https://ui.adsabs.harvard.edu/abs/2024ApJ...965..164G}
}

@ARTICLE{burkespolaor+2019,
       author = {{Burke-Spolaor}, Sarah and {Taylor}, Stephen R. and {Charisi}, Maria and {Dolch}, Timothy and {Hazboun}, Jeffrey S. and {Holgado}, A. Miguel and {Kelley}, Luke Zoltan and {Lazio}, T. Joseph W. and {Madison}, Dustin R. and {McMann}, Natasha and {Mingarelli}, Chiara M.~F. and {Rasskazov}, Alexander and {Siemens}, Xavier and {Simon}, Joseph J. and {Smith}, Tristan L.},
        title = "{The astrophysics of nanohertz gravitational waves}",
      journal = {\aapr},
         year = 2019,
        month = jun,
       volume = {27},
       number = {1},
          eid = {5},
        pages = {5},
          doi = {10.1007/s00159-019-0115-7},
archivePrefix = {arXiv},
       eprint = {1811.08826},
 primaryClass = {astro-ph.HE},
       adsurl = {https://ui.adsabs.harvard.edu/abs/2019A&ARv..27....5B}
}

@misc{enterprise,
  author       = {Justin A. Ellis and Michele Vallisneri and Stephen R. Taylor and Paul T. Baker},
  title        = {ENTERPRISE: Enhanced Numerical Toolbox Enabling a Robust PulsaR Inference SuitE},
  month        = sep,
  year         = 2020,
  howpublished = {Zenodo},
  doi          = {10.5281/zenodo.4059815},
  url          = {https://doi.org/10.5281/zenodo.4059815}
}

@INPROCEEDINGS{pint2018,
       author = {{Luo}, Jing and {Ransom}, Scott M. and {Demorest}, Paul and {Ray}, Paul S. and {Stovall}, Kevin and {Jenet}, Fredrick and {Ellis}, Justin and {van Haasteren}, Rutger and {Bachetti}, Matteo and {NANOGrav PINT developer Team}},
        title = "{PINT, A Modern Software Package for Pulsar Timing}",
    booktitle = {American Astronomical Society Meeting Abstracts \#231},
         year = 2018,
       series = {American Astronomical Society Meeting Abstracts},
       volume = {231},
        month = jan,
          eid = {453.09},
        pages = {453.09},
       adsurl = {https://ui.adsabs.harvard.edu/abs/2018AAS...23145309L}
}

@ARTICLE{pint2024,
       author = {{Susobhanan}, Abhimanyu and {Kaplan}, David L. and {Archibald}, Anne M. and {Luo}, Jing and {Ray}, Paul S. and {Pennucci}, Timothy T. and {Ransom}, Scott M. and {Agazie}, Gabriella and {Fiore}, William and {Larsen}, Bjorn and {O'Neill}, Patrick and {van Haasteren}, Rutger and {Anumarlapudi}, Akash and {Bachetti}, Matteo and {Bhakta}, Deven and {Champagne}, Chloe A. and {Cromartie}, H. Thankful and {Demorest}, Paul B. and {Jennings}, Ross J. and {Kerr}, Matthew and {Levina}, Sasha and {McEwen}, Alexander and {Shapiro-Albert}, Brent J. and {Swiggum}, Joseph K.},
        title = "{PINT: Maximum-likelihood Estimation of Pulsar Timing Noise Parameters}",
      journal = {\apj},
         year = 2024,
        month = aug,
       volume = {971},
       number = {2},
          eid = {150},
        pages = {150},
          doi = {10.3847/1538-4357/ad59f7},
archivePrefix = {arXiv},
       eprint = {2405.01977},
 primaryClass = {astro-ph.IM},
       adsurl = {https://ui.adsabs.harvard.edu/abs/2024ApJ...971..150S}
}

@ARTICLE{gundersen+cornish2025,
       author = {{Gundersen}, Aiden and {Cornish}, Neil J.},
        title = "{Rapid inference for individual binaries and a stochastic background with pulsar timing array data}",
      journal = {\prd},
         year = 2025,
        month = oct,
       volume = {112},
       number = {8},
          eid = {083035},
        pages = {083035},
          doi = {10.1103/5xh1-kgtk},
archivePrefix = {arXiv},
       eprint = {2412.13379},
 primaryClass = {gr-qc},
       adsurl = {https://ui.adsabs.harvard.edu/abs/2025PhRvD.112h3035G}
}

@ARTICLE{leja+2020,
       author = {{Leja}, Joel and {Speagle}, Joshua S. and {Johnson}, Benjamin D. and {Conroy}, Charlie and {van Dokkum}, Pieter and {Franx}, Marijn},
        title = "{A New Census of the 0.2 < z < 3.0 Universe. I. The Stellar Mass Function}",
      journal = {\apj},
         year = 2020,
        month = apr,
       volume = {893},
       number = {2},
          eid = {111},
        pages = {111},
          doi = {10.3847/1538-4357/ab7e27},
archivePrefix = {arXiv},
       eprint = {1910.04168},
 primaryClass = {astro-ph.GA},
       adsurl = {https://ui.adsabs.harvard.edu/abs/2020ApJ...893..111L}
}

@ARTICLE{rodriguezgomez+2015,
       author = {{Rodriguez-Gomez}, Vicente and {Genel}, Shy and {Vogelsberger}, Mark and {Sijacki}, Debora and {Pillepich}, Annalisa and {Sales}, Laura V. and {Torrey}, Paul and {Snyder}, Greg and {Nelson}, Dylan and {Springel}, Volker and {Ma}, Chung-Pei and {Hernquist}, Lars},
        title = "{The merger rate of galaxies in the Illustris simulation: a comparison with observations and semi-empirical models}",
      journal = {\mnras},
         year = 2015,
        month = may,
       volume = {449},
       number = {1},
        pages = {49-64},
          doi = {10.1093/mnras/stv264},
archivePrefix = {arXiv},
       eprint = {1502.01339},
 primaryClass = {astro-ph.GA},
       adsurl = {https://ui.adsabs.harvard.edu/abs/2015MNRAS.449...49R}
}

@ARTICLE{kh2013,
       author = {{Kormendy}, John and {Ho}, Luis C.},
        title = "{Coevolution (Or Not) of Supermassive Black Holes and Host Galaxies}",
      journal = {\araa},
         year = 2013,
        month = aug,
       volume = {51},
       number = {1},
        pages = {511-653},
          doi = {10.1146/annurev-astro-082708-101811},
archivePrefix = {arXiv},
       eprint = {1304.7762},
 primaryClass = {astro-ph.CO},
       adsurl = {https://ui.adsabs.harvard.edu/abs/2013ARA&A..51..511K}
}

@ARTICLE{Lang+2014,
       author = {{Lang}, Philipp and {Wuyts}, Stijn and {Somerville}, Rachel S. and {F{\"o}rster Schreiber}, Natascha M. and {Genzel}, Reinhard and {Bell}, Eric F. and {Brammer}, Gabe and {Dekel}, Avishai and {Faber}, Sandra M. and {Ferguson}, Henry C. and {Grogin}, Norman A. and {Kocevski}, Dale D. and {Koekemoer}, Anton M. and {Lutz}, Dieter and {McGrath}, Elizabeth J. and {Momcheva}, Ivelina and {Nelson}, Erica J. and {Primack}, Joel R. and {Rosario}, David J. and {Skelton}, Rosalind E. and {Tacconi}, Linda J. and {van Dokkum}, Pieter G. and {Whitaker}, Katherine E.},
        title = "{Bulge Growth and Quenching since z = 2.5 in CANDELS/3D-HST}",
      journal = {\apj},
         year = 2014,
        month = jun,
       volume = {788},
       number = {1},
          eid = {11},
        pages = {11},
          doi = {10.1088/0004-637X/788/1/11},
archivePrefix = {arXiv},
       eprint = {1402.0866},
 primaryClass = {astro-ph.GA},
       adsurl = {https://ui.adsabs.harvard.edu/abs/2014ApJ...788...11L}
}

@ARTICLE{Bluck+2014,
       author = {{Bluck}, Asa F.~L. and {Mendel}, J. Trevor and {Ellison}, Sara L. and {Moreno}, Jorge and {Simard}, Luc and {Patton}, David R. and {Starkenburg}, Else},
        title = "{Bulge mass is king: the dominant role of the bulge in determining the fraction of passive galaxies in the Sloan Digital Sky Survey}",
      journal = {\mnras},
         year = 2014,
        month = jun,
       volume = {441},
       number = {1},
        pages = {599-629},
          doi = {10.1093/mnras/stu594},
archivePrefix = {arXiv},
       eprint = {1403.5269},
 primaryClass = {astro-ph.GA},
       adsurl = {https://ui.adsabs.harvard.edu/abs/2014MNRAS.441..599B}
}

@ARTICLE{taylor+2016,
       author = {{Taylor}, S.~R. and {Huerta}, E.~A. and {Gair}, J.~R. and {McWilliams}, S.~T.},
        title = "{Detecting Eccentric Supermassive Black Hole Binaries with Pulsar Timing Arrays: Resolvable Source Strategies}",
      journal = {\apj},
         year = 2016,
        month = jan,
       volume = {817},
       number = {1},
          eid = {70},
        pages = {70},
          doi = {10.3847/0004-637X/817/1/70},
archivePrefix = {arXiv},
       eprint = {1505.06208},
 primaryClass = {gr-qc},
       adsurl = {https://ui.adsabs.harvard.edu/abs/2016ApJ...817...70T}
}

@ARTICLE{cython2011,
       author = {{Behnel}, Stefan and {Bradshaw}, Robert and {Citro}, Craig and {Dalcin}, Lisandro and {Seljebotn}, Dag Sverre and {Smith}, Kurt},
        title = "{Cython: The Best of Both Worlds}",
      journal = {Computing in Science and Engineering},
         year = 2011,
        month = mar,
       volume = {13},
       number = {2},
        pages = {31-39},
          doi = {10.1109/MCSE.2010.118},
       adsurl = {https://ui.adsabs.harvard.edu/abs/2011CSE....13b..31B}
}

@ARTICLE{astropy,
       author = {{Astropy Collaboration} and {Price-Whelan}, A.~M. and {Sip{\H{o}}cz}, B.~M. and {G{\"u}nther}, H.~M. and {Lim}, P.~L. and {Crawford}, S.~M. and {Conseil}, S. and {Shupe}, D.~L. and {Craig}, M.~W. and {Dencheva}, N. and {Ginsburg}, A. and {VanderPlas}, J.~T. and {Bradley}, L.~D. and {P{\'e}rez-Su{\'a}rez}, D. and {de Val-Borro}, M. and {Aldcroft}, T.~L. and {Cruz}, K.~L. and {Robitaille}, T.~P. and {Tollerud}, E.~J. and {Ardelean}, C. and {Babej}, T. and {Bach}, Y.~P. and {Bachetti}, M. and {Bakanov}, A.~V. and {Bamford}, S.~P. and {Barentsen}, G. and {Barmby}, P. and {Baumbach}, A. and {Berry}, K.~L. and {Biscani}, F. and {Boquien}, M. and {Bostroem}, K.~A. and {Bouma}, L.~G. and {Brammer}, G.~B. and {Bray}, E.~M. and {Breytenbach}, H. and {Buddelmeijer}, H. and {Burke}, D.~J. and {Calderone}, G. and {Cano Rodr{\'\i}guez}, J.~L. and {Cara}, M. and {Cardoso}, J.~V.~M. and {Cheedella}, S. and {Copin}, Y. and {Corrales}, L. and {Crichton}, D. and {D'Avella}, D. and {Deil}, C. and {Depagne}, {\'E}. and {Dietrich}, J.~P. and {Donath}, A. and {Droettboom}, M. and {Earl}, N. and {Erben}, T. and {Fabbro}, S. and {Ferreira}, L.~A. and {Finethy}, T. and {Fox}, R.~T. and {Garrison}, L.~H. and {Gibbons}, S.~L.~J. and {Goldstein}, D.~A. and {Gommers}, R. and {Greco}, J.~P. and {Greenfield}, P. and {Groener}, A.~M. and {Grollier}, F. and {Hagen}, A. and {Hirst}, P. and {Homeier}, D. and {Horton}, A.~J. and {Hosseinzadeh}, G. and {Hu}, L. and {Hunkeler}, J.~S. and {Ivezi{\'c}}, {\v{Z}}. and {Jain}, A. and {Jenness}, T. and {Kanarek}, G. and {Kendrew}, S. and {Kern}, N.~S. and {Kerzendorf}, W.~E. and {Khvalko}, A. and {King}, J. and {Kirkby}, D. and {Kulkarni}, A.~M. and {Kumar}, A. and {Lee}, A. and {Lenz}, D. and {Littlefair}, S.~P. and {Ma}, Z. and {Macleod}, D.~M. and {Mastropietro}, M. and {McCully}, C. and {Montagnac}, S. and {Morris}, B.~M. and {Mueller}, M. and {Mumford}, S.~J. and {Muna}, D. and {Murphy}, N.~A. and {Nelson}, S. and {Nguyen}, G.~H. and {Ninan}, J.~P. and {N{\"o}the}, M. and {Ogaz}, S. and {Oh}, S. and {Parejko}, J.~K. and {Parley}, N. and {Pascual}, S. and {Patil}, R. and {Patil}, A.~A. and {Plunkett}, A.~L. and {Prochaska}, J.~X. and {Rastogi}, T. and {Reddy Janga}, V. and {Sabater}, J. and {Sakurikar}, P. and {Seifert}, M. and {Sherbert}, L.~E. and {Sherwood-Taylor}, H. and {Shih}, A.~Y. and {Sick}, J. and {Silbiger}, M.~T. and {Singanamalla}, S. and {Singer}, L.~P. and {Sladen}, P.~H. and {Sooley}, K.~A. and {Sornarajah}, S. and {Streicher}, O. and {Teuben}, P. and {Thomas}, S.~W. and {Tremblay}, G.~R. and {Turner}, J.~E.~H. and {Terr{\'o}n}, V. and {van Kerkwijk}, M.~H. and {de la Vega}, A. and {Watkins}, L.~L. and {Weaver}, B.~A. and {Whitmore}, J.~B. and {Woillez}, J. and {Zabalza}, V. and {Astropy Contributors}},
        title = "{The Astropy Project: Building an Open-science Project and Status of the v2.0 Core Package}",
      journal = {\aj},
         year = 2018,
        month = sep,
       volume = {156},
       number = {3},
          eid = {123},
        pages = {123},
          doi = {10.3847/1538-3881/aabc4f},
archivePrefix = {arXiv},
       eprint = {1801.02634},
 primaryClass = {astro-ph.IM},
       adsurl = {https://ui.adsabs.harvard.edu/abs/2018AJ....156..123A}
}

@article{kalepy, doi = {10.21105/joss.02784}, url = {https://doi.org/10.21105/joss.02784}, year = {2021}, publisher = {The Open Journal}, volume = {6}, number = {57}, pages = {2784}, author = {Luke Zoltan Kelley}, title = {kalepy: a Python package for kernel density estimation, sampling and plotting}, journal = {Journal of Open Source Software} }

@article{numpy2011,
	adsurl = {https://arxiv.org/abs/1102.1523},
	archiveprefix = {arXiv},
	author = {St{\'{e}}fan van der Walt and S. Chris Colbert and Ga{\"{e}}l Varoquaux},
	bibsource = {dblp computer science bibliography, http://dblp.org},
	eprint = {1102.1523},
	journal = {CoRR},
	title = {The NumPy array: a structure for efficient numerical computation},
	url = {http://arxiv.org/abs/1102.1523},
	volume = {1102.1523},
	year = {2011}
 }

@article{matplotlib2007,
	author = {Hunter, J. D.},
	journal = {Computing In Science \& Engineering},
	number = {3},
	pages = {90--95},
	publisher = {IEEE COMPUTER SOC},
	title = {Matplotlib: A 2D graphics environment},
	volume = {9},
	year = 2007
 }

@ARTICLE{2020SciPy-NMeth,
  author  = {Virtanen, Pauli and Gommers, Ralf and Oliphant, Travis E. and
            Haberland, Matt and Reddy, Tyler and Cournapeau, David and
            Burovski, Evgeni and Peterson, Pearu and Weckesser, Warren and
            Bright, Jonathan and {van der Walt}, St{\'e}fan J. and
            Brett, Matthew and Wilson, Joshua and Millman, K. Jarrod and
            Mayorov, Nikolay and Nelson, Andrew R. J. and Jones, Eric and
            Kern, Robert and Larson, Eric and Carey, C J and
            Polat, {\.I}lhan and Feng, Yu and Moore, Eric W. and
            {VanderPlas}, Jake and Laxalde, Denis and Perktold, Josef and
            Cimrman, Robert and Henriksen, Ian and Quintero, E. A. and
            Harris, Charles R. and Archibald, Anne M. and
            Ribeiro, Ant{\^o}nio H. and Pedregosa, Fabian and
            {van Mulbregt}, Paul and {SciPy 1.0 Contributors}},
  title   = {{{SciPy} 1.0: Fundamental Algorithms for Scientific
            Computing in Python}},
  journal = {Nature Methods},
  year    = {2020},
  volume  = {17},
  pages   = {261--272},
  adsurl  = {https://rdcu.be/b08Wh},
  doi     = {10.1038/s41592-019-0686-2},
}

@misc{pta_replicator,
  author       = {Bence {B{\'e}csy} and Jeff Hazboun and Aaron Johnson},
  title        = {pta\_replicator},
  year         = {2025},  
  howpublished = {GitHub repository},
  note         = {Available at \url{https://github.com/bencebecsy/pta_replicator}},
  url          = {https://github.com/bencebecsy/pta_replicator}
}

@ARTICLE{mpta+2025,
       author = {{Miles}, Matthew T. and {Shannon}, Ryan M. and {Reardon}, Daniel J. and {Bailes}, Matthew and {Champion}, David J. and {Geyer}, Marisa and {Gitika}, Pratyasha and {Grunthal}, Kathrin and {Keith}, Michael J. and {Kramer}, Michael and {Kulkarni}, Atharva D. and {Nathan}, Rowina S. and {Parthasarathy}, Aditya and {Singha}, Jaikhomba and {Theureau}, Gilles and {Thrane}, Eric and {Abbate}, Federico and {Buchner}, Sarah and {Cameron}, Andrew D. and {Camilo}, Fernando and {Moreschi}, Beatrice E. and {Shaifullah}, Golam and {Shamohammadi}, Mohsen and {Possenti}, Andrea and {Krishnan}, Vivek Venkatraman},
        title = "{The MeerKAT Pulsar Timing Array: the first search for gravitational waves with the MeerKAT radio telescope}",
      journal = {\mnras},
         year = 2025,
        month = jan,
       volume = {536},
       number = {2},
        pages = {1489-1500},
          doi = {10.1093/mnras/stae2571},
archivePrefix = {arXiv},
       eprint = {2412.01153},
 primaryClass = {astro-ph.HE},
       adsurl = {https://ui.adsabs.harvard.edu/abs/2025MNRAS.536.1489M}
}

@ARTICLE{gardiner+2025,
       author = {{Gardiner}, Emiko C. and {B{\'e}csy}, Bence and {Kelley}, Luke Zoltan and {Cornish}, Neil J.},
        title = "{Characterizing Continuous Gravitational Waves from Supermassive Black Hole Binaries in Realistic Pulsar Timing Array Data}",
      journal = {\apj},
         year = 2025,
        month = aug,
       volume = {988},
       number = {2},
          eid = {222},
        pages = {222},
          doi = {10.3847/1538-4357/ade4c2},
archivePrefix = {arXiv},
       eprint = {2502.16016},
 primaryClass = {astro-ph.CO},
       adsurl = {https://ui.adsabs.harvard.edu/abs/2025ApJ...988..222G}
}

@article{turgeon+2026,
	Author = {{Turgeon}, P. X. and others},
	
         year = {in preparation}
 }

@ARTICLE{chen+2019,
       author = {{Chen}, Siyuan and {Sesana}, Alberto and {Conselice}, Christopher J.},
        title = "{Constraining astrophysical observables of galaxy and supermassive black hole binary mergers using pulsar timing arrays}",
      journal = {\mnras},
         year = 2019,
        month = sep,
       volume = {488},
       number = {1},
        pages = {401-418},
          doi = {10.1093/mnras/stz1722},
archivePrefix = {arXiv},
       eprint = {1810.04184},
 primaryClass = {astro-ph.GA},
       adsurl = {https://ui.adsabs.harvard.edu/abs/2019MNRAS.488..401C}
}

@ARTICLE{kelley+2017,
       author = {{Kelley}, Luke Zoltan and {Blecha}, Laura and {Hernquist}, Lars},
        title = "{Massive black hole binary mergers in dynamical galactic environments}",
      journal = {\mnras},
         year = 2017,
        month = jan,
       volume = {464},
       number = {3},
        pages = {3131-3157},
          doi = {10.1093/mnras/stw2452},
archivePrefix = {arXiv},
       eprint = {1606.01900},
 primaryClass = {astro-ph.HE},
       adsurl = {https://ui.adsabs.harvard.edu/abs/2017MNRAS.464.3131K}
}

@ARTICLE{sesana+2008,
       author = {{Sesana}, A. and {Vecchio}, A. and {Colacino}, C.~N.},
        title = "{The stochastic gravitational-wave background from massive black hole binary systems: implications for observations with Pulsar Timing Arrays}",
      journal = {\mnras},
         year = 2008,
        month = oct,
       volume = {390},
       number = {1},
        pages = {192-209},
          doi = {10.1111/j.1365-2966.2008.13682.x},
archivePrefix = {arXiv},
       eprint = {0804.4476},
 primaryClass = {astro-ph},
       adsurl = {https://ui.adsabs.harvard.edu/abs/2008MNRAS.390..192S}
}

@ARTICLE{ellis+2012snr,
       author = {{Ellis}, J.~A. and {Jenet}, F.~A. and {McLaughlin}, M.~A.},
        title = "{Practical Methods for Continuous Gravitational Wave Detection Using Pulsar Timing Data}",
      journal = {\apj},
         year = 2012,
        month = jul,
       volume = {753},
       number = {2},
          eid = {96},
        pages = {96},
          doi = {10.1088/0004-637X/753/2/96},
archivePrefix = {arXiv},
       eprint = {1202.0808},
 primaryClass = {astro-ph.IM},
       adsurl = {https://ui.adsabs.harvard.edu/abs/2012ApJ...753...96E}
}

@article{Finn+2000,
  title = {Gravitational waves from a compact star in a circular, inspiral orbit, in the equatorial plane of a massive, spinning black hole, as observed by LISA},
  author = {Finn, Lee Samuel and Thorne, Kip S.},
  journal = {Phys. Rev. D},
  volume = {62},
  issue = {12},
  pages = {124021},
  numpages = {20},
  year = {2000},
  month = {Nov},
  publisher = {American Physical Society},
  doi = {10.1103/PhysRevD.62.124021},
  url = {https://link.aps.org/doi/10.1103/PhysRevD.62.124021}
}

@ARTICLE{Richstone+1998_SMBHs,
       author = {{Richstone}, D. and {Ajhar}, E.~A. and {Bender}, R. and {Bower}, G. and {Dressler}, A. and {Faber}, S.~M. and {Filippenko}, A.~V. and {Gebhardt}, K. and {Green}, R. and {Ho}, L.~C. and {Kormendy}, J. and {Lauer}, T.~R. and {Magorrian}, J. and {Tremaine}, S.},
        title = "{Supermassive black holes and the evolution of galaxies.}",
      journal = {\nat},
         year = 1998,
        month = oct,
       volume = {395},
       number = {6701},
        pages = {A14},
          doi = {10.48550/arXiv.astro-ph/9810378},
archivePrefix = {arXiv},
       eprint = {astro-ph/9810378},
 primaryClass = {astro-ph},
       adsurl = {https://ui.adsabs.harvard.edu/abs/1998Natur.395A..14R}
}

@ARTICLE{Lacey+1993_galaxymergers,
       author = {{Lacey}, Cedric and {Cole}, Shaun},
        title = "{Merger rates in hierarchical models of galaxy formation}",
      journal = {\mnras},
         year = 1993,
        month = jun,
       volume = {262},
       number = {3},
        pages = {627-649},
          doi = {10.1093/mnras/262.3.627},
       adsurl = {https://ui.adsabs.harvard.edu/abs/1993MNRAS.262..627L}
}

@ARTICLE{BBR_1980,
       author = {{Begelman}, M.~C. and {Blandford}, R.~D. and {Rees}, M.~J.},
        title = "{Massive black hole binaries in active galactic nuclei}",
      journal = {\nat},
         year = 1980,
        month = sep,
       volume = {287},
       number = {5780},
        pages = {307-309},
          doi = {10.1038/287307a0},
       adsurl = {https://ui.adsabs.harvard.edu/abs/1980Natur.287..307B}
}

@ARTICLE{merritt+2005,
       author = {{Merritt}, David and {Milosavljevi{\'c}}, Milos},
        title = "{Massive Black Hole Binary Evolution}",
      journal = {Living Reviews in Relativity},
         year = 2005,
        month = nov,
       volume = {8},
        pages = {8},
          doi = {10.12942/lrr-2005-8},
archivePrefix = {arXiv},
       eprint = {astro-ph/0410364},
 primaryClass = {astro-ph},
       adsurl = {https://ui.adsabs.harvard.edu/abs/2005LRR.....8....8M}
}

@ARTICLE{mm2013,
       author = {{McConnell}, Nicholas J. and {Ma}, Chung-Pei},
        title = "{Revisiting the Scaling Relations of Black Hole Masses and Host Galaxy Properties}",
      journal = {\apj},
         year = 2013,
        month = feb,
       volume = {764},
       number = {2},
          eid = {184},
        pages = {184},
          doi = {10.1088/0004-637X/764/2/184},
archivePrefix = {arXiv},
       eprint = {1211.2816},
 primaryClass = {astro-ph.CO},
       adsurl = {https://ui.adsabs.harvard.edu/abs/2013ApJ...764..184M}
}

@ARTICLE{matt+2026evol,
       author = {{Matt}, Cayenne and {G{\"u}ltekin}, Kayhan and {Kelley}, Luke Zoltan and {Blecha}, Laura and {Simon}, Joseph and {Agazie}, Gabriella and {Anumarlapudi}, Akash and {Archibald}, Anne M. and {Arzoumanian}, Zaven and {Baier}, Jeremy G. and {Baker}, Paul T. and {B{\'e}csy}, Bence and {Brazier}, Adam and {Brook}, Paul R. and {Burke-Spolaor}, Sarah and {Burnette}, Rand and {Case}, Robin and {Casey-Clyde}, J. Andrew and {Charisi}, Maria and {Chatterjee}, Shami and {Cohen}, Tyler and {Cordes}, James M. and {Cornish}, Neil J. and {Crawford}, Fronefield and {Cromartie}, H. Thankful and {Crowter}, Kathryn and {DeCesar}, Megan E. and {Demorest}, Paul B. and {Deng}, Heling and {Dey}, Lankeswar and {Dolch}, Timothy and {Ferrara}, Elizabeth C. and {Fiore}, William and {Fonseca}, Emmanuel and {Freedman}, Gabriel E. and {Gardiner}, Emiko C. and {Garver-Daniels}, Nate and {Gentile}, Peter A. and {Gersbach}, Kyle A. and {Glaser}, Joseph and {Good}, Deborah C. and {Harris}, C.~J. and {Hazboun}, Jeffrey S. and {Jennings}, Ross J. and {Johnson}, Aaron D. and {Jones}, Megan L. and {Kaplan}, David L. and {Kerr}, Matthew and {Key}, Joey S. and {Laal}, Nima and {Lam}, Michael T. and {Lamb}, William G. and {Larsen}, Bjorn and {Lazio}, T. Joseph W. and {Lewandowska}, Natalia and {Liu}, Tingting and {Lorimer}, Duncan R. and {Luo}, Jing and {Lynch}, Ryan S. and {Ma}, Chung-Pei and {Madison}, Dustin R. and {McEwen}, Alexander and {McKee}, James W. and {McLaughlin}, Maura A. and {McMann}, Natasha and {Meyers}, Bradley W. and {Meyers}, Patrick M. and {Mingarelli}, Chiara M.~F. and {Mitridate}, Andrea and {Ng}, Cherry and {Nice}, David J. and {Ocker}, Stella Koch and {Olum}, Ken D. and {Pennucci}, Timothy T. and {Perera}, Benetge B.~P. and {Petrov}, Polina and {Pol}, Nihan S. and {Radovan}, Henri A. and {Ransom}, Scott M. and {Ray}, Paul S. and {Romano}, Joseph D. and {Runnoe}, Jessie C. and {Saffer}, Alexander and {Sardesai}, Shashwat C. and {Schmiedekamp}, Ann and {Schmiedekamp}, Carl and {Schmitz}, Kai and {Shapiro-Albert}, Brent J. and {Siemens}, Xavier and {Fiscella}, Sophia V. Sosa and {Stairs}, Ingrid H. and {Stinebring}, Daniel R. and {Stovall}, Kevin and {Susobhanan}, Abhimanyu and {Swiggum}, Joseph K. and {Taylor}, Jacob and {Taylor}, Stephen R. and {Thompson}, Mercedes S. and {Turner}, Jacob E. and {Vallisneri}, Michele and {van Haasteren}, Rutger and {Vigeland}, Sarah J. and {Wahl}, Haley M. and {Wilson}, Kevin P. and {Witt}, Caitlin A. and {Wright}, David and {Young}, Olivia},
        title = "{Inferring M$_{BH}$─M$_{bulge}$ Evolution from the Gravitational-wave Background}",
      journal = {\apj},
         year = 2026,
        month = feb,
       volume = {997},
       number = {2},
          eid = {188},
        pages = {188},
          doi = {10.3847/1538-4357/ae2480},
archivePrefix = {arXiv},
       eprint = {2508.18126},
 primaryClass = {astro-ph.HE},
       adsurl = {https://ui.adsabs.harvard.edu/abs/2026ApJ...997..188M}
}

@ARTICLE{matt+2023,
       author = {{Matt}, Cayenne and {G{\"u}ltekin}, Kayhan and {Simon}, Joseph},
        title = "{The impact of black hole scaling relation assumptions on the mass density of black holes}",
      journal = {\mnras},
         year = 2023,
        month = sep,
       volume = {524},
       number = {3},
        pages = {4403-4417},
          doi = {10.1093/mnras/stad2146},
archivePrefix = {arXiv},
       eprint = {2307.04878},
 primaryClass = {astro-ph.GA},
       adsurl = {https://ui.adsabs.harvard.edu/abs/2023MNRAS.524.4403M}
}

@ARTICLE{matt+2026scatter,
       author = {{Matt}, Cayenne and {G{\"u}ltekin}, Kayhan and {Agazie}, Gabriella and {Agarwal}, Nikita and {Anumarlapudi}, Akash and {Archibald}, Anne M. and {Arzoumanian}, Zaven and {Baier}, Jeremy G. and {Baker}, Paul T. and {B{\'e}csy}, Bence and {Blecha}, Laura and {Brazier}, Adam and {Brook}, Paul R. and {Burke-Spolaor}, Sarah and {Burnette}, Rand and {Case}, Robin and {Casey-Clyde}, J. Andrew and {Charisi}, Maria and {Chatterjee}, Shami and {Cohen}, Tyler and {Cordes}, James M. and {Cornish}, Neil J. and {Crawford}, Fronefield and {Cromartie}, H. Thankful and {Crowter}, Kathryn and {DeCesar}, Megan E. and {Demorest}, Paul B. and {Deng}, Heling and {Dey}, Lankeswar and {Dolch}, Timothy and {Doskoch}, Graham M. and {Ferrara}, Elizabeth C. and {Fiore}, William and {Fonseca}, Emmanuel and {Freedman}, Gabriel E. and {Gardiner}, Emiko C. and {Garver-Daniels}, Nate and {Gentile}, Peter A. and {Gersbach}, Kyle A. and {Glaser}, Joseph and {Good}, Deborah C. and {Harris}, C.~J. and {Hazboun}, Jeffrey S. and {Jennings}, Ross J. and {Johnson}, Aaron D. and {Jones}, Megan L. and {Kaplan}, David L. and {Kavumkandathil Sreekumar}, Anala and {Kelley}, Luke Zoltan and {Kerr}, Matthew and {Key}, Joey S. and {Laal}, Nima and {Lam}, Michael T. and {Lamb}, William G. and {Larsen}, Bjorn and {Lazio}, T. Joseph W. and {Lewandowska}, Natalia and {Liu}, Tingting and {Lorimer}, Duncan R. and {Luo}, Jing and {Lynch}, Ryan S. and {Ma}, Chung-Pei and {Madison}, Dustin R. and {Martsen}, Ashley and {McEwen}, Alexander and {McKee}, James W. and {McLaughlin}, Maura A. and {McMann}, Natasha and {Meyers}, Bradley W. and {Meyers}, Patrick M. and {Mingarelli}, Chiara M.~F. and {Mitridate}, Andrea and {Ng}, Cherry and {Nice}, David J. and {Nichols}, Shania and {Ocker}, Stella Koch and {Olum}, Ken D. and {Pennucci}, Timothy T. and {Perera}, Benetge B.~P. and {Petrov}, Polina and {Pol}, Nihan S. and {Radovan}, Henri A. and {Ransom}, Scott M. and {Ray}, Paul S. and {Romano}, Joseph D. and {Runnoe}, Jessie C. and {Saffer}, Alexander and {Sardesai}, Shashwat C. and {Schmiedekamp}, Ann and {Schmiedekamp}, Carl and {Schmitz}, Kai and {Shapiro-Albert}, Brent J. and {Siemens}, Xavier and {Simon}, Joseph and {Sosa Fiscella}, Sophia V. and {Stairs}, Ingrid H. and {Stinebring}, Daniel R. and {Stovall}, Kevin and {Susobhanan}, Abhimanyu and {Swiggum}, Joseph K. and {Taylor}, Jacob and {Taylor}, Stephen R. and {Thompson}, Mercedes S. and {Turner}, Jacob E. and {Vallisneri}, Michele and {van Haasteren}, Rutger and {Vigeland}, Sarah J. and {Wahl}, Haley M. and {Wilson}, Kevin P. and {Witt}, Caitlin A. and {Wright}, David and {Young}, Olivia},
        title = "{Gravitational Wave Measurement of the $M_\mathrm{BH}$-$M_\mathrm{bulge}$ Intrinsic Scatter at High Redshift}",
      journal = {arXiv e-prints},
         year = 2026,
        month = mar,
          eid = {arXiv:2603.11167},
        pages = {arXiv:2603.11167},
          doi = {10.48550/arXiv.2603.11167},
archivePrefix = {arXiv},
       eprint = {2603.11167},
 primaryClass = {astro-ph.HE},
       adsurl = {https://ui.adsabs.harvard.edu/abs/2026arXiv260311167M}
}

@ARTICLE{chandrasekhar1943,
       author = {{Chandrasekhar}, S.},
        title = "{Dynamical Friction. I. General Considerations: the Coefficient of Dynamical Friction.}",
      journal = {\apj},
         year = 1943,
        month = mar,
       volume = {97},
        pages = {255},
          doi = {10.1086/144517},
       adsurl = {https://ui.adsabs.harvard.edu/abs/1943ApJ....97..255C}
}

@ARTICLE{siwek+2023,
       author = {{Siwek}, Magdalena and {Weinberger}, Rainer and {Hernquist}, Lars},
        title = "{Orbital evolution of binaries in circumbinary discs}",
      journal = {\mnras},
         year = 2023,
        month = jun,
       volume = {522},
       number = {2},
        pages = {2707-2717},
          doi = {10.1093/mnras/stad1131},
archivePrefix = {arXiv},
       eprint = {2302.01785},
 primaryClass = {astro-ph.HE},
       adsurl = {https://ui.adsabs.harvard.edu/abs/2023MNRAS.522.2707S}
}

@ARTICLE{harris+2026,
       author = {{Harris}, C.~J. and {G{\"u}ltekin}, Kayhan and {Blecha}, Laura},
        title = "{Core Scouring Dynamics and Gravitational Wave Consequences: Constraints on Supermassive Black Hole Binary Hardening}",
      journal = {arXiv e-prints},
         year = 2026,
        month = jan,
          eid = {arXiv:2601.07762},
        pages = {arXiv:2601.07762},
          doi = {10.48550/arXiv.2601.07762},
archivePrefix = {arXiv},
       eprint = {2601.07762},
 primaryClass = {astro-ph.GA},
       adsurl = {https://ui.adsabs.harvard.edu/abs/2026arXiv260107762H}
}

@ARTICLE{peters1964,
       author = {{Peters}, P.~C.},
        title = "{Gravitational Radiation and the Motion of Two Point Masses}",
      journal = {Physical Review},
         year = 1964,
        month = nov,
       volume = {136},
       number = {4B},
        pages = {1224-1232},
          doi = {10.1103/PhysRev.136.B1224},
       adsurl = {https://ui.adsabs.harvard.edu/abs/1964PhRv..136.1224P}
}

@BOOK{binney+tremaine1987,
       author = {{Binney}, James and {Tremaine}, Scott},
        title = "{Galactic dynamics}",
         year = 1987,
       adsurl = {https://ui.adsabs.harvard.edu/abs/1987gady.book.....B}
}

@ARTICLE{boylankolchin+2008,
       author = {{Boylan-Kolchin}, Michael and {Ma}, Chung-Pei and {Quataert}, Eliot},
        title = "{Dynamical friction and galaxy merging time-scales}",
      journal = {\mnras},
         year = 2008,
        month = jan,
       volume = {383},
       number = {1},
        pages = {93-101},
          doi = {10.1111/j.1365-2966.2007.12530.x},
archivePrefix = {arXiv},
       eprint = {0707.2960},
 primaryClass = {astro-ph},
       adsurl = {https://ui.adsabs.harvard.edu/abs/2008MNRAS.383...93B}
}

@ARTICLE{mm2003_finalparsec,
       author = {{Milosavljevi{\'c}}, Milo{\v{s}} and {Merritt}, David},
        title = "{Long-Term Evolution of Massive Black Hole Binaries}",
      journal = {\apj},
         year = 2003,
        month = oct,
       volume = {596},
       number = {2},
        pages = {860-878},
          doi = {10.1086/378086},
archivePrefix = {arXiv},
       eprint = {astro-ph/0212459},
 primaryClass = {astro-ph},
       adsurl = {https://ui.adsabs.harvard.edu/abs/2003ApJ...596..860M}
}

@ARTICLE{bonetti+2018,
       author = {{Bonetti}, Matteo and {Sesana}, Alberto and {Barausse}, Enrico and {Haardt}, Francesco},
        title = "{Post-Newtonian evolution of massive black hole triplets in galactic nuclei - III. A robust lower limit to the nHz stochastic background of gravitational waves}",
      journal = {\mnras},
         year = 2018,
        month = jun,
       volume = {477},
       number = {2},
        pages = {2599-2612},
          doi = {10.1093/mnras/sty874},
archivePrefix = {arXiv},
       eprint = {1709.06095},
 primaryClass = {astro-ph.GA},
       adsurl = {https://ui.adsabs.harvard.edu/abs/2018MNRAS.477.2599B}
}

@ARTICLE{ostriker1999,
       author = {{Ostriker}, Eve C.},
        title = "{Dynamical Friction in a Gaseous Medium}",
      journal = {\apj},
         year = 1999,
        month = mar,
       volume = {513},
       number = {1},
        pages = {252-258},
          doi = {10.1086/306858},
archivePrefix = {arXiv},
       eprint = {astro-ph/9810324},
 primaryClass = {astro-ph},
       adsurl = {https://ui.adsabs.harvard.edu/abs/1999ApJ...513..252O}
}

@ARTICLE{liepold+ma2024,
       author = {{Liepold}, Emily R. and {Ma}, Chung-Pei},
        title = "{Big Galaxies and Big Black Holes: The Massive Ends of the Local Stellar and Black Hole Mass Functions and the Implications for Nanohertz Gravitational Waves}",
      journal = {\apjl},
         year = 2024,
        month = aug,
       volume = {971},
       number = {2},
          eid = {L29},
        pages = {L29},
          doi = {10.3847/2041-8213/ad66b8},
archivePrefix = {arXiv},
       eprint = {2407.14595},
 primaryClass = {astro-ph.GA},
       adsurl = {https://ui.adsabs.harvard.edu/abs/2024ApJ...971L..29L}
}

@ARTICLE{spzq2024_where,
       author = {{Sato-Polito}, Gabriela and {Zaldarriaga}, Matias and {Quataert}, Eliot},
        title = "{Where are the supermassive black holes measured by PTAs?}",
      journal = {\prd},
         year = 2024,
        month = sep,
       volume = {110},
       number = {6},
          eid = {063020},
        pages = {063020},
          doi = {10.1103/PhysRevD.110.063020},
       adsurl = {https://ui.adsabs.harvard.edu/abs/2024PhRvD.110f3020S}
}

@ARTICLE{goncharov+2026,
       author = {{Goncharov}, Boris and {Sato-Polito}, Gabriela and {Bi}, Xiaoming and {Zaldarriaga}, Matias},
        title = "{A Joint Optimal Search for Gravitational Waves from Resolved and Unresolved Supermassive Binary Black Holes with Pulsar Timing Arrays}",
      journal = {arXiv e-prints},
         year = 2026,
        month = jun,
          eid = {arXiv:2606.18241},
        pages = {arXiv:2606.18241},
          doi = {10.48550/arXiv.2606.18241},
archivePrefix = {arXiv},
       eprint = {2606.18241},
 primaryClass = {astro-ph.HE},
       adsurl = {https://ui.adsabs.harvard.edu/abs/2026arXiv260618241G}
}

@ARTICLE{spz2025_distribution,
       author = {{Sato-Polito}, Gabriela and {Zaldarriaga}, Matias},
        title = "{Distribution of the gravitational-wave background from supermassive black holes}",
      journal = {\prd},
         year = 2025,
        month = jan,
       volume = {111},
       number = {2},
          eid = {023043},
        pages = {023043},
          doi = {10.1103/PhysRevD.111.023043},
archivePrefix = {arXiv},
       eprint = {2406.17010},
 primaryClass = {astro-ph.CO},
       adsurl = {https://ui.adsabs.harvard.edu/abs/2025PhRvD.111b3043S}
}

@ARTICLE{izquierdo+2022,
       author = {{Izquierdo-Villalba}, David and {Sesana}, Alberto and {Bonoli}, Silvia and {Colpi}, Monica},
        title = "{Massive black hole evolution models confronting the n-Hz amplitude of the stochastic gravitational wave background}",
      journal = {\mnras},
         year = 2022,
        month = jan,
       volume = {509},
       number = {3},
        pages = {3488-3503},
          doi = {10.1093/mnras/stab3239},
archivePrefix = {arXiv},
       eprint = {2108.11671},
 primaryClass = {astro-ph.GA},
       adsurl = {https://ui.adsabs.harvard.edu/abs/2022MNRAS.509.3488I}
}

@ARTICLE{spz2025_uncertainties,
       author = {{Sato-Polito}, Gabriela and {Zaldarriaga}, Matias},
        title = "{Uncertainties in the supermassive black hole abundance and implications for the GW background}",
      journal = {arXiv e-prints},
         year = 2025,
        month = sep,
          eid = {arXiv:2509.08041},
        pages = {arXiv:2509.08041},
          doi = {10.48550/arXiv.2509.08041},
archivePrefix = {arXiv},
       eprint = {2509.08041},
 primaryClass = {astro-ph.GA},
       adsurl = {https://ui.adsabs.harvard.edu/abs/2025arXiv250908041S}
}

@ARTICLE{deNicola+2025,
       author = {{de Nicola}, Stefano and {Thomas}, Jens and {Saglia}, Roberto P. and {Kluge}, Matthias and {Snigula}, Jan and {Bender}, Ralf},
        title = "{Eight New Ultramassive Black Hole Masses confirm Best Correlation with Galaxy Core Sizes}",
      journal = {arXiv e-prints},
         year = 2025,
        month = dec,
          eid = {arXiv:2512.04178},
        pages = {arXiv:2512.04178},
          doi = {10.48550/arXiv.2512.04178},
archivePrefix = {arXiv},
       eprint = {2512.04178},
 primaryClass = {astro-ph.GA},
       adsurl = {https://ui.adsabs.harvard.edu/abs/2025arXiv251204178D}
}

@ARTICLE{soltan1982,
       author = {{Soltan}, A.},
        title = "{Masses of quasars.}",
      journal = {\mnras},
         year = 1982,
        month = jul,
       volume = {200},
        pages = {115-122},
          doi = {10.1093/mnras/200.1.115},
       adsurl = {https://ui.adsabs.harvard.edu/abs/1982MNRAS.200..115S}
}

@ARTICLE{driver+2022,
       author = {{Driver}, Simon P. and {Bellstedt}, Sabine and {Robotham}, Aaron S.~G. and {Baldry}, Ivan K. and {Davies}, Luke J. and {Liske}, Jochen and {Obreschkow}, Danail and {Taylor}, Edward N. and {Wright}, Angus H. and {Alpaslan}, Mehmet and {Bamford}, Steven P. and {Bauer}, Amanda E. and {Bland-Hawthorn}, Joss and {Bilicki}, Maciej and {Bravo}, Mat{\'\i}as and {Brough}, Sarah and {Casura}, Sarah and {Cluver}, Michelle E. and {Colless}, Matthew and {Conselice}, Christopher J. and {Croom}, Scott M. and {de Jong}, Jelte and {D'Eugenio}, Franceso and {De Propris}, Roberto and {Dogruel}, Burak and {Drinkwater}, Michael J. and {Dvornik}, Andrej and {Farrow}, Daniel J. and {Frenk}, Carlos S. and {Giblin}, Benjamin and {Graham}, Alister W. and {Grootes}, Meiert W. and {Gunawardhana}, Madusha L.~P. and {Hashemizadeh}, Abdolhosein and {H{\"a}u{\ss}ler}, Boris and {Heymans}, Catherine and {Hildebrandt}, Hendrik and {Holwerda}, Benne W. and {Hopkins}, Andrew M. and {Jarrett}, Tom H. and {Heath Jones}, D. and {Kelvin}, Lee S. and {Koushan}, Soheil and {Kuijken}, Konrad and {Lara-L{\'o}pez}, Maritza A. and {Lange}, Rebecca and {L{\'o}pez-S{\'a}nchez}, {\'A}ngel R. and {Loveday}, Jon and {Mahajan}, Smriti and {Meyer}, Martin and {Moffett}, Amanda J. and {Napolitano}, Nicola R. and {Norberg}, Peder and {Owers}, Matt S. and {Radovich}, Mario and {Raouf}, Mojtaba and {Peacock}, John A. and {Phillipps}, Steven and {Pimbblet}, Kevin A. and {Popescu}, Cristina and {Said}, Khaled and {Sansom}, Anne E. and {Seibert}, Mark and {Sutherland}, Will J. and {Thorne}, Jessica E. and {Tuffs}, Richard J. and {Turner}, Ryan and {van der Wel}, Arjen and {van Kampen}, Eelco and {Wilkins}, Steve M.},
        title = "{Galaxy And Mass Assembly (GAMA): Data Release 4 and the z < 0.1 total and z < 0.08 morphological galaxy stellar mass functions}",
      journal = {\mnras},
         year = 2022,
        month = jun,
       volume = {513},
       number = {1},
        pages = {439-467},
          doi = {10.1093/mnras/stac472},
archivePrefix = {arXiv},
       eprint = {2203.08539},
 primaryClass = {astro-ph.GA},
       adsurl = {https://ui.adsabs.harvard.edu/abs/2022MNRAS.513..439D}
}

@ARTICLE{conselice+2022,
       author = {{Conselice}, Christopher J. and {Mundy}, Carl J. and {Ferreira}, Leonardo and {Duncan}, Kenneth},
        title = "{A Direct Measurement of Galaxy Major and Minor Merger Rates and Stellar Mass Accretion Histories at Z < 3 Using Galaxy Pairs in the REFINE Survey}",
      journal = {\apj},
         year = 2022,
        month = dec,
       volume = {940},
       number = {2},
          eid = {168},
        pages = {168},
          doi = {10.3847/1538-4357/ac9b1a},
archivePrefix = {arXiv},
       eprint = {2207.03984},
 primaryClass = {astro-ph.GA},
       adsurl = {https://ui.adsabs.harvard.edu/abs/2022ApJ...940..168C}
}

@ARTICLE{Mayer+2007,
       author = {{Mayer}, L. and {Kazantzidis}, S. and {Madau}, P. and {Colpi}, M. and {Quinn}, T. and {Wadsley}, J.},
        title = "{Rapid Formation of Supermassive Black Hole Binaries in Galaxy Mergers with Gas}",
      journal = {Science},
         year = 2007,
        month = jun,
       volume = {316},
       number = {5833},
        pages = {1874},
          doi = {10.1126/science.1141858},
archivePrefix = {arXiv},
       eprint = {0706.1562},
 primaryClass = {astro-ph},
       adsurl = {https://ui.adsabs.harvard.edu/abs/2007Sci...316.1874M}
}

@ARTICLE{criswell+2026,
       author = {{Criswell}, Alexander W. and {Banagiri}, Sharan and {Delfavero}, Vera and {Bustamante-Rosell}, Maria Jose and {Taylor}, Stephen R. and {Rosati}, Robert},
        title = "{A Foundation for Gravitational-Wave Population Inference within the LISA Global Fit}",
      journal = {arXiv e-prints},
         year = 2026,
        month = apr,
          eid = {arXiv:2604.03390},
        pages = {arXiv:2604.03390},
          doi = {10.48550/arXiv.2604.03390},
archivePrefix = {arXiv},
       eprint = {2604.03390},
 primaryClass = {astro-ph.IM},
       adsurl = {https://ui.adsabs.harvard.edu/abs/2026arXiv260403390C}
}

@ARTICLE{enoki+2007,
       author = {{Enoki}, M. and {Nagashima}, M.},
        title = "{The Effect of Orbital Eccentricity on Gravitational Wave Background Radiation from Supermassive Black Hole Binaries}",
      journal = {Progress of Theoretical Physics},
         year = 2007,
        month = feb,
       volume = {117},
       number = {2},
        pages = {241-256},
          doi = {10.1143/PTP.117.241},
archivePrefix = {arXiv},
       eprint = {astro-ph/0609377},
 primaryClass = {astro-ph},
       adsurl = {https://ui.adsabs.harvard.edu/abs/2007PThPh.117..241E}
}

@ARTICLE{lapi+2026,
       author = {{Lapi}, Andrea and {Shankar}, Francesco and {Bosi}, Michele and {Roberts}, Daniel and {Fu}, Hao and {Varadarajan}, Karthik M. and {Boco}, Lumen},
        title = "{Semi-empirical framework of supermassive black hole evolution: highlighting a possible tension between demographics and gravitational wave background}",
      journal = {\jcap},
         year = 2026,
        month = feb,
       volume = {2026},
       number = {2},
          eid = {001},
        pages = {001},
          doi = {10.1088/1475-7516/2026/02/001},
archivePrefix = {arXiv},
       eprint = {2507.15436},
 primaryClass = {astro-ph.CO},
       adsurl = {https://ui.adsabs.harvard.edu/abs/2026JCAP...02..001L}
}

@ARTICLE{ng+26_cnm,
       author = {{Agarwal}, Nikita and {Agazie}, Gabriella and {Amosso}, Alessandra and {Anumarlapudi}, Akash and {Archibald}, Anne M. and {Arzoumanian}, Zaven and {Ashok}, Anjana and {Baier}, Jeremy G. and {Baker}, Paul T. and {Becsy}, Bence and {Blecha}, Laura and {Brazier}, Adam and {Brook}, Paul R. and {Burke-Spolaor}, Sarah and {Burnette}, Rand and {Case}, Robin and {Casey-Clyde}, J. Andrew and {Chang}, Yu-Ting and {Charisi}, Maria and {Chatterjee}, Shami and {Cohen}, Tyler and {Cordes}, James M. and {Cornish}, Neil J. and {Crawford}, Fronefield and {Cromartie}, H. Thankful and {Crowter}, Kathryn and {DeCesar}, Megan E. and {Demorest}, Paul B. and {Deng}, Heling and {Dey}, Lankeswar and {Dolch}, Timothy and {Doskoch}, Graham M. and {Ferrara}, Elizabeth C. and {Fiore}, William and {Fonseca}, Emmanuel and {Freedman}, Gabriel E. and {Gardiner}, Emiko C. and {Garver-Daniels}, Nate and {Gentile}, Peter A. and {Gersbach}, Kyle A. and {Glaser}, Joseph and {Good}, Deborah C. and {Gultekin}, Kayhan and {Gundersen}, Aiden and {Harris}, C.~J. and {Hashemi Asl}, Doa and {Hazboun}, Jeffrey S. and {Jennings}, Ross J. and {Johnson}, Aaron D. and {Jones}, Megan L. and {Kaplan}, David L. and {Sreekumar}, Anala K. and {Kelley}, Luke Zoltan and {Kerr}, Matthew and {Key}, Joey S. and {Laal}, Nima and {Lam}, Michael T. and {Lamb}, William G. and {Larsen}, Bjorn and {Lazio}, T. Joseph W. and {Lewandowska}, Natalia and {Liu}, Tingting and {Lorimer}, Duncan R. and {Luo}, Jing and {Lynch}, Ryan S. and {Ma}, Chung-Pei and {Madison}, Dustin R. and {Martsen}, Ashley and {Matt}, Cayenne and {McEwen}, Alexander and {McKee}, James W. and {McLaughlin}, Maura A. and {McMann}, Natasha and {Meyers}, Bradley W. and {Meyers}, Patrick M. and {Miles}, Matthew T. and {Mingarelli}, Chiara M.~F. and {Mitridate}, Andrea and {Ng}, Cherry and {Nice}, David J. and {Nichols}, Shania and {Ocker}, Stella K. and {Oliver}, Daniel J. and {Olum}, Ken D. and {Pennucci}, Timothy T. and {Perera}, Benetge B.~P. and {Petrov}, Polina and {Pol}, Nihan S. and {Radovan}, Henri A. and {Ransom}, Scott M. and {Ray}, Paul S. and {Romano}, Joseph D. and {Runnoe}, Jessie C. and {Saffer}, Alexander and {Sardesai}, Shashwat C. and {Schmiedekamp}, Ann and {Schmiedekamp}, Carl and {Schmitz}, Kai and {Schult}, Levi and {Shapiro-Albert}, Brent J. and {Siemens}, Xavier and {Simon}, Joseph and {Sosa Fiscella}, Sophia V. and {Stairs}, Ingrid H. and {Stinebring}, Daniel R. and {Stovall}, Kevin and {Strahler}, Robin and {Susobhanan}, Abhimanyu and {Swiggum}, Joseph K. and {Taylor}, Jacob and {Taylor}, Stephen R. and {Thompson}, Mercedes S. and {Turner}, Jacob E. and {Vallisneri}, Michele and {van Haasteren}, Rutger and {Verbiest}, Joris P.~W. and {Vigeland}, Sarah J. and {Wahl}, Haley M. and {Wayt}, Kalista and {Wilson}, Kevin P. and {Witt}, Caitlin A. and {Wright}, David and {Young}, Olivia},
        title = "{The NANOGrav 15 yr Data Set: Impacts of Customized Chromatic Noise Models on Gravitational Wave Analyses}",
      journal = {arXiv e-prints},
         year = 2026,
        month = jun,
          eid = {arXiv:2606.28554},
        pages = {arXiv:2606.28554},
          doi = {10.48550/arXiv.2606.28554},
archivePrefix = {arXiv},
       eprint = {2606.28554},
 primaryClass = {astro-ph.CO},
       adsurl = {https://ui.adsabs.harvard.edu/abs/2026arXiv260628554A}
}

@ARTICLE{huber+2025,
       author = {{Huber}, Maggie C. and {Simon}, Joseph and {Comerford}, Julia M.},
        title = "{The Impact of Applying Black Hole─Host Galaxy Scaling Relations to Large Galaxy Populations}",
      journal = {\apj},
         year = 2025,
        month = jul,
       volume = {988},
       number = {1},
          eid = {90},
        pages = {90},
          doi = {10.3847/1538-4357/ade30a},
archivePrefix = {arXiv},
       eprint = {2506.08102},
 primaryClass = {astro-ph.GA},
       adsurl = {https://ui.adsabs.harvard.edu/abs/2025ApJ...988...90H}
}

@ARTICLE{huber+2026,
       author = {{Huber}, Maggie C. and {Comerford}, Julia M. and {Simon}, Joseph},
        title = "{Overmassive Supermassive Black Holes in SDSS Close Galaxy Pairs}",
      journal = {\apj},
         year = 2026,
        month = sep,
       volume = {1008},
       number = {2},
          eid = {204},
        pages = {204},
          doi = {10.3847/1538-4357/ae975b},
archivePrefix = {arXiv},
       eprint = {2608.27730},
 primaryClass = {astro-ph.GA},
       adsurl = {https://ui.adsabs.harvard.edu/abs/2026ApJ..1008..204H}
}

@ARTICLE{ng+26_targeted,
       author = {{Agarwal}, Nikita and {Agazie}, Gabriella and {Anumarlapudi}, Akash and {Archibald}, Anne M. and {Arzoumanian}, Zaven and {Baier}, Jeremy G. and {Baker}, Paul T. and {B{\'e}csy}, Bence and {Blecha}, Laura and {Brazier}, Adam and {Brook}, Paul R. and {Burke-Spolaor}, Sarah and {Burnette}, Rand and {Case}, Robin and {Casey-Clyde}, J. Andrew and {Chang}, Yu-Ting and {Charisi}, Maria and {Chatterjee}, Shami and {Cohen}, Tyler and {Coppi}, Paolo and {Cordes}, James M. and {Cornish}, Neil J. and {Crawford}, Fronefield and {Cromartie}, H. Thankful and {Crowter}, Kathryn and {Decesar}, Megan E. and {Demorest}, Paul B. and {Deng}, Heling and {Dey}, Lankeswar and {Dolch}, Timothy and {D'Orazio}, Daniel J. and {Eisenberg}, Ellis and {Ferrara}, Elizabeth C. and {Doskoch}, Graham and {Fiore}, William and {Fonseca}, Emmanuel and {Freedman}, Gabriel E. and {Gardiner}, Emiko C. and {Garver-Daniels}, Nate and {Gentile}, Peter A. and {Gersbach}, Kyle A. and {Glaser}, Joseph and {Graham}, Matthew J. and {Good}, Deborah C. and {G{\"u}ltekin}, Kayhan and {Harris}, C.~J. and {Hazboun}, Jeffrey S. and {Hutchison}, Forrest and {Jennings}, Ross J. and {Johnson}, Aaron D. and {Jones}, Megan L. and {Kaplan}, David L. and {Kelley}, Luke Zoltan and {Kerr}, Matthew and {Key}, Joey S. and {Laal}, Nima and {Lam}, Michael T. and {Lamb}, William G. and {Larsen}, Bjorn and {Lazio}, T. Joseph W. and {Lewandowska}, Natalia and {Liu}, Tingting and {Lorimer}, Duncan R. and {Luo}, Jing and {Lynch}, Ryan S. and {Ma}, Chung-Pei and {Madison}, Dustin R. and {Matt}, Cayenne and {McEwen}, Alexander and {McKee}, James W. and {McLaughlin}, Maura A. and {McMann}, Natasha and {Meyers}, Bradley W. and {Meyers}, Patrick M. and {Mingarelli}, Chiara M.~F. and {Mitridate}, Andrea and {Natarajan}, Priyamvada and {Ng}, Cherry and {Nice}, David J. and {Nichols}, Shania and {Ocker}, Stella Koch and {Olum}, Ken D. and {Pennucci}, Timothy T. and {Perera}, Benetge B.~P. and {Petrov}, Polina and {Pol}, Nihan S. and {Radovan}, Henri A. and {Ransom}, Scott M. and {Ray}, Paul S. and {Romano}, Joseph D. and {Runnoe}, Jessie C. and {Saffer}, Alexander and {Sardesai}, Shashwat C. and {Schmiedekamp}, Ann and {Schmiedekamp}, Carl and {Schmitz}, Kai and {Semenzato}, Federico and {Shapiro-Albert}, Brent J. and {Shivakumar}, Rohan and {Siemens}, Xavier and {Simon}, Joseph and {Sosa Fiscella}, Sophia V. and {Stairs}, Ingrid H. and {Stinebring}, Daniel R. and {Stovall}, Kevin and {Susobhanan}, Abhimanyu and {Swiggum}, Joseph K. and {Taylor}, Jacob A. and {Taylor}, Stephen R. and {Thompson}, Mercedes S. and {Turner}, Jacob E. and {Vallisneri}, Michele and {van Haasteren}, Rutger and {Vigeland}, Sarah J. and {Wahl}, Haley M. and {Willson}, London and {Wilson}, Kevin P. and {Witt}, Caitlin A. and {Wright}, David and {Young}, Olivia and {Zheng}, Qinyuan and {Nanograv Collaboration}},
        title = "{The NANOGrav 15 yr Dataset: Targeted Searches for Supermassive Black Hole Binaries}",
      journal = {\apjl},
         year = 2026,
        month = feb,
       volume = {998},
       number = {1},
          eid = {L11},
        pages = {L11},
          doi = {10.3847/2041-8213/ae3719},
archivePrefix = {arXiv},
       eprint = {2508.16534},
 primaryClass = {astro-ph.HE},
       adsurl = {https://ui.adsabs.harvard.edu/abs/2026ApJ...998L..11A}
}

@ARTICLE{blecha2026,
       author = {{Blecha}, Laura},
        title = "{An ``inside-out'' approach to modeling supermassive black hole binary inspiral}",
      journal = {arXiv e-prints},
         year = 2026,
        month = aug,
          eid = {arXiv:2608.06269},
        pages = {arXiv:2608.06269},
          doi = {10.48550/arXiv.2608.06269},
archivePrefix = {arXiv},
       eprint = {2608.06269},
 primaryClass = {astro-ph.HE},
       adsurl = {https://ui.adsabs.harvard.edu/abs/2026arXiv260806269B}
}

@ARTICLE{lauer+2007,
       author = {{Lauer}, Tod R. and {Tremaine}, Scott and {Richstone}, Douglas and {Faber}, S.~M.},
        title = "{Selection Bias in Observing the Cosmological Evolution of the M$_{{\ensuremath{\bullet}}}$-{\ensuremath{\sigma}} and M$_{{\ensuremath{\bullet}}}$-L Relationships}",
      journal = {\apj},
         year = 2007,
        month = nov,
       volume = {670},
       number = {1},
        pages = {249-260},
          doi = {10.1086/522083},
archivePrefix = {arXiv},
       eprint = {0705.4103},
 primaryClass = {astro-ph},
       adsurl = {https://ui.adsabs.harvard.edu/abs/2007ApJ...670..249L}
}

@ARTICLE{snyder+2017,
       author = {{Snyder}, Gregory F. and {Lotz}, Jennifer M. and {Rodriguez-Gomez}, Vicente and {Guimar{\~a}es}, Renato da Silva and {Torrey}, Paul and {Hernquist}, Lars},
        title = "{Massive close pairs measure rapid galaxy assembly in mergers at high redshift}",
      journal = {\mnras},
         year = 2017,
        month = jun,
       volume = {468},
       number = {1},
        pages = {207-216},
          doi = {10.1093/mnras/stx487},
archivePrefix = {arXiv},
       eprint = {1610.01156},
 primaryClass = {astro-ph.GA},
       adsurl = {https://ui.adsabs.harvard.edu/abs/2017MNRAS.468..207S}
}

@ARTICLE{quinlan1996,
       author = {{Quinlan}, Gerald D.},
        title = "{The dynamical evolution of massive black hole binaries I. Hardening in a fixed stellar background}",
      journal = {\na},
         year = 1996,
        month = jul,
       volume = {1},
       number = {1},
        pages = {35-56},
          doi = {10.1016/S1384-1076(96)00003-6},
archivePrefix = {arXiv},
       eprint = {astro-ph/9601092},
 primaryClass = {astro-ph},
       adsurl = {https://ui.adsabs.harvard.edu/abs/1996NewA....1...35Q}
}

@ARTICLE{sesana+khan2015,
       author = {{Sesana}, Alberto and {Khan}, Fazeel Mahmood},
        title = "{Scattering experiments meet N-body - I. A practical recipe for the evolution of massive black hole binaries in stellar environments}",
      journal = {\mnras},
         year = 2015,
        month = nov,
       volume = {454},
       number = {1},
        pages = {L66-L70},
          doi = {10.1093/mnrasl/slv131},
archivePrefix = {arXiv},
       eprint = {1505.02062},
 primaryClass = {astro-ph.GA},
       adsurl = {https://ui.adsabs.harvard.edu/abs/2015MNRAS.454L..66S}
}

@ARTICLE{khan+2018,
       author = {{Khan}, Fazeel Mahmood and {Berczik}, Peter and {Just}, Andreas},
        title = "{Gravitational wave driven mergers and coalescence time of supermassive black holes}",
      journal = {\aap},
         year = 2018,
        month = jul,
       volume = {615},
          eid = {A71},
        pages = {A71},
          doi = {10.1051/0004-6361/201730489},
archivePrefix = {arXiv},
       eprint = {1803.11394},
 primaryClass = {astro-ph.GA},
       adsurl = {https://ui.adsabs.harvard.edu/abs/2018A&A...615A..71K}
}

@ARTICLE{holleybockelmann+2025,
       author = {{Holley-Bockelmann}, Kelly and {Khan}, Fazeel Mahmood and {Williams}, Isaiah and {Roth}, Jaelyn and {Rizzo Smith}, Michael and {Porter}, Kaitlin and {Bellovary}, Jillian and {Derdzinski}, Andrea and {Macci{\`o}}, Andrea V.},
        title = "{Handy Relation between Binary Black Hole Merger Times and Host Galaxy Properties}",
      journal = {\apjl},
         year = 2025,
        month = dec,
       volume = {995},
       number = {1},
          eid = {L32},
        pages = {L32},
          doi = {10.3847/2041-8213/ae1ccd},
archivePrefix = {arXiv},
       eprint = {2508.14253},
 primaryClass = {astro-ph.GA},
       adsurl = {https://ui.adsabs.harvard.edu/abs/2025ApJ...995L..32H}
}

@ARTICLE{khan+2016,
       author = {{Khan}, Fazeel Mahmood and {Fiacconi}, Davide and {Mayer}, Lucio and {Berczik}, Peter and {Just}, Andreas},
        title = "{Swift Coalescence of Supermassive Black Holes in Cosmological Mergers of Massive Galaxies}",
      journal = {\apj},
         year = 2016,
        month = sep,
       volume = {828},
       number = {2},
          eid = {73},
        pages = {73},
          doi = {10.3847/0004-637X/828/2/73},
archivePrefix = {arXiv},
       eprint = {1604.00015},
 primaryClass = {astro-ph.GA},
       adsurl = {https://ui.adsabs.harvard.edu/abs/2016ApJ...828...73K}
}

@ARTICLE{yu2002,
       author = {{Yu}, Qingjuan},
        title = "{Evolution of massive binary black holes}",
      journal = {\mnras},
         year = 2002,
        month = apr,
       volume = {331},
       number = {4},
        pages = {935-958},
          doi = {10.1046/j.1365-8711.2002.05242.x},
archivePrefix = {arXiv},
       eprint = {astro-ph/0109530},
 primaryClass = {astro-ph},
       adsurl = {https://ui.adsabs.harvard.edu/abs/2002MNRAS.331..935Y}
}

@article{BayesHopper,
   title={Joint search for isolated sources and an unresolved confusion background in pulsar timing array data},
   volume={37},
   ISSN={1361-6382},
   url={http://dx.doi.org/10.1088/1361-6382/ab8bbd},
   DOI={10.1088/1361-6382/ab8bbd},
   number={13},
   journal={Classical and Quantum Gravity},
   publisher={IOP Publishing},
   author={B{\'e}csy, Bence and Cornish, Neil J.},
   year={2020},
   month=June, pages={135011} }

@ARTICLE{laal+2026,
       author = {{Laal}, Nima and {Taylor}, Stephen R. and {Matt}, Cayenne and {G{\"u}ltekin}, Kayhan},
        title = "{Multimessenger Probes of the Supermassive Black Hole Binary Population: The Role of Pulsar Timing Arrays}",
      journal = {\apj},
         year = 2026,
        month = jun,
       volume = {1004},
       number = {1},
          eid = {90},
        pages = {90},
          doi = {10.3847/1538-4357/ae6cea},
archivePrefix = {arXiv},
       eprint = {2512.11981},
 primaryClass = {astro-ph.HE},
       adsurl = {https://ui.adsabs.harvard.edu/abs/2026ApJ..1004...90L}
}

@ARTICLE{houlden+2026,
       author = {{Houlden}, Blaze and {Thrane}, Eric and {Auchettl}, Katie},
        title = "{The properties of the first continuous gravitational waves: Optimizing pulsar timing observations for supermassive black hole binary detection}",
      journal = {arXiv e-prints},
         year = 2026,
        month = aug,
          eid = {arXiv:2608.25715},
        pages = {arXiv:2608.25715},
          doi = {10.48550/arXiv.2608.25715},
archivePrefix = {arXiv},
       eprint = {2608.25715},
 primaryClass = {astro-ph.HE},
       adsurl = {https://ui.adsabs.harvard.edu/abs/2026arXiv260825715H}
}

@ARTICLE{lamb+2026,
       author = {{Lamb}, William G. and {Wachter}, Jeremy M. and {Mitridate}, Andrea and {Sardesai}, Shashwat C. and {B{\'e}csy}, Bence and {Hagen}, Emily L. and {Taylor}, Stephen R. and {Kelley}, Luke Zoltan},
        title = "{Finite populations and finite time: The non-Gaussianity of a gravitational wave background}",
      journal = {\prd},
         year = 2026,
        month = jun,
       volume = {113},
       number = {12},
          eid = {123065},
        pages = {123065},
          doi = {10.1103/96zk-qtck},
archivePrefix = {arXiv},
       eprint = {2511.09659},
 primaryClass = {gr-qc},
       adsurl = {https://ui.adsabs.harvard.edu/abs/2026PhRvD.113l3065L}
}

@ARTICLE{xue+2025,
       author = {{Xue}, Xiao and {Pan}, Zhen and {Dai}, Liang},
        title = "{Non-Gaussian statistics of nanohertz stochastic gravitational waves}",
      journal = {\prd},
         year = 2025,
        month = feb,
       volume = {111},
       number = {4},
          eid = {043022},
        pages = {043022},
          doi = {10.1103/PhysRevD.111.043022},
archivePrefix = {arXiv},
       eprint = {2409.19516},
 primaryClass = {astro-ph.CO},
       adsurl = {https://ui.adsabs.harvard.edu/abs/2025PhRvD.111d3022X}
}

@ARTICLE{iemoto+2026,
       author = {{Iemoto}, Kanyuni and {Goncharov}, Boris and {Sato-Polito}, Gabriela and {Bi}, Xiaoming},
        title = "{Bridging the Population Synthesis of Supermassive Binary Black Holes and the Gravitational Wave Background}",
      journal = {arXiv e-prints},
         year = 2026,
        month = sep,
          eid = {arXiv:2609.07703},
        pages = {arXiv:2609.07703},
archivePrefix = {arXiv},
       eprint = {2609.07703},
 primaryClass = {astro-ph.HE},
       adsurl = {https://ui.adsabs.harvard.edu/abs/2026arXiv260907703I}
}

@ARTICLE{laal+2025,
       author = {{Laal}, Nima and {Taylor}, Stephen R. and {Kelley}, Luke Zoltan and {Simon}, Joseph and {G{\"u}ltekin}, Kayhan and {Wright}, David and {B{\'e}csy}, Bence and {Casey-Clyde}, J. Andrew and {Chen}, Siyuan and {Cingoranelli}, Alexander and {D'Orazio}, Daniel J. and {Gardiner}, Emiko C. and {Lamb}, William G. and {Matt}, Cayenne and {Siwek}, Magdalena S. and {Wachter}, Jeremy M.},
        title = "{Deep Neural Emulation of the Supermassive Black Hole Binary Population}",
      journal = {\apj},
         year = 2025,
        month = mar,
       volume = {982},
       number = {1},
          eid = {55},
        pages = {55},
          doi = {10.3847/1538-4357/adb4ef},
archivePrefix = {arXiv},
       eprint = {2411.10519},
 primaryClass = {astro-ph.IM},
       adsurl = {https://ui.adsabs.harvard.edu/abs/2025ApJ...982...55L}
}

@ARTICLE{mingarelli+2017,
       author = {{Mingarelli}, Chiara M.~F. and {Lazio}, T. Joseph W. and {Sesana}, Alberto and {Greene}, Jenny E. and {Ellis}, Justin A. and {Ma}, Chung-Pei and {Croft}, Steve and {Burke-Spolaor}, Sarah and {Taylor}, Stephen R.},
        title = "{The local nanohertz gravitational-wave landscape from supermassive black hole binaries}",
      journal = {Nature Astronomy},
         year = 2017,
        month = nov,
       volume = {1},
        pages = {886-892},
          doi = {10.1038/s41550-017-0299-6},
archivePrefix = {arXiv},
       eprint = {1708.03491},
 primaryClass = {astro-ph.GA},
       adsurl = {https://ui.adsabs.harvard.edu/abs/2017NatAs...1..886M}
}

@article{lamb2023rapid,
  title={Rapid refitting techniques for Bayesian spectral characterization of the gravitational wave background using pulsar timing arrays},
  author={Lamb, William G and Taylor, Stephen R and van Haasteren, Rutger},
  journal={Physical Review D},
  volume={108},
  number={10},
  pages={103019},
  year={2023},
  publisher={APS}
}

@article{kl_divergence,
 ISSN = {00034851},
 URL = {http://www.jstor.org/stable/2236703},
 author = {S. Kullback and R. A. Leibler},
 journal = {The Annals of Mathematical Statistics},
 number = {1},
 pages = {79--86},
 publisher = {Institute of Mathematical Statistics},
 title = {On Information and Sufficiency},
 urldate = {2026-09-22},
 volume = {22},
 year = {1951}
}

@ARTICLE{tomson+2026,
       author = {{Tomson}, Sharon Mary and {van Haasteren}, Rutger and {Goncharov}, Boris},
        title = "{Hierarchical Inference of the Supermassive Black Hole Binary Merger Rates from Joint Searches using Pulsar Timing Arrays}",
      journal = {arXiv e-prints},
         year = 2026,
        month = sep,
          eid = {arXiv:2609.09086},
        pages = {arXiv:2609.09086},
          doi = {10.48550/arXiv.2609.09086},
archivePrefix = {arXiv},
       eprint = {2609.09086},
 primaryClass = {astro-ph.HE},
       adsurl = {https://ui.adsabs.harvard.edu/abs/2026arXiv260909086T}
}
